\documentclass[a4paper,10pt]{revtex4}
\usepackage{mathrsfs}
\usepackage{graphicx}
\usepackage{latexsym}
\usepackage{amsmath}
\usepackage{amssymb}
\usepackage{textcomp}
\usepackage{amsbsy}
\usepackage{graphics}
\usepackage{epstopdf}
\usepackage{color}

\allowdisplaybreaks[4]

\begin{document}

\tolerance=5000

\title{Antisymmetric tensor fields in modified gravity: a summary}

\author{Tanmoy~Paul$^{1,2}$\,\thanks{pul.tnmy9@gmail.com}} \affiliation{ $^{1)}$ Department of Physics, Chandernagore College, Hooghly - 712 136.\\
$^{(2)}$ Department of Theoretical Physics,\\
Indian Association for the Cultivation of Science,\\
2A $\&$ 2B Raja S.C. Mullick Road Kolkata - 700 032, India }

\tolerance=5000

\begin{abstract}
 We provide various aspects of second rank antisymmetric Kalb-Ramond (KR) field in modified theories of gravity. 
 The KR field energy density is found to decrease with the expansion of our universe 
 at a faster rate in comparison to radiation and matter components. 
 Thus as the Universe evolves and cools down, the contribution of the KR field on the evolutionary process reduces significantly, 
 and at present it almost does not affect the universe evolution. However the KR field has a significant 
 contribution during early universe, in particular, it affects the beginning of inflation as well as 
 increases the amount of primordial gravitational radiation and hence enlarges the value of tensor to scalar ratio in respect to the case 
 when the KR field is absent. In regard to the KR field couplings, it turns out that in four dimensional higher curvature inflationary 
 model the couplings of the KR field to other matter fields 
 is given by $1/M_{Pl}$ (where $M_{Pl}$ is known as the ``reduced Planck mass'' defined by 
 $M_{Pl} = \frac{1}{\sqrt{8\pi G}}$ with $G$ is the ``Newton's 
 constant'') i.e same as the usual gravity-matter coupling; however in the context of higher dimensional higher curvature model 
 the KR couplings get an additional suppression over $1/M_{Pl}$. Thus in comparison to the four dimensional model, the higher curvature braneworld scenario 
 gives a better explanation of why the present universe carries practically no footprint of the Kalb-Ramond field. The 
 higher curvature term in the higher dimensional gravitational action acts as a suitable stabilizing agent in the dynamical stabilization 
 mechanism of the extra dimensional modulus field from the perspective of effective on-brane theory. Based on the evolution of KR field, 
 one intriguing question can be - ``sitting in present day universe, how do we confirm the existence of the Kalb-Ramond field 
 which has considerably low energy density 
 (with respect to the other components) in our present universe but has a significant impact during early universe ?'' We try to answer this question 
 by the phenomena ``cosmological quantum entanglement'' which indeed carries the information of early universe. Finally, we briefly 
 discuss some future perspectives of Kalb-Ramond cosmology at the end of the paper.
\end{abstract}

\maketitle

\section{Introduction}
Current theoretical cosmology is faced with two problems - what is inflation and what is dark energy era or in other words, 
why does the universe undergo through an accelerated stage 
in early as well as in late time era ? What is the reason that our universe behaves in a similar way at very large and at very low curvature scale ? 
Apart from these problems, the initial Big-Bang singularity problem also hinges the cosmological sector. Although the inflationary scenario 
is successful to resolve the horizon, flatness problems and also in generating an almost scale invariant primordial power spectrum 
\cite{Guth:1980zm,Linde:1981mu,Albrecht:1982wi,Kinney:2003xf,Langlois:2004de,Riotto:2002yw,Barrow:1993ah,Barrow:1994nx,Mimoso:1994wn,Baumann:2009ds,
Nojiri:2007as,Sriramkumar:2009kg,Langlois:2010xc,Brandenberger:1997yf,Wang:2013zva}, it 
does not provide any satisfactory explanation in regard to this singularity problem and thus the bouncing cosmology gets a lot of attention, 
in which case the evolution of the universe becomes free of singularity with a scale invariant perturbation spectra (in some cases depending on the models) 
\cite{Brandenberger:2012zb,Brandenberger:2016vhg,Battefeld:2014uga,Novello:2008ra,Cai:2014bea,Ijjas:2018qbo,Cai:2013vm,Battefeld:2004mn,
Peter:2016kan,Cai:2012va,Odintsov:2015zza,Cai:2016thi,Cai:2017tku,Qiu:2010ch,Koehn:2015vvy}. 
However the mechanism of the bouncing cosmology has its own problems like 
the violation of energy condition, instability of the primordial perturbation etc.\\
Modified theories of gravity play a crucial role in resolving the above mentioned problems to a certain extend. In general, the modifications 
of Einstein's gravity can be broadly classified as - (1) by inclusion of higher curvature term in addition to the Einstein-Hilbert term in the gravitational 
action (see \cite{Nojiri:2010wj,Nojiri:2017ncd,Capozziello:2011et} for general reviews on higher curvature gravity theories), 
(2) by introducing some extra spatial dimension over our usual four dimensional spacetime 
\cite{Csaki:2004ay,Csaki:2005vy,Brax:2003fv,Maartens:2010ar,Whisker:2008kk,Brax:2004xh,Kim:2003pc} and (3) by introducing additional matter fields. 
One may interpret the higher curvature terms as the quantum corrections terms, the way they appear in String theory. 
Among the higher curvature models proposed so far, F(R) model, Gauss-Bonnet model or more generally Lanczos-Lovelock gravity catch a lot attraction 
due to their power in describing the early and late stage of the universe concomitantly. 
Some of the higher curvature models which are well known
to provide such a unified description of early and late-time
acceleration can be found in Refs.
\cite{Nojiri:2010wj,Nojiri:2017ncd,Capozziello:2011et,Artymowski:2014gea,
Nojiri:2003ft,Odintsov:2019mlf,Johnson:2019vwi,Pinto:2018rfg,Odintsov:2019evb,Nojiri:2019riz,Nojiri:2019fft,Lobo:2008sg,
Gorbunov:2010bn,Li:2007xn,Odintsov:2020nwm,Odintsov:2020iui,Appleby:2007vb,Elizalde:2010ts,Cognola:2007zu}
in the context of $F(R)$ gravity,
\cite{Li:2007jm,Odintsov:2018nch,Carter:2005fu,Nojiri:2019dwl,Elizalde:2010jx,Makarenko:2016jsy,delaCruzDombriz:2011wn,Chakraborty:2018scm,
Kanti:2015pda,Kanti:2015dra,Odintsov:2018zhw,Saridakis:2017rdo,Cognola:2006eg}
in the $f(R,\mathcal{G})$ gravity etc. Recently, the axion-$F(R)$
gravity model has been proposed in
\cite{Odintsov:2019evb,Odintsov:2020nwm}, where the axion field
mimics the dark matter evolution and hence the model provides a
description of dark matter along with the unification of early and
late-time acceleration eras. The higher curvature models also achieves a rich success in the context of bouncing cosmology as well 
\cite{Nojiri:2017ncd,Bamba:2014mya,Bamba:2014zoa,Odintsov:2015zza,Amoros:2014tha,Nojiri:2016ygo,Odintsov:2015ynk,Bamba:2013fha,
Haro:2015oqa,Helling:2009ia,Elizalde:2020zcb}. 
In general, inclusion of higher curvature terms 
lead to certain instability such as the Ostragradsky instability or the matter instability in the astrophysical sector. However the Gauss-Bonnet 
theory, a special case of the Lanczos-Lovelock gravity, is free from the Ostragradsky instability due to the special choice of the 
coefficients in front of the Ricci scalar, Ricci tensor and Reimann tensor respectively. Unlike to the Gauss-Bonnet theory, the F(R) model consists 
to any analytic function of Ricci scalar and may lead to the Ostragradsky instability, however there exists some F(R) models, in particular 
the models with $F'(R) > 0$ are free from such Ostragradsky instability. On other hand, a F(R) theory can be equivalently transformed to a 
scalar-tensor theory by a conformal transformation of the metric and the instability free condition of the F(R) model (i.e $F'(R) > 0$) is 
transferred to the scalar-tensor side by the condition of a positive scalar field kinetic energy. Regarding the modification by 
extra spatial dimension, spacetime having more than four dimensions is a natural conjecture in String theory. The higher dimensional model 
also arises in order to resolve the hierarchy problem i.e an apparent mismatch between the fundamental Planck scale and the 
electroweak symmetry breaking scale \cite{ArkaniHamed:1998rs,Antoniadis:1998ig,Randall:1999ee,Randall:1999vf,ArkaniHamed:1998nn,ArkaniHamed:1999gq}. 
In this regard, the Randall-Sundrum (RS) braneworld scenario \cite{Randall:1999ee} earned special attention as it resolves the 
gauge hierarchy problem without introducing any intermediate scale in the theory. The RS scenario consists an extra spatial 
dimension in addition to our usual four dimensional spacetime where the extra dimension is $S^1/Z_2$ compactified and moreover the orbifolded 
fixed points are identified with two 3-branes \cite{Randall:1999ee}. In RS braneworld model, the gravity is allowed to propagate 
in the bulk while the Standard model fields are generally considered to be confined on the visible 3-brane. Due to the intervening gravity, the branes 
must be collapsed and thus the modulus stabilization becomes necessary and an important aspect to address. According to the Goldberger-Wise 
mechanism, a stable potential term for the modulus field can fulfill the purpose of stabilization \cite{Goldberger:1999uk,Goldberger:1999un}. 
Here it may be mentioned that a bulk massive scalar field or higher curvature term(s) in the bulk can generate 
such stable modulus potential and thus may act as suitable stabilizing agent in the braneworld model \cite{Chakraborty:2016gpg,Das:2017htt}. 
This resulted into a large volume of work on phenomenological and cosmological implications of warped geometry model \cite{Csaki:2000zn,
DeWolfe:1999cp,Lesgourgues:2000tj,Csaki:1999mp,Binetruy:1999ut,Csaki:1999jh,Cline:1999yq,Nojiri:2000gv,Nojiri:2001ae,Banerjee:2017lxi,Das:2017jrl,
Davoudiasl:1999jd,Das:2015zxa,Tang:2012pv,Arun:2014dga,Das:2013lqa,Banerjee:2018kcz,Chakraborty:2013ipa}. 
However in the context of cosmology 
in the braneworld scenario, one needs to incorporate dynamical stabilization mechanism of the interbrane separation. One can indeed 
generalize the RS model by invoking a non-zero cosmological constant on the brane \cite{Das:2007qn}, unlike to the original RS one where the branes 
are assumed to be flat (some variants of RS model and its modulus stabilization are discussed in \cite{Csaki:2000zn,Banerjee:2017jyk,Paul:2016itm}). 
As mentioned earlier, apart 
from the two ways, the Einstein gravity can also be modified by some additional matter fields, in particular we will be mainly concerned 
about the effects of antisymmetric tensor fields \cite{Kalb:1974yc,Callan:1985ia,
Buchbinder:2008jf,Majumdar:1999jd,Mukhopadhyaya:2002jn,Mukhopadhyaya:2007jn,Das:2014asa,Das:2010xx,
DiGrezia:2003fe,Chakraborty:2014fva,Elizalde:2018rmz,Elizalde:2018now,Das:2018jey,Paul:2018ycm,Aashish:2020mlw,Aashish:2019ykb,
Aashish:2019zsy,Aashish:2018lhv,Aashish:2018aqn,Aashish:2018jzo,Do:2020ojg,Do:2018zac,DeRisi:2007dn,DeRisi:2008qw}. 
In general, antisymmetric tensor fields or equivalently p-forms, constitute the 
field content of all superstring models, and in effect these can actually have a realistic impact 
in the low-energy limit of the theory. In this context, the second rank antisymmetric 
tensor field, known as the Kalb-Ramond (KR) field \cite{Kalb:1974yc}, has been studied extensively. The Kalb-Ramond field 
arises as a massless representation of the underlying Lorentz group in the context of field theory and transforms as a two-form. 
The KR field can be thought as a sort of a generalization of electrodynamics: the gauge vector field in electrodynamics gets 
replaced by a second rank antisymmetric tensor field with a corresponding third rank antisymmetric 
field strength. Similar to the gauge invariance of the electromagnetic field action (i.e under 
$A_{\mu} \rightarrow A_{\mu} + \partial_{\mu}\omega$, where $A_{\mu}$ is the electromagnetic field and $\omega(x^{\mu})$ 
is an arbitrary function of spacetime 
coordinates), the rank-2 Kalb-Ramond field ($B_{\mu\nu}$) action 
is invariant under the transformation $B_{\mu\nu} \rightarrow B_{\mu\nu} + \partial_{[\mu}\Omega_{\nu]}$ (with $\Omega = \Omega(x^{\mu})$ being arbitrary). 
Interestingly, the corresponding third rank antisymmetric tensor field appearing as field 
strength of the Kalb-Ramond field has conceptual as well as mathematical similarity with the appearance 
of spacetime torsion and can be related 
to particle spins \cite{Hehl:1976kj,deSabbata:1994wi}. In particular if one assumes spacetime torsion to be antisymmetric in all the 
three indices then the decomposition of the Ricci scalar into parts dependent and independent of torsion 
leads to Einstein-Hilbert action with an additional term coinciding with the action for Kalb-Ramond 
field coupled to gravity. Here we would like to mention that terms (known as contorsion field) similar to the antisymmetric Kalb-Ramond ones 
can also be obtained in the context of torsional modified gravity (see \cite{Kofinas:2014owa,Kofinas:2014daa}), 
which is expected since the teleparallel/torsional 
formulation of gravity is already a gauge formulation of gravity. 
Apart from being a massless representation of the Lorentz group, the KR field also appears 
as a massless closed string mode in a heterotic string model. 
Beside the string theory point of view, the KR field also plays significant role in many other places as well, some of them are given by :
\begin{itemize}
\item Modified theories of gravity, formulated using twistors, require the inclusion of this antisymmetric tensor field \cite{Howe:1997pn,Howe:1996kj}.

\item Attempts to unify gravity and electromagnetism necessitates the inclusion of Kalb-Ramond field in higher-dimensional theories 
\cite{Kubyshin:1993wm,German:1993bq}.

\item Spacetime endowed with Kalb-Ramond field becomes optically active exhibiting birefringence \cite{Kar:2000ct,Kar:2001eb}. 

\item In \cite{Majumdar:1999jd} an antisymmetric tensor field $B_{\mu\nu}$ identified to be the Kalb-Ramond field was shown to act as the 
source of spacetime torsion. 

\end{itemize}

However a surprising feature in the present Universe is that there is no noticeable observable effect of antisymmetric tensor 
fields in any natural phenomena. More explicitly, our present universe carries observable footprints of scalar, fermion and vector 
degrees of freedom along with spin 2 symmetric tensor field (in the form of gravity), 
while there is no noticeable observable effect of any massless antisymmetric tensor modes. This observation raises some important questions like :
\begin{itemize}
 \item Why the present universe is practically free from the observable footprints of the higher rank antisymmetric tensor fields despite getting the 
 signatures of scalar, fermion, vector and spin-2 massless graviton, while they all originate from the same underlying Lorentz group ?
 
 \item Have the higher rank antisymmetric tensor fields a considerable impact during early stage of the universe, in particular during inflation 
 and on the inflationary parameters ?
\end{itemize}

In the present paper, we address these questions from modified gravity theories, both in four and five dimensional higher curvature spacetime. 
The higher dimensional model we consider is a generalized Randall-Sundrum braneworld scenario 
where the modulus becomes stabilized due to the presence of the higher curvature term in the gravitational action. In both four and five dimensional 
spacetime, we employ non-dynamical (where the fields have no dynamics at all) and dynamical (where the fields have ceratin dynamics which can be 
obtained by the corresponding solutions in FRW spacetime) method respectively. It turns out that 
the 2nd rank Kalb-Ramond (KR) field has a significant contribution during early universe, in particular, 
it affects the beginning of inflation as well as increases the amount of primordial gravitational radiation \cite{Elizalde:2018rmz,Elizalde:2018now,
Das:2010xx,Aashish:2018lhv}. However as the Universe expands, the contribution of the KR field
on the evolutionary process reduces significantly, and at present it almost does not affect the universe evolution. 
Based on this evolution of the KR field, one immediate question that will arise in mind will be:
\begin{itemize}
 \item Sitting in present day universe, how do we confirm the existence of the Kalb-Ramond field 
 which has considerably low energy density 
 (with respect to the other components) in our present universe but has a significant impact during early universe ?
\end{itemize}
It is clear that the answer of this question is encrypted within such a late time phenomena which indeed carries the information 
of early universe. One such phenomena is ``cosmological quantum entanglement'' of a scalar field coupled with the KR field \cite{Paul:2020bdy}. 
Motivated by this, we also address the cosmological particle production and the quantum entanglement of the scalar field in the present review.\\ 

\section{Antisymmetric tensor fields in 4D higher curvature gravity}
The field strength tensor of any massless rank $n$ antisymmetric tensor field $X_{a_{1}a_{2}....a_{n}}$ can be expressed as,
\begin{eqnarray}
 Y_{a_{1}a_{2}....a_{n+1}} = \partial_{[a_{n+1}}X_{a_{1}a_{2}....a_{n}]}
 \nonumber
\end{eqnarray}
Thereby in four dimensional spacetime, the rank of antisymmetric tensor field can at most be 3, beyond which the corresponding 
field strength tensor will vanish identically. Similarly, in the five dimensional case, the utmost rank of the antisymmetric tensor field will be $4$. 
We begin our discussion with rank two antisymmetric Kalb-Ramond (KR) field in four dimensional spacetime where, as mentioned earlier, we employ 
non-dynamical and dynamical methods to discuss various aspects of the KR field.\\

\subsection{Suppression of antisymmetric tensor fields: A non-dynamical way}\label{sec_non_dynamical_4d}
The subject of the present section is to provide a possible explanation of 
why the present universe is practically free from any perceptible signatures 
of the Kalb-Ramond field by a non-dynamical way in a F(R) gravity theory. 
For this purpose, the KR field is considered to be coupled with other matter fields, in particular 
with fermion and U(1) gauge field; actually these interaction terms play the key role to determine the 
observable signatures of the KR field in our universe. The action of massless KR field 
together with spin $\frac{1}{2}$ fermion and $U(1)$ gauge field in the background of $F(R)$ gravity in four dimension 
is given by \cite{Das:2018jey},
\begin{eqnarray}
 S = \int d^4x \sqrt{-g}\bigg[\frac{F(R)}{2\kappa^2} - \frac{1}{4}H_{\mu\nu\rho}H^{\mu\nu\rho} 
 + \bar{\Psi}i\gamma^{\mu}\partial_{\mu}\Psi 
 - \frac{1}{4}F_{\mu\nu}F^{\mu\nu} - \frac{1}{M_{Pl}}\bar{\Psi}\gamma^{\mu}\sigma^{\nu\rho}H_{\mu\nu\rho}\Psi 
 - \frac{1}{M_{Pl}}A^{[\mu}F^{\nu\rho]}H_{\mu\nu\rho}\bigg]
 \label{action1}
\end{eqnarray}
where $H_{\mu\nu\rho} = \partial_{[\mu}B_{\nu\rho]}$ is the field strength tensor of the Kalb-Ramond field $B_{\nu\rho}$ and 
$\kappa^2 = 8\pi G = \frac{1}{M_{Pl}^2}$ (where $G$ is the Newton's constant and $M_{Pl} \sim 10^{18}\mathrm{GeV}$ is known as the reduced Planck mass). 
The third and fourth terms of the action denote the kinetic Lagrangians of spin $\frac{1}{2}$ fermions $\Psi$ and $U(1)$ gauge 
field $A_{\mu}$, while the last two terms represent the interactions of KR field with the fermion and the gauge field respectively. 
The interaction between KR and U(1) electromagnetic field can be thought as it originates from Chern-Simons term which is incorporated to 
make the theory free from gauge anomaly. The above action or in particular the interactions of the KR field with the other matter fields 
are invariant under electromagnetic ($A_{\mu} \rightarrow A_{\mu} + \partial_{\mu}\omega$) and KR 
($B_{\mu\nu} \rightarrow B_{\mu\nu} + \partial_{[\mu}\Omega_{\nu]}$) gauge transformation. Such kind of interaction terms can also be found in 
\cite{Majumdar:1999jd,Mukhopadhyaya:2002jn,Das:2014asa}.\\
Performing a transformation of the metric as
\begin{equation}
 g_{\mu\nu}(x) \longrightarrow \widetilde{g}_{\mu\nu}(x) = e^{-\sqrt{\frac{2}{3}}\kappa \xi(x)}g_{\mu\nu}(x)
 \label{conformal}
\end{equation}
the above action can be transformed to a scalar-tensor action where the scalar field is endowed within a potential depending on the form of 
F(R). In particular, the scalar-tensor action is given by,
\begin{eqnarray}
 S&=&\int d^4x \sqrt{-\tilde{g}}\bigg[\frac{\widetilde{R}}{2\kappa^2} + \frac{1}{2}\tilde{g}^{\mu\nu}\partial_{\mu}\xi \partial_{\nu}\xi 
 - \bigg(\frac{ZF'(Z) - F(Z)}{2\kappa^2F'(Z)^2}\bigg) 
 - \frac{1}{4}e^{-\sqrt{\frac{2}{3}}\kappa \xi} H_{\mu\nu\rho}H_{\alpha\beta\delta} 
 \tilde{g}^{\mu\alpha}\tilde{g}^{\nu\beta}\tilde{g}^{\rho\delta} 
 - \frac{1}{4}F_{\mu\nu}F_{\alpha\beta} \tilde{g}^{\mu\alpha}\tilde{g}^{\nu\beta}\nonumber\\
 &+&e^{\sqrt{\frac{2}{3}}\kappa \xi} {\Psi}^{+}\tilde{\gamma}^{0}i\tilde{\gamma}^{\mu}\partial_{\mu}\Psi 
 - \frac{1}{M_{Pl}} {\Psi}^{+}\tilde{\gamma}^{0}\tilde{\gamma}^{\mu}\tilde{\sigma}^{\nu\rho}H_{\mu\nu\rho}\Psi 
 - \frac{1}{M_{Pl}} e^{-\sqrt{\frac{2}{3}}\kappa \xi} A_{[\alpha}F_{\beta\delta]}H_{\mu\nu\rho} 
 \tilde{g}^{\mu\alpha}\tilde{g}^{\nu\beta}\tilde{g}^{\rho\delta}\bigg]
 \label{action3}
\end{eqnarray}
where $\widetilde{R}$ is the Ricci scalar formed by $\widetilde{g}_{\mu\nu}$. To derive the above equation, we use the transformation of 
$\gamma^{\mu} \longrightarrow \widetilde{\gamma}^{\mu} = e^{\frac{1}{2}\sqrt{\frac{2}{3}}\kappa \xi} \gamma^{\mu}$ and 
$\sigma^{\nu\rho} \longrightarrow \widetilde{\sigma}^{\nu\rho} = e^{\sqrt{\frac{2}{3}}\kappa \xi} \sigma^{\nu\rho}$ induced by 
the conformal transformation of the metric. Actually such transformations of and $\gamma^{\mu}$ and $\sigma^{\nu\rho}$ follows from the reason 
that the gamma matrices satisfy the algebra $\{\gamma^{\mu},\gamma^{\nu}\} = 2g^{\mu\nu}$ and 
$\sigma^{\nu\rho}$ is the commutator of gamma matrices, in particular $\sigma^{\nu\rho} = \frac{i}{2}[\gamma^{\nu},\gamma^{\rho}]$. 
The field $\xi(x)$ in the action (\ref{action3}) acts a scalar field 
endowed with a potential $\frac{ZF'(Z) - F(Z)}{2\kappa^2F'(Z)^2}$ ($= V(Z(\xi))$, say) where 
$Z(x)$ is related to $\xi(x)$ as $F'(Z) = e^{-\sqrt{\frac{2}{3}}\kappa\xi}$. Being the dimension of 
the potential $V(Z)$ is [+4], the field $Z(x)$ has dimension [+2] in natural units. The appearance of the scalaron field 
$\xi(x)$ results the kinetic terms of the fermion and the KR field non-canonical, while the electromagnetic field is still canonical. The canonicity 
of electromagnetic field is due to the fact that the em field is conformally invariant in four dimensional spacetime. However, 
in order to make the kinetic terms canonical,we redefine the fields as, 
\begin{eqnarray}
 B_{\mu\nu} \longrightarrow \widetilde{B}_{\mu\nu} = e^{-\frac{1}{2}\sqrt{\frac{2}{3}}\kappa \xi} B_{\mu\nu}~~~~~~,~~~~~
 \Psi \longrightarrow \widetilde{\Psi} = e^{\frac{1}{2}\sqrt{\frac{2}{3}}\kappa \xi} \Psi~~~~~,~~~~~
 A_{\mu} \longrightarrow \widetilde{A}_{\mu} = A_{\mu}
 \label{ft_1}
 \end{eqnarray}
In terms of these redefined fields, the scalar-tensor action becomes canonical and is given by,
\begin{eqnarray}
 S&=&\int d^4x \sqrt{-\tilde{g}}\bigg[\frac{\widetilde{R}}{2\kappa^2} + \frac{1}{2}\tilde{g}^{\mu\nu}\partial_{\mu}\xi \partial_{\nu}\xi 
 - V(\xi) - \frac{1}{4} \tilde{H}_{\mu\nu\rho}\tilde{H}^{\mu\nu\rho}  
 - \frac{1}{4}\tilde{F}_{\mu\nu}\tilde{F}^{\mu\nu} + \bar{\tilde{\Psi}}i\tilde{\gamma}^{\mu}\partial_{\mu}\tilde{\Psi}\nonumber\\ 
 &-&\frac{1}{M_{Pl}} e^{(-\frac{1}{2}\sqrt{\frac{2}{3}}\kappa \xi)} 
 \bar{\tilde{\Psi}}\tilde{\gamma}^{\mu}\tilde{\sigma}^{\nu\rho}\tilde{H}_{\mu\nu\rho}\tilde{\Psi} 
 - \frac{1}{M_{Pl}} e^{(-\frac{1}{2}\sqrt{\frac{2}{3}}\kappa \xi)} \tilde{A}_{[\alpha}\tilde{F}_{\beta\delta]}\tilde{H}_{\mu\nu\rho} 
 + terms~proportional~to~\partial_{\mu}\xi\bigg]
 \label{action4}
\end{eqnarray}
It may be observed that the interaction terms (between $\tilde{B}_{\mu\nu}$ and $\tilde{\Psi}$, $\tilde{A_{\mu}}$) 
in the canonical scalar-tensor action (see Eq.(\ref{action4})) carries an exponential factor 
$e^{(-\frac{1}{2}\sqrt{\frac{2}{3}}\kappa \xi)}$ over the usual gravity-matter coupling $1/M_{Pl}$. Our main goal is to investigate whether 
such exponential factor (in front of the interaction terms) suppresses the coupling strengths of KR field with various matter fields 
which in turn may explain the invisibility of the Kalb-Ramond field on our present universe. For this purpose, we need a certain form of the scalar 
field potential $V(\xi)$ and recall, $V(\xi)$ in turn depends on the form of $F(R)$. However, the stability 
of $V(\xi)$ follows from the following two conditions on $F(R)$:
\begin{equation}
 \bigg[2F(R) - RF'(R)\bigg]_{\langle R\rangle} = 0~~~~~~~~~,~~~~~~~~\bigg[\frac{F'(R)}{F''(R)} - R\bigg]_{\langle R\rangle} > 0
 \label{stability condition1}
\end{equation}
In order to 
achieve an explicit expression of a stable scalar potential, we first consider the form of $F(R)$ as an exponential analytic function of 
Ricci scalar,
\begin{eqnarray}
 F(R) = R + \alpha \big(e^{-\beta R} - 1\big)
 \label{form}
\end{eqnarray}
where $\alpha$ and $\beta$ are the free parameters (with 'mass dimensions' $[\beta] = -2$ and $[\alpha] = +2$) of the theory. 
The above F(R) model is free of ghosts for $\alpha\beta < 1$ and it satisfies $\lim_{R\rightarrow0}[F(R)-R] = 0$, which indicates that there exists 
a flat spacetime solution. Moreover such exponential F(R) 
model is known to provide a good dark energy model for suitable parametric regimes of $\alpha$ and $\beta$ 
\cite{Odintsov:2017qif,Cognola:2007zu,Elizalde:2010ts}. 
The exponential correction over the usual Einstein-Hilbert action becomes important at cosmological 
scales and at late-times, providing an alternative to the dark energy problem. In addition, the exponential F(R) (\ref{form}) can be 
extended in such a way that is able to produce an early inflation as well (see \cite{Odintsov:2017qif} for the unified description of inflation 
and dark energy epochs from exponential F(R) gravity). Furthermore 
one more extension of this type of exponential gravity with log-corrected $R^2$ term was proposed to 
explain the unified universe history (see Ref.\cite{Odintsov:2017hbk}).\\ 
However, as we will illustrate later that apart from the $F(R)$ (\ref{form}), 
some other generic forms of $F(R)$ models are also able to fulfill the purpose in the present context. One of the F(R) forms is given by 
$F(R) = R + \omega R^2 + \rho\big[e^{-\sigma R}-1\big]$ which is, due to the presence of the term $\omega R^2$, a kind of generalization 
of the model considered in Eq.(\ref{form}).\\
For the specific choice of $F(R)$ shown in Eq.(\ref{form}), the potential $V(\xi)$ becomes,
\begin{eqnarray}
 V(\xi) = \frac{\alpha}{2\kappa^2} e^{(2\sqrt{\frac{2}{3}}\kappa\xi)} \bigg[\frac{(1-e^{-\sqrt{\frac{2}{3}}\kappa\xi})}{\alpha\beta} 
 \bigg(\ln{\bigg[\frac{1-e^{-\sqrt{\frac{2}{3}}\kappa\xi}}{\alpha\beta}}\bigg] + 1\bigg) + 1\bigg]
 \label{potential}
\end{eqnarray}
This potential has a minimum at 
\begin{eqnarray}
 \langle\xi\rangle = \frac{1}{\kappa} \sqrt{\frac{3}{2}}\ln{\bigg[\frac{1}{1-\alpha\beta}\bigg]}
 \label{vev}
\end{eqnarray}
Eq.(\ref{vev}) indicates that for a wide range of values of the product $\alpha\beta$ (between $0$ and $1$), 
the vacuum expectation value (vev) of the 
scalar field $\xi(x)$ becomes of the order of $\frac{1}{\kappa} = M_{Pl}$. This suggests that at the early epoch of universe, 
when the energy scale was high, the scalar field $\xi$ was a dynamical scalar degree of freedom giving rise to new 
interaction vertices with fermion and gauge fields. However as the universe evolved into a lower energy scale 
due to cosmological expansion, $\xi$ finally froze into its vacuum expectation value $\langle\xi\rangle$ as given in eqn.(\ref{vev}). 
The action in Eq.(\ref{action4}) therefore turns out to be,
\begin{eqnarray}
 S&=&\int d^4x \sqrt{-\tilde{g}}\bigg[\frac{\widetilde{R}}{2\kappa^2} - \frac{1}{4} \tilde{H}_{\mu\nu\rho}\tilde{H}^{\mu\nu\rho}  
 - \frac{1}{4}\tilde{F}_{\mu\nu}\tilde{F}^{\mu\nu} 
 + \bar{\tilde{\Psi}}i\tilde{\gamma}^{\mu}\partial_{\mu}\tilde{\Psi}\nonumber\\
 &-&\frac{1}{M_{Pl}} e^{(-\frac{1}{2}\sqrt{\frac{2}{3}}\kappa \langle\xi\rangle)} 
 \bar{\tilde{\Psi}}\tilde{\gamma}^{\mu}\tilde{\sigma}^{\nu\rho}\tilde{H}_{\mu\nu\rho}\tilde{\Psi} 
 - \frac{1}{M_{Pl}} e^{(-\frac{1}{2}\sqrt{\frac{2}{3}}\kappa \langle\xi\rangle)} \tilde{A}_{[\alpha}\tilde{F}_{\beta\delta]}\tilde{H}_{\mu\nu\rho}\bigg]
 \label{action5}
\end{eqnarray}
where $L_{int}\big(\partial_{\mu}\xi, \tilde{B}_{\mu\nu}, \partial_{\alpha}\tilde{B}_{\mu\nu}, \tilde{A}_{\mu}, \tilde{F}_{\mu\nu}, 
 \tilde{\Psi}, \partial_{\mu}\tilde{\Psi}\big)$ vanishes as $\xi(x)$ is frozen 
 at its vev. The last two terms in the above expression of action give the coupling of KR field to fermion, 
 $U(1)$ gauge field and are given by
 \begin{eqnarray}
  \lambda_{KR-fermion}&=&\frac{1}{M_{Pl}} e^{(-\frac{1}{2}\sqrt{\frac{2}{3}}\kappa \langle\xi\rangle)} = \frac{1}{M_{Pl}} \sqrt{1-\alpha\beta}\nonumber\\
  \lambda_{KR-U(1)}&=&\frac{1}{M_{Pl}} e^{(-\frac{1}{2}\sqrt{\frac{2}{3}}\kappa \langle\xi\rangle)} = \frac{1}{M_{Pl}} \sqrt{1-\alpha\beta}
  \label{coupling1_nd}
 \end{eqnarray}
 respectively. The above two equations clearly demonstrate that the product $\alpha\beta$ must be less than unity, otherwise the couplings 
 of KR field become imaginary, an unphysical situation. However the condition $\alpha\beta<1$ is also supported by the fact that 
 the higher curvature terms may have their origin from quantum corrections which from dimensional argument are suppressed 
 by Planck scale.\\ 
 Eq.(\ref{coupling1_nd}) indicate that the coupling strengths 
 of KR field to matter fields are suppressed over the 
 usual gravity-matter coupling strength $1/M_{Pl}$ by the factor $\sqrt{1-\alpha\beta}$ and hence the suppression increases 
 as the product $\alpha\beta$ approaches to unity. This may explain why the present universe is dominated by spacetime curvature and carries 
 practically no observable signature of the rank 2 antisymmetric Kalb-Ramond field (or equivalently the torsion field).\\
 Let us now consider the rank 3 antisymmetric tensor field $X_{\alpha\beta\rho}$ with the corresponding 'field 
 strength tensor' $Y_{\alpha\beta\rho\delta}$ ($= \partial_{[\alpha}X_{\beta\rho\delta]}$). The action for such a field 
 in four dimension is  
 \begin{eqnarray}
  S[X] = \int d^4x \sqrt{-g} Y_{\alpha\beta\rho\delta}Y^{\alpha\beta\rho\delta}
  \nonumber
 \end{eqnarray}
 Adopting the same procedure as for KR field, one can land with the coupling of the field $X$ to matter fields in the canonical 
 scalar-tensor action as: 
 \begin{eqnarray}
 \Omega_{X-fermion} = \frac{1}{M_{Pl}} \sqrt{1-\alpha\beta}~~~,\nonumber\\ 
 \Omega_{X-U(1)} = \frac{1}{M_{Pl}} \big(1-\alpha\beta\big)
 \label{coupling_higher_nd}
 \end{eqnarray}
 where $\Omega_{X-fermion}$ and $\Omega_{X-U(1)}$ denote the coupling between $X$-fermion and $X$-U(1) gauge field respectively. 
 Like the case of second rank Kalb-Ramond field, $\Omega_{X-fermion}$ and $\Omega_{X-U(1)}$ are also suppressed in comparison to $1/M_{Pl}$. 
 However it may be observed from Eqns.(\ref{coupling1_nd}), (\ref{coupling_higher_nd}) that 
 $\lambda_{KR-fermion}$ and $\Omega_{X-fermion}$ carry the same suppression factor while the interaction with the electromagnetic 
 field become progressively smaller with the increasing rank of the tensor field. For example, with $\alpha\beta = 0.99$, the couplings of higher rank 
 antisymmetric tensor fields to fermion are given by $\lambda_{KR-fermion} = \Omega_{X-fermion} = 10^{-1}/M_{Pl}$, while the corresponding couplings 
 to U(1) gauge field are $\lambda_{KR-U(1)} = 10^{-1}/M_{Pl}$ and $\Omega_{X-U(1)} = 10^{-2}/M_{Pl}$ i.e the interaction strengths of $X$-U(1) gauge field 
 is more suppressed in comparison to that of the $KR-U(1)$ gauge field. Therefore the visibility of an antisymmetric 
 tensor field in our present universe becomes lesser with the increasing rank of the tensor field.\\
 Eqns.(\ref{coupling1_nd}) and (\ref{coupling_higher_nd}) indicate that the suppression on the 
 coupling strengths of higher rank antisymmetric tensor fields 
 (i.e Kalb-Ramond and $X_{\alpha\beta\rho}$) to fermion, U(1) gauge fields (over $\frac{1}{M_{Pl}}$) depend on the factor $1-\alpha\beta$ and moreover 
 the suppression increases as $\alpha\beta$ approaches to unity. As an example, Eq.(\ref{coupling1_nd}) shows that for $1-\alpha\beta = 10^{-2}$, 
 the coupling of KR-U(1) gauge field is given by $10^{-1}/M_{Pl}$ while for $1-\alpha\beta = 10^{-4}$, $\lambda_{KR-U(1)}$ comes as 
 $= 10^{-2}/M_{Pl}$ i.e the KR-U(1) gauge field interaction strength becomes more suppressed as $\alpha\beta$ approaches to unity; 
 the similar argument also holds 
 for the other couplings of higher rank antisymmetric fields. Using Eqns.(\ref{coupling1_nd}) and (\ref{coupling_higher_nd}), we present the following 
 Table[\ref{Table-1}] which depicts $\lambda_{KR-fermion}$, $\lambda_{KR-U(1)}$, $\Omega_{X-fermion}$ and $\Omega_{X-U(1)}$ for a considerable range of 
 values of $\alpha\beta$ \cite{Das:2018jey}.\\
 
 \begin{table}[!h]
  \centering
  \resizebox{\columnwidth}{1.3 cm}{%
  \begin{tabular}{|c| c| c| c| c|}
   \hline 
   $(1-\alpha\beta)$  & $\lambda_{KR-fermion}$ & $\lambda_{KR-U(1)}$ & $\Omega_{X-fermion}$ & $\Omega_{X-U(1)}$\\
   \hline
   $10^{-2}$ & $10^{-1}/M_{Pl}$ & $10^{-1}/M_{Pl}$ & $10^{-1}/M_{Pl}$ & $10^{-2}/M_{Pl}$ \\
   $10^{-4}$ & $10^{-2}/M_{Pl}$ & $10^{-2}/M_{Pl}$ & $10^{-2}/M_{Pl}$ & $10^{-4}/M_{Pl}$\\
   $10^{-8}$ & $10^{-4}/M_{Pl}$ & $10^{-4}/M_{Pl}$ & $10^{-4}/M_{Pl}$ & $10^{-8}/M_{Pl}$\\
    \hline
  \end{tabular}%
  }
  \caption{The couplings of rank 2 (onwards) antisymmetric tensor field(s) to matter fields for a considerable range of $\alpha\beta$.}
  \label{Table-1}
 \end{table}
 At this stage, it deserves mentioning that beside the model presented in Eq.(\ref{form}), there exist several other $F(R)$ models 
 for which the intrinsic scalar degree of freedom suppresses the coupling strengths of antisymmetric tensor fields with matter fields 
 by an additional reduced factor over the gravity-matter coupling $1/M_{Pl}$. In fact any generic form of $F(R)$ 
 which corresponds to a stable scalar potential with a vev of Planck order will lead to the suppression of antisymmetric tensor fields 
 in the present context. Some such $F(R)$ models are : 
 (1) $F(R) = R - \gamma \ln{(R/\mu^2)} - \delta R^2$ \cite{Nojiri:2010wj} where $\gamma$, $\mu$ and $\delta$ are constants; 
 (2) $F(R) = a\big[e^{bR} - 1\big]$ with $a$, $b$ are constants; 
 (3) $F(R) = R + \omega R^2 + \rho\big[e^{-\sigma R}-1\big]$ ($\omega$, $\rho$, $\sigma$ 
 are constants). With appropriate choices of the parameters, these $F(R)$ models also lead to 
 the suppression on the coupling strengths between antisymmetric tensor fields 
 and various matter fields.\\
 In conclusion, we have discussed a non-dynamical way which provides a possible explanation for the negligible footprints of the antisymmetric 
 tensor fields in our present universe. In particular, the intrinsic scalar degree of freedom (also known as scalaron) 
 of F(R) gravity acquires a stable potential for suitable forms of F(R), 
 as for example, $F(R) = R + \alpha\big[e^{-\beta R} - 1\big]$, $F(R) = R - \gamma \ln{(R/\mu^2)} - \delta R^2$, 
 $F(R) = R + \omega R^2 + \rho\big[e^{-\sigma R}-1\big]$ etc. lead to a stable scalar field potential. It turns out that at present energy scale, the 
 scalaron field becomes frozen at its vacuum expectation value and as a result the coupling 
 of antisymmetric tensor fields to the matter fields get suppressed over the gravity-matter coupling strength $1/M_{Pl}$. 
 The suppression actually increases with the rank of the tensor field. This may well serve as an explanation why the present universe 
 carries practically no observable signatures of antisymmetric tensor fields.\\
 
 \subsection{A different non-dynamical method for the suppression of antisymmetric tensor fields by the ``scalaron tunneling''}\label{sec_nd2}
 
 Apart from the above procedure, a different non-dynamical method, in particular by the effect of the quantum tunneling of the scalaron field, was 
 proposed in the context of F(R) gravity to explain the imperceptible signatures of the antisymmetric tensor fields in the present universe 
 \cite{Paul:2018ycm}. The demonstration goes as follows : the scalar degree of freedom associated with F(R) gravity, known as scalaron field, is generally 
 endowed within a potential which in turn depends on the form of F(R). Recall, the stability of 
 the scalaron potential can be determined from $\big[2F(R) - RF'(R)\big]_{\langle R \rangle} = 0$ and the 
 maximum or minimum depend on the condition whether 
 the expression $\big[\frac{F'(R)}{F''(R)} - R\big]_{\langle R \rangle}$ is greater or less than zero. 
 Thus for a general class of polynomial 
 \begin{eqnarray}
 F(R) = R - \alpha R^2 + \beta R^m
 \label{form_nd2}
 \end{eqnarray}
 model with odd values of $m$ and $m \geq 3$, the scalaron potential contains three extrema. 
 In particular, for odd $m$, the scalaron potential corresponds to this F(R) has 
 two minima ($\langle\xi\rangle_\mathrm{-}$ and $\langle\xi\rangle_\mathrm{+}$) and one maximum ($\langle\xi\rangle_\mathrm{max}$ 
 in-between the two minima) at
 \begin{eqnarray}
  \langle\xi\rangle_\mathrm{-} = \sqrt{\frac{3}{2\kappa^2}} 
  \ln{\bigg[\frac{1}{\frac{2m-2}{m-2} + 2\alpha \big(\frac{1}{\beta(m-2)}\big)^{\frac{1}{m-1}}}\bigg]}~~~,~~~
  \langle\xi\rangle_\mathrm{+} = \sqrt{\frac{3}{2\kappa^2}} 
  \ln{\bigg[\frac{1}{\frac{2m-2}{m-2} - 2\alpha \big(\frac{1}{\beta(m-2)}\big)^{\frac{1}{m-1}}}\bigg]}~~~,~~~\langle\xi\rangle_\mathrm{max} = 0\nonumber\\
  \label{general_vev1}
 \end{eqnarray}
respectively with the parameters satisfy the condition $\alpha \big(\frac{1}{\beta(m-2)}\big)^{\frac{1}{m-1}} < (m-1)/(m-2)$. To obtain the 
above extrema of $V(\xi)$ in terms of $\xi$, one needs to use the relation $e^{-\sqrt{\frac{2}{3}}\kappa\xi} = F'(R)$. 
Moreover the scalaron potential has two zero at $\xi_1 = 0$ and 
$\xi_2 = \sqrt{\frac{3}{2\kappa^2}} \ln{\bigg[\frac{1}{1 - \alpha\big(\frac{m-2}{m-1}\big)\big(\frac{\alpha}{\beta(m-1)}\big)^{\frac{1}{m-2}}}\bigg]}$. 
Furthermore due to the aforementioned constraint $\alpha \big(\frac{1}{\beta(m-2)}\big)^{\frac{1}{m-1}} < (m-1)/(m-2)$, 
$V(\langle\xi\rangle_\mathrm{-})$ and $V(\langle\xi\rangle_\mathrm{+})$ 
become negative and also $\big|V(\langle\xi\rangle_\mathrm{+})\big| > \big|V(\langle\xi\rangle_\mathrm{-})\big|$. 
Thereby from stability criteria, if $\xi$ is at $\langle\xi\rangle_\mathrm{-}$, 
then in order to have a lower energy configuration, the scalar field has a non-zero probability to tunnel 
 from $\xi = \langle\xi\rangle_\mathrm{-}$ to $\xi = \langle\xi\rangle_\mathrm{+}$, which occur due to quantum fluctuation. 
 The tunneling probability has been determined and it is found to increase with the larger value of higher curvature parameter. 
 However it turns out that before the tunneling of the scalar field, 
 the interactions of antisymmetric tensor fields with matter fields (i.e with fermion and U(1) gauge field) 
 are same as usual gravity-matter coupling strength $1/M_{Pl}$, while after the tunneling, the couplings get severely suppressed over 
 $\frac{1}{M_{Pl}}$. Moreover by suitable parametric choices, the time scale for the scalaron 
 tunneling from $\langle\xi\rangle_\mathrm{-}$ to $\langle\xi\rangle_\mathrm{+}$ becomes at the order of the present age of the universe i.e 
\begin{eqnarray}
 \frac{\kappa}{T} \simeq 10^{17} sec. \simeq 10^{41} (GeV)^{-1}
 \nonumber
\end{eqnarray}
where $T$ is the tunneling probability the conversion $1$sec. $= 10^{24}\mathrm{GeV}^{-1}$ may be useful. 
For example, in $F(R) = R - \alpha R^2 + \beta R^3$ 
(i.e $m = 3$ in the general $F(R) = R - \alpha R^2 + \beta R^m$ form mentioned in Eq.(\ref{form_nd2})) model, 
$\frac{\sqrt{\beta}}{\alpha} = 0.53$ and $\beta = 2.43\times10^{-2}\kappa^4$ make 
$T^{-1} \simeq 10^{60}$ leading to the time scale of the scalaron tunneling at the order of the present age of our universe. 
Thereby one can argue that the effect of antisymmetric fields may be dominant in the early universe but as the universe evolves, 
the scalar field tunnels from $\xi = \langle\xi\rangle_\mathrm{-}$ to $\xi = \langle\xi\rangle_\mathrm{+}$ which in turn induces a 
suppression on the interaction strengths of antisymmetric fields in comparison to $1/M_{Pl}$. 
Moreover it is also found that the suppression increases with the increasing rank of the tensor field. This may provide a natural 
explanation why the present universe is free from any observable signatures of antisymmetric tensor fields.\\
Despite the success, this non-dynamical 
method (i.e the suppression of antisymmetric tensor fields in effect scalaron tunneling) has some $problems$, given by - 
in this method, the antisymmetric tensor fields become invisible due to the $F(R) = R - \alpha R^2 + \beta R^m$, let us consider 
$F(R) = R - \alpha R^2 + \beta R^3$ (i.e $m = 3$ case) involving a second and third powers of the curvature scalar with coefficients $\alpha$ and 
$\beta$. The $R^2$ term plays a role for renormalizability and a popular term in the context of F(R) gravity for its various cosmological and 
phenomenological implications. However the other one i.e the $R^3$ term is an unpopular one and 
never appears in the ordinary higher derivative extensions of 
General Relativity. In this sense, the appearance of the $R^3$ term hinges the motivation of the considered F(R) model. 
The scenario is different in the earlier non-dynamical method discussed in Sec.[\ref{sec_non_dynamical_4d}] 
where the F(R) has the form as considered in Eq.(\ref{form}) which is viable 
and have been used in the earlier literature of F(R) gravity \cite{Odintsov:2017qif,Cognola:2007zu,Elizalde:2010ts}. Moreover, although the 
$F(R) = R - \alpha R^2 + \beta R^m$ model explains the invisibility of the KR field in the present universe by a non-dynamical method, 
however from the perspective of the cosmological evolution of the universe, it is a bit strange to intuitively understand the result. 
In regard to the cosmological evolution of the universe, the term $R^m$ (where $m \geq 3$) becomes negligible in the low curvature regime, 
in particular $R^m$ goes as $H^{2m}$ (where $H$ is the Hubble parameter) and thus has almost no effect in the low curvature regime 
of the present universe; unlike to the F(R) considered in Eq.(\ref{form}) which indeed has considerable effects in the dark energy epoch of the 
universe.\\

Based on the above arguments, we may argue that the non-dynamical method (to explain the suppression of antisymmetric tensor fields 
in the present universe) discussed in Sec.[\ref{sec_non_dynamical_4d}] is more viable in comparison to that of discussed in Sec.[\ref{sec_nd2}].\\

\subsection{Cosmological scenario}\label{sec_dynamical4d}

\begin{figure}[!h]
\begin{center}
 \centering
 \includegraphics[width=3.5in,height=2.0in]{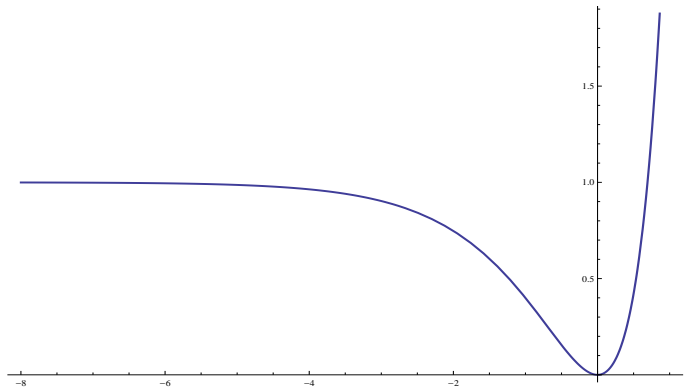}
 \caption{$V(\xi)$ vs $\xi$}
 \label{plot potential}
\end{center}
\end{figure}

\begin{figure}[!h]
\begin{center}
 \centering
 \includegraphics[width=3.8in,height=2.8in]{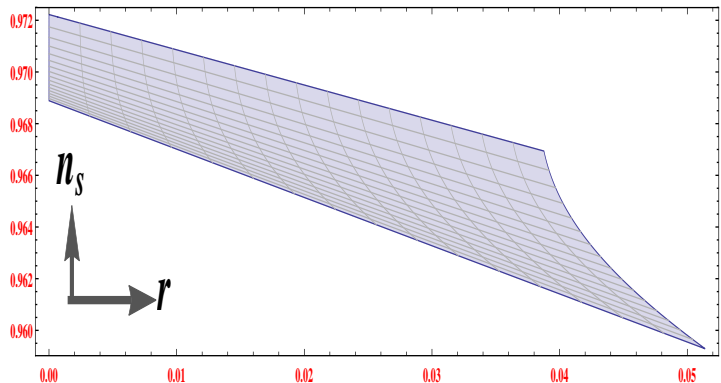}
 \caption{$n_s$ vs $r$ for $10 \leq |\sigma_0| \leq 14$ and $0.003 \leq \frac{\kappa^2h_0}{m^2} \leq 0.004$.}
 \label{plot dynamical1}
\end{center}
\end{figure}

\begin{figure}[!h]
\begin{center}
 \centering
 \includegraphics[width=3.3in,height=2.5in]{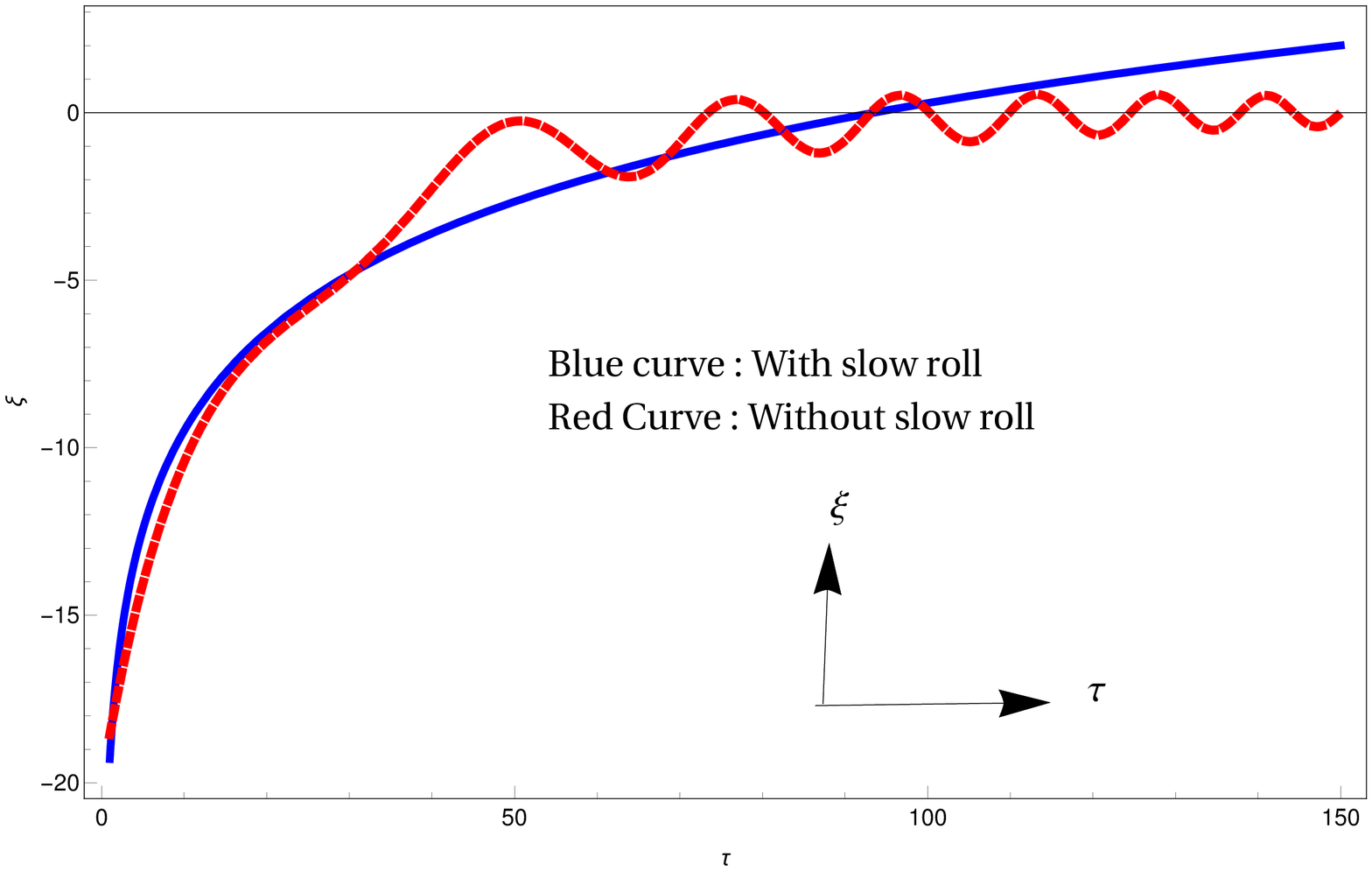}
 \includegraphics[width=3.3in,height=2.5in]{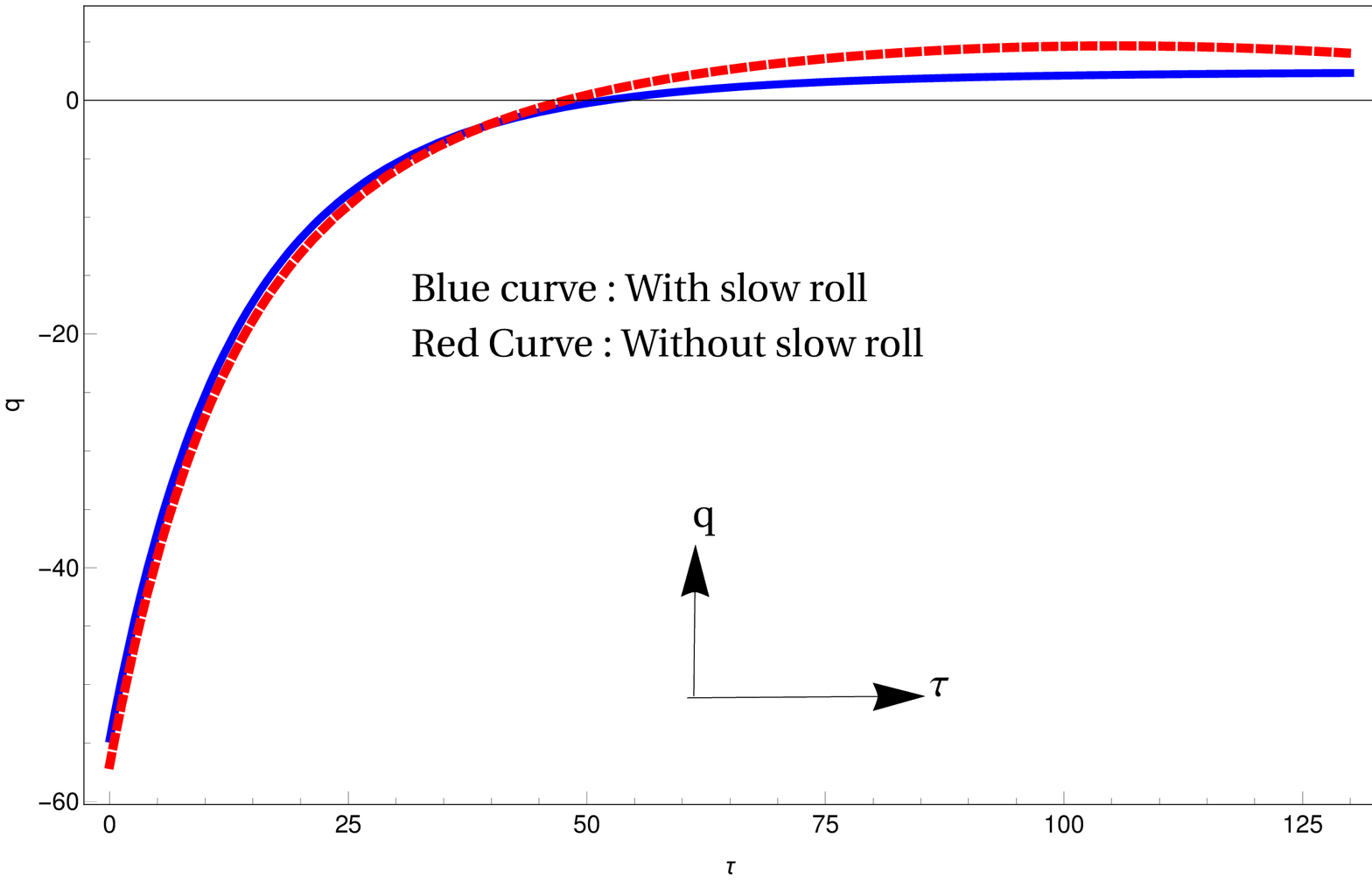}
 \caption{$Left~Figure$: $\xi$ vs $\tau$, $Right~Figure$: $q$ vs $\tau$. In both the figures, we take 
 $\frac{\kappa^2h_0}{m^2} = 0.0035$, $\sigma_0 = -10$ and $m = 10^{-5}$ (in reduced Planckian unit)}
 \label{plot field}
\end{center}
\end{figure}

\begin{figure}[!h]
\begin{center}
 \centering
 \includegraphics[width=3.5in,height=2.0in]{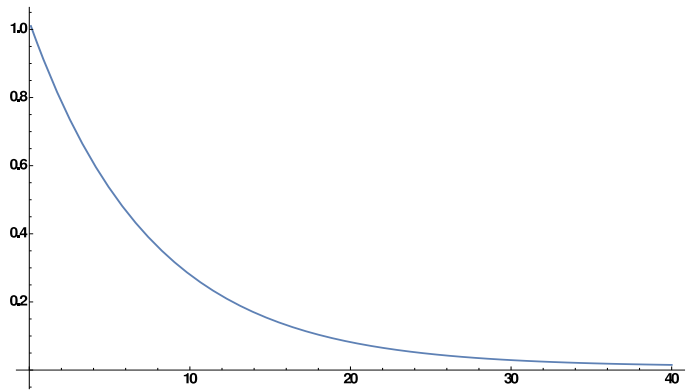}
 \caption{$\rho_{KR}$ vs $\tau$ for $\frac{\kappa^2h_0}{m^2} = 0.0035$, $\sigma_0 = -10$ and $m = 10^{-5}$ (in reduced Planckian unit)}
 \label{plot KR field suppression}
\end{center}
\end{figure}

As mentioned earlier, our motivation is to investigate the questions like -
(a) why the present universe is practically free from the observable footprints of the Kalb-Ramond field despite 
having the signatures of scalar, fermion, vector and spin-2 massless graviton, while they all
originate from the same underlying Lorentz group ?, (b) what are the possible roles of the Kalb-Ramond field during early phase of the universe ?; in 
the framework of modified theories of gravity, in particular from $F(R)$ gravity theory. Motivated by these questions, we consider the 
Starobinsky + Kalb Ramond field model as the starting model. Thus the action we consider is given by \cite{Elizalde:2018rmz},
\begin{eqnarray}
 S&=&\int d^4x \sqrt{-g}\bigg[\frac{F(R)}{2\kappa^2} - \frac{1}{12}H_{\mu\nu\rho}H^{\mu\nu\rho}\bigg]\nonumber\\
 &=&\int d^4x \sqrt{-g}\bigg[\frac{1}{2\kappa^2}\bigg(R + \frac{R^2}{m^2}\bigg) - \frac{1}{12}H_{\mu\nu\rho}H^{\mu\nu\rho}\bigg]
 \label{action1_sec1_cos}
\end{eqnarray}
where the F(R) is taken as the Starobinsky model and $m$ is a parameter having mass dimension [+1]. Thus the above action describes 
the Starobinsky F(R) model along with the Kalb-Ramond field. It is well known that the original Starobinsky model is able to produce a good inflationary 
scenario consistent with the Planck observations. However we always try to understand how far the Starobinsky model can be deviated, 
which still produces a viable inflationary scenario in the early universe. It may be observed that in the present context, 
the Starobinsky model is deviated 
by the presence of the KR field, so besides the explanation for the suppression of the KR field in the present universe, 
it is also important to investigate the possible effects of the KR field 
in the early inflation, in particular whether the Starobinsky + Kalb-Ramond field model is viable in respect to the Planck 2018 constraints 
\cite{Akrami:2018odb}. For this purpose, we need to solve the field equations for the action (\ref{action1_sec1_cos}) in a FRW spacetime. 
The solutions of the field equations will be obtained here by using the conformal correspondence of F(R) theory to a scalar-tensor theory. In particular, 
we first solve the field equations in the conformally connected scalar-tensor theory and then by using these solutions we will 
determine the solutions in the corresponding F(R) theory in view of the inverse conformal transformation. Due to the metric transformation 
$g_{\mu\nu}(x) \longrightarrow \widetilde{g}_{\mu\nu}(x) = e^{-\sqrt{\frac{2}{3}}\kappa \xi(x)}g_{\mu\nu}(x)$, the F(R) action (\ref{action1_sec1_cos}) 
is transformed to the following scalar-tensor theory,
\begin{eqnarray}
 S = \int d^4x \sqrt{-\tilde{g}}\bigg[\frac{\widetilde{R}}{2\kappa^2} + \frac{1}{2}\tilde{g}^{\mu\nu}\partial_{\mu}\xi \partial_{\nu}\xi 
 - V(\xi) - \frac{1}{4} \tilde{H}_{\mu\nu\rho}\tilde{H}^{\mu\nu\rho}\bigg]
 \label{action4}
\end{eqnarray}
where the tilde quantities are reserved for the ST theory, for example $\tilde{R}$ is formed by $\tilde{g}_{\mu\nu}$. Moreover 
we consider $\kappa\tilde{B}_{\mu\nu} < 1$ where 
$\tilde{B}_{\mu\nu}$ is the canonical field in the ST model. Due to the 
appearance of $\xi(x)$, the KR field becomes non-canonical and thus in order to make it canonical we redefine the field as 
$B_{\mu\nu} \longrightarrow \widetilde{B}_{\mu\nu} = e^{-\frac{1}{2}\sqrt{\frac{2}{3}}\kappa \xi} B_{\mu\nu}$ 
(recall the same transformation on the KR field is also applied in the 
previous scenario where we tried to explain the suppression of the KR field from a non-dynamical way). The scalaron potential $V(\xi)$, 
for $F(R) = R + \frac{R^2}{m^2}$, is obtained as $V(\xi) = \frac{m^2}{8\kappa^2}\bigg(1 - e^{\sqrt{\frac{2}{3}}\kappa\xi}\bigg)^2$ which has the plot 
in Fig.[\ref{plot potential}].\\
Having described the scalar-tensor action, now we are are going to determine the field equations and the corresponding solutions of the 
ST model. The spatially flat FRW metric ansatz fulfills our purpose, where the time dependent quantity 
is the scale factor which also depicts the expansion of the universe. The metric ansatz is given by,
\begin{eqnarray}
 \tilde{ds}^2 = \tilde{g}_{\mu\nu}(x) dx^{\mu}dx^{\nu} = -dt^2 + a^2(t)\big[dx^2 + dy^2 + dz^2\big]
 \label{4d metric}
\end{eqnarray}
The KR field strength tensor has, in general, 64 components in four dimensional spacetime, however due to the antisymmetric nature, the 
number of independent components are four and they can be symbolized as $\tilde{H}_{012} = h_1$, $\tilde{H}_{013} = h_2$, 
$\tilde{H}_{023} = h_3$ and $\tilde{H}_{123} = h_4$ respectively. Due to the FRW spacetime, the off-diagonal components of the Einstein tensor are zero. 
Moreover the energy-momentum tensor of the scalar field also vanish along the off-diagonal direction, while that of the KR field, the off-diagonal 
energy-momentum tensor has a non-zero contribution. Thereby the off-diagonal Einstein equation becomes 
$h_4h^3 = h_4h^2 = h_4h^1 = h_2h^3 = h_1h^3 = h_1h^2 = 0$ which has a solution as 
$h_1 = h_2 = h_3 = 0$ and $h_4 \neq 0$. With these solutions, the energy density and pressure of the matter fields 
($\xi$, $\tilde{B}_{\mu\nu}$) as 
$\rho_T = \bigg[\frac{1}{2}\dot{\xi}^2 + V(\xi) + \frac{1}{2}h_4h^4\bigg]$ and 
$p_T = \bigg[\frac{1}{2}\dot{\xi}^2 - V(\xi) + \frac{1}{2}h_4h^4\bigg]$ respectively (where 
an overdot denotes $\frac{d}{dt}$). As a result, the diagonal Friedmann equations turn out to be,
\begin{eqnarray}
 H^2 = \frac{\kappa^2}{3}\bigg[\frac{1}{2}\dot{\xi}^2 + \frac{m^2}{8\kappa^2}\big(1 - e^{\sqrt{\frac{2}{3}}\kappa\xi}\big)^2 
 + \frac{1}{2}h_4h^4\bigg]
 \label{einstein equation1}
\end{eqnarray}
and
\begin{eqnarray}
 2\dot{H} + 3H^2 + \kappa^2\bigg[\frac{1}{2}\dot{\xi}^2 
 - \frac{m^2}{8\kappa^2}\bigg(1 - e^{\sqrt{\frac{2}{3}}\kappa\xi}\bigg)^2 + \frac{1}{2}h_4h^4] = 0
 \label{einstein equation2}
\end{eqnarray}
where $H=\frac{\dot{a}}{a}$ symbolizes the Hubble parameter in the ST model. Further, the field equations for the KR field ($\tilde{B}_{\mu\nu}$) 
and the scalar field ($\xi$) are given by,
\begin{eqnarray}
 \tilde{\nabla}_{\mu}\tilde{H}^{\mu\nu\lambda} = \frac{1}{a^3(t)}\partial_{\mu}\bigg[a^3(t)\tilde{H}^{\mu\nu\lambda}\bigg] = 0
 \label{KR equation}
\end{eqnarray}
and
\begin{eqnarray}
 \ddot{\xi} + 3H\dot{\xi} - \sqrt{\frac{2}{3}}\frac{m^2}{4\kappa}e^{\sqrt{\frac{2}{3}}\kappa\xi}\big(1 - e^{\sqrt{\frac{2}{3}}\kappa\xi}\big)  = 0
 \label{scalar equation}
\end{eqnarray}
respectively. The information that can be obtained from Eq.(\ref{KR equation}) is the only non-zero component of the KR field i.e $h_4$ depends 
on $t$ only, which is also confirmed from the gravitational equations. Differentiating Eq.(\ref{einstein equation1}) and using 
Eq.(\ref{einstein equation2}), one finally lands with the evolution of the KR field energy density as 
$\frac{d}{dt}(h_4h^4) = -6H h_4h^4$ which can be easily solved (in terms of the scale factor) and given by,
\begin{eqnarray}
 h_4h^4 = \frac{h_0}{a^6}
 \label{solution of KR energy density}
\end{eqnarray}
where $h_0$ is an integration constant which must take only positive values in order to get a real valued solution of 
$h_4(t)$. Eq.(\ref{solution of KR energy density}) clearly indicates that the KR field energy density decreases with the expansion of the universe 
at a faster rate in comparison to normal matter and radiation (recall, the normal matter decreases as $1/a^3$, while the radiation 
energy density goes as $1/a^4$). However Eq.(\ref{solution of KR energy density}) also demonstrates that the KR field energy density 
is large and may play a significant role during early universe. Thus in order to explain the dynamical suppression of the KR field, we have to investigate 
the evolution of the KR field from the early universe when it is also important to examine whether the universe passes through an inflationary stage 
or not. Being the EoS parameter is unity, the KR field produces a decelerating effect in the expansion of the universe and thus in the 
context of Starobinsky + KR model, it is not confirmed, a priori, that the early universe passes through an inflationary stage. Hence determination of 
the early time scale factor solution is necessary and for this purpose we assume the slow roll condition where the potential of the scalar field is 
considered to be much greater than that of the kinetic energy i.e $V(\xi) \gg \frac{1}{2}\dot{\xi}^2$. The slow roll condition is natural 
to consider for the potential plotted in Fig.[\ref{plot potential}] if the scalar field starts from negative regime where the potential 
becomes approximately constant. Due to slow roll assumption, the kinetic energy and the acceleration term of the scalar field 
can be ignored from Eqs.(\ref{einstein equation1}) and (\ref{scalar equation}) respectively. As a result, the equations 
can be solved for $\xi(t)$ and $a(t)$ and the solutions are given by,
\begin{eqnarray}
 \xi(t) = \sqrt{\frac{3}{2\kappa^2}}\bigg[\ln{\bigg(\frac{9}{-\sqrt{6}m(t-t_0) + 9C}\bigg)} + \frac{\kappa^2h_0}{m^2}P(t)\bigg]
 \label{solution scalar field ST}
\end{eqnarray}
and
\begin{eqnarray}
 a(t) = D\bigg(1 - \frac{2m(t-t_0)}{3\sqrt{6}C}\bigg)^{3/4} 
 \bigg(1 + \frac{\kappa^2h_0}{m^2}Q(t)\bigg) \exp{\bigg[\frac{m(t-t_0)}{2\sqrt{6}}\bigg]}
 \label{solution scale factor ST}
\end{eqnarray}
It may be mentioned that for $h_0 = 0$ (i.e without the KR field), the solutions of $\xi(t)$ and $a(t)$ match with that of the Starobinsky model. 
The functions $P(t)$ and $Q(t)$ have the following expressions,
\begin{eqnarray}
 P(t) = \frac{(-\sqrt{6}m(t-t_0) + 9C)^2}{(-\sqrt{6}m(t-t_0) + 9C + 9)^3}
 \label{U}
\end{eqnarray}
and
\begin{eqnarray}
 Q(t)= \frac{\big(5 + 9C(1+3C)\big)\big(1 - \sqrt{\frac{3}{2}}m(t-t_0)\big) + \sqrt{6}\big(1 + 6C + 81C^2\big)
 \big(-m(t-t_0) + \sqrt{\frac{3}{2}}m^2(t-t_0)^2\big)}{\bigg(1 - \frac{2m(t-t_0)}{3\sqrt{6}C}\bigg)^{7/2}}
 \label{V}
\end{eqnarray}
respectively. To determine the above solutions, we use Eq.(\ref{solution of KR energy density}) and also 
consider $\frac{\kappa^2h_0}{m^2} < 1$. In a later part, we will show that the condition $\frac{\kappa^2h_0}{m^2} < 1$ is consistent 
to make the observable quantities like the spectral index, tensor to scalar ratio compatible with the Planck constraints. 
Further $C$ and $D$ are integration constants related to the initial values of $\xi(t)$ and $a(t)$ (the initial condition will be symbolized as 
$\xi(t_0) = \xi_0$ and $a(t_0) = a_0$). 
Differentiating both sides of Eq.(\ref{solution scalar field ST}) with respect to $t$ (twice) in the limit $t \rightarrow t_0$, one finally lands with 
the following expression of the early universe acceleration as,
\begin{eqnarray}
 \frac{\ddot{a}}{a}\bigg|_{t\rightarrow t_0} = \big(\frac{m}{2\sqrt{6}}\big)^2 (1-e^{-|\sigma_0|}) \bigg[\big(1-e^{-|\sigma_0|}\big) 
 - 4\frac{\kappa^2h_0}{m^2}\big(11 + 45e^{|\sigma_0|} + 513e^{2|\sigma_0|}\big)\bigg]
 \label{limiting acceleration ST1}
\end{eqnarray}
where $\sigma_0 = \sqrt{\frac{2}{3}}\kappa\xi_0$. Recall, $\xi_0$ be the initial value of the scalar field which is considered to be negative to 
ensure the slow roll conditions. It may be noticed that for the condition,
\begin{eqnarray}
 \frac{m^2(1 - e^{-|\sigma_0|})}{4\big(11 + 45e^{|\sigma_0|} + 513e^{2|\sigma_0|}\big)} > \kappa^2h_0
 \label{condition ST}
\end{eqnarray}
the early universe undergoes through an accelerating stage, while for the other condition i.e for 
$\frac{m^2(1 - e^{-|\sigma_0|})}{4\big(11 + 45e^{|\sigma_0|} + 513e^{2|\sigma_0|}\big)} < \kappa^2h_0$, $\ddot{a}(t \rightarrow t_0)$ becomes less 
than zero. The parameters $m$ and $h_0$ actually controls the strength of the scalaron and the KR field respectively. The scalaron field produces 
an accelerating effect in the early universe when the slow roll conditions are valid and on the other hand, the KR field triggers a decelerating effect 
in the universe evolution (in particular, 
in a model like $S = \int d^4x\sqrt{-\tilde{g}}\big[\tilde{R} - \frac{1}{12}\tilde{H}_{\mu\nu\rho}\tilde{H}^{\mu\nu\rho}\big]$, the scale 
factor evolves as $a(t)\sim t^{1/3}$). Thereby the interplay between the scalar and the KR fields fixes whether the early universe 
passes through an accelerating stage or not and such interplay is reflected from the condition (\ref{condition ST}). Thus as a whole, 
the early inflationary era is ensured for a Starobinsky + KR model universe, if the parameters of the model satisfy the condition (\ref{condition ST}).\\
Having determined the solutions in the scalar-tensor model, now we turn our focus to the solutions of the original F(R) model. Recall, the 
F(R) model action is depicted by Eq.(\ref{action1_sec1_cos}) and connected to the scalar-tensor model by the aforementioned conformal transformation 
of the spacetime metric. Thus the line element of the F(R) model can be obtained from that of the ST model by using the inverse 
conformal transformation i.e given by,
\begin{eqnarray}
 ds^2 = e^{\sqrt{\frac{2}{3}}\kappa\xi(t)} \bigg[-dt^2 + a^2(t)\big(dx^2 + dy^2 + dz^2\big)\bigg] 
 = -d\tau^2 + s^2(\tau)\big(dx^2 + dy^2 + dz^2\big) 
\end{eqnarray}
Thus the F(R) spacetime also behaves as FRW one where the cosmic time ($\tau$) and the scale factor ($s(\tau)$) have the following 
expressions:
\begin{eqnarray}
\tau(t) = \int dt e^{[\frac{1}{2}\sqrt{\frac{2}{3}}\kappa\xi(t)]}~~~~~~~~,~~~~~~~~s(\tau(t)) = e^{[\frac{1}{2}\sqrt{\frac{2}{3}}\kappa\xi(t)]} a(t)
\label{cosmic time1 F(R)}
\end{eqnarray}
with $\xi(t)$ is obtained in Eq.(\ref{solution scalar field ST}). Due to the conformal transformation, the cosmic time and the scale factor 
in the two frames get changed, as expected. It may be observed that $\tau(t)$ is a monotonic increasing function of $t$. Moreover, the 
KR field energy density in the F(R) model is given by $\rho_{KR} = \frac{1}{2}H_{123}H^{123}$. In view of the conformal connection between 
the F(R) and the ST model, the KR field energy density of the F(R) model is related to that of the ST model as 
$\rho_{KR} = e^{-2\sqrt{\frac{2}{3}}\kappa\xi}\tilde{\rho}_{KR}$ (where we also use the transformation of $B_{\mu\nu}$ and 
$\tilde{B}_{\mu\nu}$). We will need this relation later. By using 
the solution of $\xi(t)$, one can integrate the left expression of Eq.(\ref{cosmic time1 F(R)}) to get the explicit the functional behaviour 
of $\tau = \tau(t)$ as,
\begin{eqnarray}
 \tau - \tau_0&=&\frac{1}{4m}\sqrt{\frac{3}{2}} \bigg[\bigg(8\sqrt{9C} - 8\sqrt{9C - \sqrt{6}m(t-t_0)}\bigg) 
 + \frac{\kappa^2h_0}{2m^2}\bigg(\tan^{-1}\big(\frac{3}{\sqrt{9C - \sqrt{6}m(t-t_0)}}\big) - \tan^{-1}(1/\sqrt{C})\bigg)\nonumber\\
 &+&\frac{\kappa^2h_0}{2m^2}\bigg(\frac{(27 + 45C - 5\sqrt{6}m(t-t_0))\sqrt{9C - \sqrt{6}m(t-t_0)}}{(9 + 9C - \sqrt{6}m(t-t_0))^2} 
 - \frac{(3+5C)\sqrt{C}}{3(1+C)^2}\bigg)\bigg]
 \label{cosmic time2 F(R)}
\end{eqnarray}
where the integration constant is adjusted by the condition $\tau(t_0) = \tau_0$. Plugging the solutions of $\xi(t)$ and $a(t)$ into the right expression 
of Eq.(\ref{cosmic time1 F(R)}) immediately yields the F(R) frame scale factor as follows,
\begin{eqnarray}
 s(\tau(t)) = D\bigg(1 - \frac{2m(t-t_0)}{3\sqrt{6}C}\bigg)^{3/4} 
 \bigg(1 + \frac{\kappa^2h_0}{m^2}\big(Q(t) + \frac{1}{2}P(t)\big)\bigg) 
 \exp{\bigg[\frac{m(t-t_0)}{2\sqrt{6}} + \ln{\bigg(\frac{9}{-\sqrt{6}m(t-t_0) + 9C}\bigg)}\bigg]}
 \label{scale factor2 F(R)}
\end{eqnarray}
with $P(t)$ and $Q(t)$ are shown in Eqs.(\ref{U}) and (\ref{V}) respectively.  The scale factor $s(\tau)$ corresponds to an early inflationary era 
(with the onset at $\tau = \tau_0$) if the condition 
\begin{eqnarray}
 \frac{m^2(1 - e^{-|\sigma_0|})}{4\big(11 + \frac{1219}{27}e^{|\sigma_0|} + \frac{13849}{27}e^{2|\sigma_0|}\big)} > \kappa^2h_0
 \label{condition F(R)}
\end{eqnarray}
holds true, otherwise $\frac{d^2s}{d\tau^2}$ becomes negative leading to a decelerating universe. 
In absence of the KR field i.e for $h_0 = 0$, the above condition is trivially satisfied, which indicates an accelerating 
expansion of the universe - this is expected because without the KR field, the model becomes the Starobinsky model which indeed leads to an 
inflationary universe. However, due to the presence of the KR field, the acceleration of the scale factor gets modified 
by the term proportional to $\frac{\kappa^2h_0}{m^2}$. Similar to the ST model, the relative strengths 
of the parameters $m$ and $h_0$ fix whether the early universe passes through an inflationary stage or not. In order to solve the horizon and 
flatness problems, the inflationary scenario is necessary and that's why we stick to the condition (\ref{condition F(R)}). Having 
getting the inflationary condition, the next move is to check whether the inflation has an end within a finite time. The end point of inflation is 
defined as $\frac{d^2s}{d\tau^2} = 0$. Using Eq.(\ref{cosmic time1 F(R)}), one obtains,
\begin{eqnarray}
 \frac{d^2s}{d\tau^2} = e^{[-\frac{1}{2}\sqrt{\frac{2}{3}\kappa\xi(t)}]} a(t) \bigg[\dot{H} + H^2 + \frac{\kappa}{\sqrt{6}}H\dot{\xi}\bigg]
 \label{end1}
\end{eqnarray}
where $H$ is the Hubble parameter in the ST model and an overdot symbolizes $\frac{d}{dt}$. Thus the inflation termination condition 
becomes $\dot{H} + H^2 + \frac{\kappa}{\sqrt{6}}H\dot{\xi} = 0$. Using the field equations, this condition can be expressed in terms of the scalar field as,
 \begin{eqnarray}
  \frac{1}{4} - \frac{2}{3}e^{[\sqrt{\frac{2}{3}}\kappa\xi(t_f)]} + \frac{1}{9}e^{[2\sqrt{\frac{2}{3}}\kappa\xi(t_f)]} 
  + \frac{8}{81}e^{[4\sqrt{\frac{2}{3}}\kappa\xi(t_f)]} = 0
  \nonumber
 \end{eqnarray}
 with $t_f - t_0$ being the duration of inflation in F(R) model (in terns of $t$). 
 Solving the above algebraic equation, we get $\xi(t_f) = \xi_{f} \simeq \sqrt{\frac{3}{2\kappa^2}} \ln{\big(\frac{3}{5}\big)} < 0$. Thus 
 the inflationary era of the universe continues as long as the scalar field is lesser than $\xi_f$ (recall, the scalar field starts from 
 negative regime to validate the slow roll condition). Correspondingly, by using the scalar field solution and the functional form 
 of $\tau = \tau(t)$, the inflationary duration in the F(R) model ($\tau_f - \tau_0$) comes as,
 \begin{eqnarray}
  \tau_f - \tau_0&=&\frac{1}{4m}\sqrt{\frac{3}{2}} \bigg[\bigg(8\sqrt{9C} - 8\sqrt{d}\bigg) 
  + \frac{\kappa^2h_0}{2m^2}\bigg(\frac{(27 + 5d)\sqrt{d}}{(9 + d)^2} 
 - \frac{(3+5C)\sqrt{C}}{3(1+C)^2}\bigg)\nonumber\\ 
 &+&\frac{\kappa^2h_0}{2m^2}\bigg(\tan^{-1}\big(\frac{3}{\sqrt{q}}\big) - \tan^{-1}(1/\sqrt{C})\bigg)\bigg]
 \label{duration2}
 \end{eqnarray}
 with $d = e^{\sigma_f}\bigg(1 + \frac{\kappa^2h_0}{9m^2}e^{-|\sigma_0|}\bigg)$, $\sigma_0 = \sqrt{\frac{2\kappa^2}{3}}\xi_0$ 
 and $\sigma_f = \sqrt{\frac{2\kappa^2}{3}}\xi_f$. It may be noticed that the duration of inflation depends on the parameters 
 $\sigma_0$ and $\frac{\kappa^2h_0}{m^2}$. Thus to get an estimation of $\tau_f - \tau_0$, we first need to estimate the parameters, which can 
 be done once we confront the model with the Planck observations. 
 In order to test the broad inflationary paradigm as well as particular models 
 against precision observations, we need to calculate the value of spectral index($n_s$), 
 tensor to scalar ratio ($r$) and in the present context, they are defined as \cite{Hwang:2005hb,Noh:2001ia,Hwang:2002fp},
 \begin{eqnarray}
  n_s = \big[1 - 4\epsilon_F - 2\epsilon_2 + 2\epsilon_3 - 2\epsilon_4\big]\bigg|_{\tau_0}~~~~~~,~~~~~~
  r = 8\kappa^2 \frac{\varTheta}{F'(R)}\bigg|_{\tau_0}
  \label{spectral index1}
 \end{eqnarray}
 respectively, where the slow roll parameters ($\epsilon_F$, $\epsilon_2$, $\epsilon_3$, $\epsilon_4$) have the following expressions,
\begin{eqnarray}
 \epsilon_F&=&-\frac{1}{H_F^2} \frac{dH_F}{d\tau}~~~~~~~~~~~,~~~~~~~~~~\epsilon_2 = \frac{1}{2\rho_{KR}H_F} \frac{d\rho_{KR}}{d\tau}\nonumber\\
 \epsilon_3&=&\frac{1}{2F'(R)H_F} \frac{dF'(R)}{d\tau}~~~~~~,~~~~~~~~~~\epsilon_4 = \frac{1}{2EH_F} \frac{dE}{d\tau}
 \label{various slow roll parameters}
\end{eqnarray}
with $\varTheta$ and $E$ are given by,
\begin{eqnarray}
 \varTheta = \frac{\rho_{KR}}{F'(R)H_F^2}\bigg[F'(R) + \frac{3}{2\kappa^2\rho_{KR}}\bigg(\frac{d}{d\tau}F'(R)\bigg)^2\bigg]
 ~~~~~~~,~~~~~~~E = \frac{\varTheta F'(R)H_F^2}{\rho_{KR}}
 \label{vartheta}
\end{eqnarray}
where $H_F = \frac{1}{s}\frac{ds}{d\tau}$ and $\rho_{KR}$ ($= H_{123}H^{123}$) are the Hubble parameter and 
the energy density of the KR field respectively, both defined in $F(R)$ model. Here it may be mentioned that the speeds for both the 
scalar and tensor perturbation variables in F(R) model are unity. The unit speed of the gravitational wave confirms the compatibility of the 
F(R) model with GW170817 according to which the gravitational and the electromagnetic waves propagate with same speed which is indeed unity 
in the natural unit system. However this is not the case of all higher curvature gravity theories, 
for example in scalar-Einstein-Gauss-Bonnet gravity theory, 
the speed of the gravitational wave is not unity and the deviation from unity is proportional to the Gauss-Bonnet coupling function. Still, 
the Gauss-Bonnet model can be made compatible with GW170817 by considering a particular form of the scalar coupling function, but such coupling 
function leads to the speed of the scalar perturbation variable different from unity. Such scenario of Gauss-Bonnet 
model where the tensor perturbation speed is unity, but the scalar perturbation variable 
has a different speed than unity has been investigated recently in inflation as well as in bouncing spacetime \cite{Elizalde:2020zcb,Odintsov:2020sqy}. 
One comment deserves to mention here that 
in the case of F(R) model, the gravitational wave speed is always unity irrespective of the form of F(R), however in the scalar coupled 
Einstein-Gauss-Bonnet model, as mentioned earlier, a certain form of the Gauss-Bonnet coupling function can do so. Coming to our present model, 
by using the slow roll field equations and the conformal correspondence between the ST and the F(R) model, the final expressions of $n_s$ and $r$ in 
the Starobinsky + KR model are given by,
\begin{eqnarray}
 n_s&=&1 - 2\bigg[\frac{-\frac{\kappa^2h_0}{2m^2} + \frac{1}{9}\bigg(\frac{e^{2\sigma_0}(1-e^{\sigma_0})(1-2e^{\sigma_0})}
 {(1-e^{\sigma_0})^2 + \frac{4\kappa^2h_0}{m^2}}\bigg)}{\frac{3\kappa^2h_0}{2m^2} + \frac{1}{12}e^{\sigma_0}(1-e^{\sigma_0})}\bigg] 
 + 2\bigg[\frac{-\frac{3\kappa^2h_0}{m^2} - \frac{1}{9}\bigg(\frac{e^{2\sigma_0}(1-e^{2\sigma_0})}
 {(1-e^{\sigma_0})^2 + \frac{4\kappa^2h_0}{m^2}}\bigg)}{\frac{\kappa^2h_0}{2m^2} + \frac{1}{8}(1-e^{\sigma_0})^2 
 + \frac{1}{6}e^{\sigma_0}(1-e^{\sigma_0})}\bigg]\nonumber\\ 
 &+&\frac{9}{16}\frac{\kappa^2h_0/m^2}
 {\bigg[\frac{1}{8}(1 - e^{\sigma_0})^2 + \frac{\kappa^2h_0}{2m^2} + \frac{1}{6}e^{\sigma_0}(1 - e^{\sigma_0})\bigg]
 \bigg[\frac{3\kappa^2h_0}{2m^2} + \frac{1}{12}e^{\sigma_0}(1 - e^{\sigma_0})\bigg]} 
 - \frac{3}{8}\frac{\kappa^2h_0/m^2}{\bigg[\frac{3\kappa^2h_0}{2m^2} + \frac{1}{12}e^{\sigma_0}(1 - e^{\sigma_0})\bigg]^2}
 \label{spectral index2}
\end{eqnarray}
and
\begin{eqnarray}
  r = \frac{3\frac{\kappa^2h_0}{2m^2} + 48\bigg[\frac{3\kappa^2h_0}{2m^2} + \frac{1}{12}e^{\sigma_0}(1 - e^{\sigma_0})\bigg]^2}
  {\bigg[\frac{1}{8}(1 - e^{\sigma_0})^2 + \frac{\kappa^2h_0}{2m^2} + \frac{1}{6}e^{\sigma}(1 - e^{\sigma})\bigg]^2}
  \label{ratio2}
 \end{eqnarray}
respectively. In absence of the KR field, the $n_s$ becomes $n_s = 1 - 2\frac{\epsilon_F'}{\epsilon_F H_F}$ and the tensor 
to scalar ratio gets the expression as $r = 48\epsilon_F^2$. 
However the presence of the KR field modifies $n_s$ and $r$ by the term proportional to 
$\rho_{KR}$. With the above expressions of $n_s$ and $r$, we can confront the model with the Planck results which put constraints as 
$n_s = 0.9649 \pm 0.0042$ and $r < 0.064$. It may be observed from Eqs.(\ref{spectral index2}) and (\ref{ratio2}) that in the present context, 
the spectral index and the tensor to scalar ratio depend on the dimensionless parameters $\sigma_0$ and $\frac{\kappa^2h_0}{m^2}$. For the model 
in hand, the $n_s$ and $r$ become simultaneously compatible with the Planck results for the parametric regime : 
$10 \leq |\sigma_0| \leq 14$ and $0.003 \lesssim \frac{\kappa^2h_0}{m^2} \lesssim 0.004$. The simultaneous compatibility of $n_s$ and $r$ is plotted 
in Fig.[\ref{plot dynamical1}]. With $\frac{\kappa^2h_0}{m^2} = 0.0035$ and $|\sigma_0| = 10$, the duration of inflation comes as 
$\Delta\tau = \tau_f - \tau_0 = 10^{-12}\mathrm{GeV}^{-1}$ if the mass parameter is taken as $m = 10^{-5}$ in reduced Planckian unit. 
With this value of $m$, 
we calculate the number of e-foldings $N = \int_{0}^{\Delta\tau}H_Fd\tau$ numerically and finally lands with $N \simeq 56$ (for 
$|\sigma_0| = 10$).\\ 
As mentioned earlier, in order to validate the model 
with the Planck constraints, the maximum allowed value of the parameter $\frac{\kappa^2h_0}{m^2} = 0.004$. Considering 
$m = 10^{-5}$ (in reduced Planckian unit), one easily gets $h_0^{max} = 10^{63}\mathrm{GeV}^{4}$. 
Recall, the energy density of the KR field in the F(R) model 
is given by $\rho_{KR} = e^{-2\sigma_0}h_0$ (at the point of horizon crossing) and thus the present model along with the Planck observations 
give an upper bound of the KR field energy density in the early universe as 
$\rho_{KR} = 10^{70}\mathrm{GeV}^{4}$ ($|\sigma_0| = 10$). Despite this amount of initial energy density, the KR field has practically no 
observable signatures in the present universe. Thus it is important to explain how the KR field energy density (starting 
with $10^{70}\mathrm{GeV}^{4}$ from the early universe) gets suppressed and leads to a negligible footprint on the present universe and to explain this 
phenomena, what we need is the expression of the KR field energy density in the F(R) model, which is given by 
$\rho_{KR} = e^{-2\sqrt{\frac{2}{3}}\kappa\xi(t)}\frac{h_0}{a^6}$. The solutions of $\xi(t)$ and $a(t)$ obtained earlier are based on the slow roll 
conditions, however to address the effect of the KR field in the present universe, the late time behaviour of $\xi(t)$ and $a(t)$ are required, where 
the slow roll conditions may not valid. Keeping this in mind, we relax the slow roll assumptions and solve the field equations 
numerically for $\xi(t)$ and $a(t)$ (in particular the plot is given for the deceleration parameter $q(\tau)$), 
 as shown in Fig.[\ref{plot field}] by using the relation of $\tau(t)$ (see eqn.(\ref{cosmic time2 F(R)})). 
 Fig. [5] demonstrates that the plotted result of $\xi(\tau)$ based on solving 
 the slow roll equations and the plotted result of $\xi(\tau)$ based on solving the full Friedmann equations 
 are almost same during the inflation. But after the inflation the acceleration term of $\xi(\tau)$ 
 starts to contribute and as a result the two solutions (with and without slow roll conditions) differ from each other. 
 Similar argument holds for the deceleration parameter. These numerical solutions immediately lead to the evolution of the KR field energy density 
 in the F(R) model, as shown in Fig.[\ref{plot KR field suppression}]. 
 
As it is evident from Fig.[\ref{plot KR field suppression}] that the energy density of KR field gradually 
decreases with the cosmic time ($\tau$) and the decaying 
time scale ($\tau=40$) is less than the exit time of inflation ($\tau=56$). Apart from the description of the decaying nature 
of the KR field energy density, it is also necessary to determine the couplings of the KR field with other matter fields to discuss 
the observable signatures of the KR field in the present universe. In the four dimensional context, the KR field couplings with the matter fields 
turn out to be $\frac{1}{M_{Pl}}$ which is same as the usual gravity-matter coupling. Thereby it is expected that the KR field may show its signatures 
in some present day gravity experiment. But as mentioned earlier, there has been no experimental evidence of the footprint of the KR field 
on the present universe and an example is the Gravity Probe B experiment. Thereby the four dimensional model (\ref{action1_sec1_cos}) may not serve the full 
explanation why the present universe practically carries no footprint of the Kalb-Ramond field. However in the context of higher dimensional 
higher curvature scenario, in particular a warped braneworld scenario, the KR field couplings get an additional suppression over the 
gravity-matter coupling $1/M_{Pl}$. This along with the cosmological counterpart will be discussed in the next two sections.\\

\begin{figure}[!h]
\begin{center}
 \centering
 \includegraphics[width=3.5in,height=2.0in]{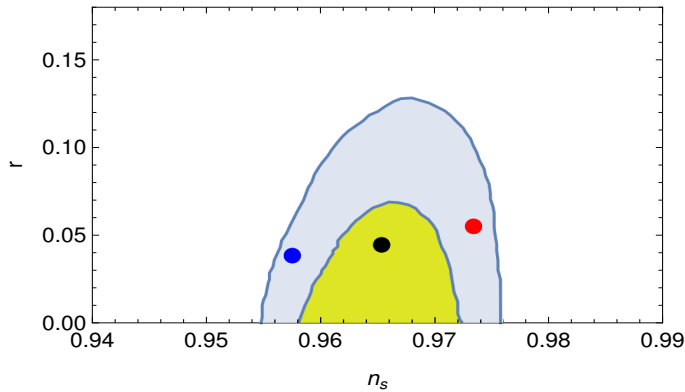}
 \caption{$1\sigma$ (yellow) and $2\sigma$ (light blue) contours for Planck 
 2018 results \cite{Akrami:2018odb}, on $n_s-r$ plane. 
 Additionally, we present the predictions of the present bounce scenario with $(\sigma_0, \kappa^2h_0\sqrt{\alpha}) = (-5, 0.003)$ (blue point), 
 $(\sigma_0, \kappa^2h_0\sqrt{\alpha}) = (-5, 0.03)$ (black point) and $(\sigma_0, \kappa^2h_0\sqrt{\alpha}) = (-5, 0.1)$ (red point).}
 \label{plot cubic quantities}
\end{center}
\end{figure}

Before moving to the next section, we would like to mention that the presence of the KR field 
actually enhances the value of the tensor to scalar ratio in the 
Starobinsky + KR model, in comparison to the Starobinsky model where the KR field is absent. The same kind of effect of the KR field is also 
observed in a different F(R) model, in particular for a logarithmic corrected $R^2$ gravity model, where the action has the form \cite{Elizalde:2018now},
\begin{eqnarray}
 S = \int d^4x \sqrt{-g}\bigg[\frac{1}{2\kappa^2}\bigg(R + \alpha R^2 + \beta R^2\ln{\big(\beta R\big)}\bigg) - \frac{1}{12}H_{\mu\nu\rho}H^{\mu\nu\rho}\bigg]
 \label{action2_sec1_cos}
\end{eqnarray}
In the regime $\alpha$, $\beta > 0$, $F'(R)$ becomes positive which in turn renders the model ghost free. Similar to the previous model, the amplitude of the 
KR field in the logarithmic corrected $R^2$ model is large during early universe, but these reduce as the 
universe expands, at a rate $\sim a^{-6}$ , and in effect, radiation and matter dominate the post-inflationary phase of our universe. The spectral 
index and the tensor to scalar ratio have been also determined under the slow roll and the constant roll conditions. As a result, it was found that 
under both the conditions, the presence of the KR field increases the value of the tensor to scalar ratio in comparison to the case when the 
KR field is absent. However the ratio $\frac{\beta}{\alpha}$ is constrained in different ways in the slow roll and constant roll cases: 
with the  slow roll constraint being $0.20 < \frac{\beta}{\alpha} < 0.25$ while the constant roll one being $0.25 < \frac{\beta}{\alpha} < 0.30$. 
In addition, the theoretical framework along with the Planck observations put an upper bound on the KR field energy density 
in the early universe as $\rho_{KR}^{max} \sim 10^{71}\mathrm{GeV}^{4}$. Thereby the findings of the models (\ref{action1_sec1_cos}) 
and (\ref{action2_sec1_cos}) are 
more-or-less same, with the noticeable effect of the KR field on the 
inflationary phenomenology of $R^2$ Starobinsky inflation and logarithmic $R^2$ gravity is that 
it increases the amount of the primordial gravitational radiation. Thus the impact of the KR field in inflationary phenomenology 
is considerable. One more example where the importance of the KR field is significant is \cite{Elizalde:2018rmz},
\begin{eqnarray}
 S = \int d^4x \sqrt{-g}\bigg[\frac{1}{2\kappa^2}\bigg(R + \frac{1}{3}\alpha R^3\bigg) - \frac{1}{12}H_{\mu\nu\rho}H^{\mu\nu\rho}\bigg]
 \label{action3}
\end{eqnarray}
The cubic curvature vacuum F(R) gravity, i.e., $F(R) = R + \frac{1}{3}\alpha R^3$ model does not produce a viable inflation, in particular, the 
theoretical expectations of spectral index ($n_s$) and tensor to scalar ratio ($r$) do not match with the Planck 2018 
constraints, however in the presence of the Kalb-Ramond field the cubic gravity model becomes compatible with the 
Planck constraints \cite{Elizalde:2018rmz}. 
The demonstration of this argument can be shown by the method applied for the Starobinsky model in the beginning of this section, 
in particular we first determine 
the slow roll field equations of the scalar tensor model corresponds to the $F(R) = R + \frac{\alpha}{3}R^3$ model and then we transform back to the original 
$F(R)$ model by the inverse conformal transformation of the spacetime metric. Using the metric transformation 
$g_{\mu\nu}(x) \longrightarrow \widetilde{g}_{\mu\nu}(x) = e^{-\sqrt{\frac{2}{3}}\kappa \xi(x)}g_{\mu\nu}(x)$, the F(R) action (\ref{action3}) 
is mapped to the following scalar-tensor (ST) theory,
\begin{eqnarray}
 S = \int d^4x \sqrt{-\tilde{g}}\bigg[\frac{\widetilde{R}}{2\kappa^2} + \frac{1}{2}\tilde{g}^{\mu\nu}\partial_{\mu}\xi \partial_{\nu}\xi 
 - V(\xi) - \frac{1}{4} \widetilde{H}_{\mu\nu\rho}\widetilde{H}^{\mu\nu\rho}\bigg]
 \label{action4}
\end{eqnarray}
where recall, the tilde quantities are reserved for the ST theory and 
$\widetilde{H}_{\mu\nu\rho} = e^{-\frac{1}{2}\sqrt{\frac{2}{3}}\kappa \xi} H_{\mu\nu\rho}$ is the canonical Kalb-Ramond field tensor in the scalar tensor 
theory. The scalaron potential $V(\xi)$ for $F(R) = R + \frac{\alpha}{3}R^3$ model takes the following expression,
\begin{eqnarray}
 V(\xi) = \frac{1}{3\kappa^2\sqrt{\alpha}} \big[e^{\sigma/3} - e^{4\sigma/3}\big]^{3/2}
 \label{revision_cubic_potential}
\end{eqnarray}
where $\sigma = \sqrt{\frac{2}{3}}\kappa\xi$. With a FRW metric ansatz, the evolution of the KR field energy density ($\widetilde{\rho}_{KR}$) 
in the ST theory is obtained as
\begin{eqnarray}
 \frac{d}{dt}\big(\widetilde{\rho}_{KR}\big) + 6H\widetilde{\rho}_{KR} = 0
 \label{revsision_cubic_evolution KR}
\end{eqnarray}
where $t$ and $H (= \dot{a}/a)$ represent cosmic time and Hubble parameter in the ST theory. Eq.(\ref{revsision_cubic_evolution KR}) can be 
solved to obtain $\widetilde{\rho}_{KR}$ in terms of the scale factor 
and is given by $\widetilde{\rho}_{KR} = \frac{h_0}{a^6}$ with $h_0$ (having mass dimension [+4]) being an integration constant. Thereby the 
KR field energy density decreases at a faster rate in comparison to radiation and matter components with the expansion of the universe, which in turn 
may explain the suppression of the KR field at the present energy scale of our universe. The solution of $\widetilde{\rho}_{KR}$ along with 
the scalaron potential given in Eq.(\ref{revision_cubic_potential}) immediately lead to the slow roll field equations for the ST theory as,
\begin{eqnarray}
 H^2 = \frac{1}{3\sqrt{\alpha}}\bigg[\frac{1}{3}\big(e^{\sigma/3} - e^{4\sigma/3}\big)^{3/2} + \frac{\kappa^2h_0\sqrt{\alpha}}{2a^6}\bigg]
 \label{revision_slow roll1}
\end{eqnarray}
and
\begin{eqnarray}
 3H\dot{\xi} + \frac{1}{6\kappa\sqrt{\alpha}}\sqrt{\frac{2}{3}}\big(e^{\sigma/3} - e^{4\sigma/3}\big)^{1/2}\big(e^{\sigma/3} - 4e^{4\sigma/3}\big) = 0
 \label{revision_slow roll2}
\end{eqnarray}
respectively. Having determined the slow roll field equations in the ST theory, now we move forward to find the primordial observable quantities 
like the spectral index ($n_s$), tensor to scalar ratio ($r$) for the original $F(R)$ model (\ref{action3}). Recall, as presented in Eq.(\ref{spectral index1}), 
$n_s$ and $r$ for a $F(R)$ model are defined by $n_s = \big[1 - 4\epsilon_F - 2\epsilon_2 + 2\epsilon_3 - 2\epsilon_4\big]\bigg|_{H.C}$ and 
$r = 8\kappa^2 \frac{\varTheta}{F'(R)}\bigg|_{H.C}$ respectively, where the suffix 'H.C' denotes the horizon crossing time and furthermore the 
slow roll parameters $\epsilon_{i}$ ($i = F, 2, 3, 4$) are given earlier in Eq.(\ref{various slow roll parameters}). Using the explicit expressions of 
$\epsilon_i$(s) and with some simplifications, we land with the following expressions of the spectral index and tensor to scalar ratio:
\begin{eqnarray}
 n_s = 1 - \frac{2\epsilon_F'}{\epsilon_F~H_F} + \frac{\kappa^2\rho_{KR}}{12F'(R)\epsilon_F^2H_F^2}\bigg(3 - \frac{\dot{\sigma}}{2H_F}e^{-\sigma/2} 
 - \frac{H_F'}{H_F^2}\bigg)
 \label{revision_cubic_spectral1}
\end{eqnarray}
and
\begin{eqnarray}
 r = 48\epsilon_F^2 + 8\kappa^2\frac{\rho_{KR}}{F'(R)H_F^2}~~,
 \label{revision_cubic_tensor1}
\end{eqnarray}
where $H_F$ and $\rho_{KR}$ are the Hubble parameter and the KR field energy density 
in the $F(R)$ frame, a prime and an overdot symbolize the differentiation with respect to cosmic time 
in $F(R)$ and ST frame respectively. It may be observed that in absence of the KR field, 
the $n_s$ becomes $n_s = 1 - 2\frac{\epsilon_F'}{\epsilon_F H_F}$ and the tensor 
to scalar ratio gets the expression as $r = 48\epsilon_F^2$. However the presence of the KR field modifies $n_s$ and $r$ by the term proportional to 
$\rho_{KR}$. The slow roll field Eqns.(\ref{revision_slow roll1}) and (\ref{revision_slow roll2}) allow us to obtain the following expressions like, 
\begin{eqnarray}
 H_F^2&=&\frac{1}{3\sqrt{\alpha}}e^{-\sigma}\bigg[\frac{1}{3}\big(e^{\sigma/3} - e^{4\sigma/3}\big)^{3/2} + \frac{\kappa^2h_0\sqrt{\alpha}}{2a^6}\bigg]\nonumber\\
 \frac{\dot{\sigma}}{H_F}e^{-\sigma/2}&=&-\frac{\big(e^{\sigma/3} - e^{4\sigma/3}\big)^{1/2}\big(e^{\sigma/3} - 4e^{4\sigma/3}\big)}
 {9\bigg[\frac{1}{3}\big(e^{\sigma/3} - e^{4\sigma/3}\big)^{3/2} + \frac{\kappa^2h_0\sqrt{\alpha}}{2a^6}\bigg]}
 \label{revision_cubic_intermediate}
\end{eqnarray}
which yield the final forms of $n_s$ and $r$ (for the model (\ref{action3})) as,
\begin{eqnarray}
 n_s = 1 - \frac{2\epsilon_F'}{\epsilon_F~H_F} + \frac{\kappa^2h_0\sqrt{\alpha}~e^{\sigma}}
 {4F'(R)\epsilon_F^2\bigg[\frac{1}{3}\big(e^{\sigma/3} - e^{4\sigma/3}\big)^{3/2} + \frac{\kappa^2h_0\sqrt{\alpha}}{2}\bigg]}\times
 \bigg\{3 + \epsilon_F + \frac{\big(e^{\sigma/3} - e^{4\sigma/3}\big)^{1/2}\big(e^{\sigma/3} - 4e^{4\sigma/3}\big)}
 {9\bigg[\frac{1}{3}\big(e^{\sigma/3} - e^{4\sigma/3}\big)^{3/2} + \frac{\kappa^2h_0\sqrt{\alpha}}{2}\bigg]}\bigg\}
 \label{revision_cubic_spectral2}
\end{eqnarray}
and
\begin{eqnarray}
 r= 48\epsilon_F^2 + \frac{24\kappa^2h_0\sqrt{\alpha}~e^{\sigma}}
 {F'(R)\bigg[\frac{1}{3}\big(e^{\sigma/3} - e^{4\sigma/3}\big)^{3/2} + \frac{\kappa^2h_0\sqrt{\alpha}}{2}\bigg]}
 \label{revision_cubic_tensor2}
\end{eqnarray}
respectively, with $\epsilon_F$ given by,
\begin{eqnarray}
 \epsilon_F&=&-\frac{H_F'}{H_F^2}\nonumber\\
 &=&\frac{\big(e^{\sigma/3} - e^{4\sigma/3}\big)^{1/2}\big(e^{\sigma/3} - 4e^{4\sigma/3}\big)
 \bigg[\frac{1}{6}\big(e^{\sigma/3} - e^{4\sigma/3}\big)^{1/2}\big(3e^{\sigma/3} - 2e^{4\sigma/3}\big) + \frac{\kappa^2h_0\sqrt{\alpha}}{2}\bigg]}
 {6\bigg[\frac{1}{3}\big(e^{\sigma/3} - e^{4\sigma/3}\big)^{3/2} + \frac{\kappa^2h_0\sqrt{\alpha}}{2}\bigg]}~~.
 \label{revision_cubic_SR parameter}
\end{eqnarray} 
It may be observed from Eqs.(\ref{revision_cubic_spectral2}) and (\ref{revision_cubic_tensor2}) that in the present context, the spectral 
index and the tensor to scalar ratio depend on the dimensionless parameters $\sigma_0$ and $\kappa^2h_0\sqrt{\alpha}$, where 
$e^{-\sigma_0} = 1 + \alpha R_0^2$ with $R_0$ being the Ricci scalar at the time of horizon crossing 
and $h_0$ denotes the KR field energy density at horizon crossing. With this information, we 
now directly confront the theoretical expectations of spectral index and tensor-to-scalar ratio of the model (\ref{action3}) 
with the Planck 2018 constraints \cite{Akrami:2018odb}. In Fig.[\ref{plot cubic quantities}] we present the estimated spectral index and tensor-to-scalar ratio
of the present scenario for three choices of ($\sigma_0$, $\kappa^2h_0\sqrt{\alpha}$), 
on top of the $1\sigma$ and $2\sigma$ contours of the Planck 2018 results \cite{Akrami:2018odb}. As 
we observe, the agreement with observations is efficient, and in particular well inside the $1\sigma$ as well as in the $2\sigma$ region. At this stage 
it deserves mentioning that the cubic $F(R)$ model in absence of the Kalb-Ramond field, in which case $n_s$ and $r$ follow 
Eqns.(\ref{revision_cubic_spectral2}) and (\ref{revision_cubic_tensor2}) respectively with $h_0 = 0$, 
is not viable in respect to the Planck 2018 observations. Therefore the $R + R^3 + KR$ field model has advantages from two sides : (1) the presence 
of the higher curvature term triggers an early inflationary scenario and explains the suppression of the KR field at the present universe, and (2) the 
KR field energy density is significant during early universe and helps to make the inflationary model viable with respect to Planck 2018 constraints.\\

The inflationary phenomena in presence of Kalb-Ramond field (without higher curvature term(s)) 
has also been discussed in \cite{Aashish:2018lhv} where the authors 
investigated the possibility of inflation with models of antisymmetric tensor field having minimal 
and non-minimal couplings to gravity. As a result, in the minimal model with the action,
\begin{eqnarray}
 S = \int d^4x\sqrt{-g}\bigg[\frac{R}{2\kappa^2} - \frac{1}{12}H_{\mu\nu\lambda}H^{\mu\nu\lambda} - \frac{1}{2}m^2B_{\mu\nu}B^{\mu\nu}\bigg]
 \label{S1}
\end{eqnarray}
the acceleration of the scale factor becomes $\frac{\ddot{a}}{a} = -\frac{\dot{H}^2}{2}$. 
Clearly, the acceleration of $a(t)$ is negative and hence the minimal model does not support the possibility of inflation. As a next step, let us consider 
the non-minimal model,
\begin{eqnarray}
 S = \int d^4x\sqrt{-g}\bigg[\frac{R}{2\kappa^2} - \frac{1}{12}H_{\mu\nu\lambda}H^{\mu\nu\lambda} - \frac{1}{2}m^2B_{\mu\nu}B^{\mu\nu} + L_{nm}\bigg]
 \label{S2}
\end{eqnarray}
where $L_{nm}$ is the non-minimal coupling term. The two cases with $L_{nm} = \frac{1}{2\kappa^2}\vartheta B_{\mu\nu}B^{\mu\nu}R$ and 
$L_{nm} = \frac{1}{2\kappa^2}\zeta B^{\lambda\nu}B^{\mu}_{\nu}R_{\lambda\mu}$ are discussed separately. As a consequence, it has been found 
that for both the $L_{nm}$, a de-Sitter solution can be achieved by the non-minimal model. However the constraints on model parameter in the two 
cases become different: in particular, for the first case the dS solution is obtained for $\vartheta > \frac{\kappa^2}{6}$, while the other coupling 
function allows a dS expansion of the universe for $\zeta > \frac{2\kappa^2}{3}$. The de-Sitter solution is indeed a stable one and also the field 
values remain sub-Planckian. Moreover such non-minimal models lead to the propagating speed of the gravitational wave as unity and thus 
the model becomes compatible with GW170817 \cite{Aashish:2020mlw}. Here it may be mentioned that the constraints on the model parameters 
in order to achieve a de-Sitter solution match with that coming from the demand that the gravitational wave speed is unity.\\

The F(R) inflationary model in presence of Kalb-Ramond field has an equivalent holographic origin where the generalized holographic cut-off can be 
expressed in terms of an integral \cite{Paul:2019hys}. For example, in the $R+R^2+KR$ model, the holographic cut-off ($L_{IR}$) is given by,
\begin{equation}
L_\mathrm{IR} = -\frac{1}{6\alpha\dot{H}^2a^6} \int dt a^6\dot{H} \left(1 - \frac{\kappa^2\rho_0}{a^{6}H^2}\right) \, ,
\label{quadratic_integral form2}
\end{equation}
where $\rho_0$ is a constant which can be equated to the energy density of the KR field at the horizon crossing. Inserting the above expression into 
the holographic Friedmann equation $H = \frac{1}{L_\mathrm{IR}}$ and after some simple algebra, one lands with the known field equations of the 
Starobinsky+KR model. For $\rho_0 = 0$, the above cut-off converges to that for the Starobinsky F(R) model \cite{Nojiri:2019kkp}. 
Furthermore, for $R+R^3+KR$ model, the corresponding holographic cut-off has the following expression, 
\begin{equation}
L_\mathrm{IR} = -\frac{\int dt a^{15/2}\dot{H} \left(1 - \frac{\kappa^2h_0}{a^{6}H^2}\right)}
{36\left[2\alpha\dot{H}^3a^{15/2} - 4\alpha H\int dt a^{15/2}H\dot{H}\left(H^3-9H\dot{H}-3\ddot{H}\right)\right]} \, .
\label{cubic_integral form2}
\end{equation}
which along with the Holographic Friedmann equation can reproduce the cosmological field equations for the cubic curvature F(R) in presence of KR field 
model. Similarly, in the context of exponential F(R) model i.e for $e^{\beta R} + KR$ model, the holographic cut-off 
can be determined as,
\begin{equation}
L_\mathrm{IR} = \frac{\int dt a^4\left(1 - 6\beta\dot{H}^2/H^2 - \frac{\kappa^2\rho_0}{a^6H^2}\right)}
{6\beta a^4\dot{H} + H\int dt a^4\left(\frac{1}{6\beta H^2} - \frac{\dot{H}}{H^2}\right)}\, .
\label{exp_integral form2}
\end{equation}
Thereby, the cosmology of F(R) gravity with Kalb-Ramond field can be realized from holographic origin. Apart from an integral form, the 
holographic cut-offs can also be expressed by some different ways like $L_\mathrm{IR}$ can be written in terms of the particle horizon and their derivatives 
or in terms of the future horizon and their derivatives \cite{Nojiri:2020wmh}. However, 
our understanding for the choice of fundamental viable cut-off still remains to be lacking. The comparison of such 
cut-offs for realistic description of the universe evolution in unified manner may help in better understanding of holographic principle.\\

\section{Kalb-Ramond field in Randall-Sundrum braneworld scenrio}
Let us discuss the Kalb-Ramond field from higher dimensional perspective, in particular we consider the Randall-Sundrum (RS) scenario where 
the rank-2 Kalb-Ramond field exits in the bulk, unlike to the case of electromagnetic field which is generally considered to be confined in the brane. 
The RS model consists of an extra spatial 
dimension in addition of usual four dimension.  The bulk spacetime is AdS in nature and the extra dimension is $S^1/Z_2$ compactified where the 
orbifolded points are identified with two 3-branes. If $\varphi$ is taken to be 
the extra dimensional angular coordinate, then $\varphi = 0$ (hidden brane) and $\varphi = \pi$ (visible brane) are 
the two branes, while the latter one is identified with our visible universe. The hidden brane tension is positive, while that in the visible brane 
is negative. In such scenario, one can solve the five dimensional Einstein equation and as a result, the RS metric is given by \cite{Randall:1999ee}, 
\begin{eqnarray}
 ds^2 = e^{-2kr_c|\varphi|} \eta_{\mu\nu} dx^{\mu} dx^{\nu} + r_c^2d\varphi^2
 \label{RS1_metric}
\end{eqnarray}
where $r_c$ is the interbrane separation and the factor $k$ is related to the five dimensional Planck scale $M$ as 
$k = \sqrt{\frac{-\Lambda}{24M^3}}$ with $\Lambda$ ($< 0$) being the bulk cosmological constant. 
The exponential factor is known as the warp factor and related to the effective four dimensional reduced Planck mass ($M_{Pl}$) as 
$M_{Pl}^2 = \frac{M^3}{k}\big[1 - e^{-2kr_c\pi}\big]$. For $kr_c = 12$ the exponential factor produces TeV scale mass parameters 
from the Planck scale when one considers projections on the ‘standard model’ brane and thus resolves the gauge hierarchy problem. The KR field action in 
such five dimensional model is (up to a multiplicative constant),
\begin{eqnarray}
 S_{KR} = \int d^4x~\int d\varphi \sqrt{-G} H_{MNL}H^{MNL}
 \label{RS1_KR action}
\end{eqnarray}
with $M$ runs from 0 to 4 and $H_{MNL} = \partial_{[M}B_{NL]}$. The KR field action is invariant under the gauge transformation 
$B_{MN} \rightarrow B_{MN} + \partial_{M}W_{N}$ (where $W_{N}$ being an arbitrary function of spacetime coordinates), which in turn allows us to 
set $B_{\varphi\mu} = 0$. Being an extra dimensional bulk field, the KR field has Kaluza-Klein mode decomposition as 
\begin{eqnarray}
 B_{\mu\nu}(x,\varphi) = \sum_{n=0}^{\infty}~B_{\mu\nu}^{(n)}(x)\frac{\chi^{(n)}(\varphi)}{\sqrt{r_c}}
 \label{RS1_KK decomposition}
\end{eqnarray}
where $B_{\mu\nu}^{(n)}(x)$ and $\chi^{(n)}(x,\varphi)$ are nth 
mode of the on-brane KR field and the extra dimensional KR wave function respectively. Plugging back such Kaluza-Klein mode decomposition into 
$S_{KR}$ and integrating over the extra dimensional coordinate yields the four dimensional effective theory of the KR field 
as follows \cite{Mukhopadhyaya:2002jn},
\begin{eqnarray}
 S_{KR} = \int d^4x~\sum_{n=0}^{\infty}~\big[\eta^{\mu\alpha}\eta^{\nu\beta}\eta^{\lambda\gamma}H_{\mu\nu\lambda}^{(n)}H_{\alpha\beta\gamma}^{(n)} 
 + 3m_n^2\eta^{\mu\alpha}\eta^{\nu\beta}B_{\mu\nu}^{(n)}B_{\alpha\beta}^{(n)}\big]
 \label{RS1_4d action}
\end{eqnarray}
provided $\chi^{(n)}(x,\phi)$ satisfies the following equation of motion, 
\begin{eqnarray}
 -\frac{1}{r_c^2}\frac{d^2\chi^{(n)}}{d\varphi^2} = m_n^2\chi^{(n)}e^{2kr_c\varphi}
 \label{RS1_eq1}
\end{eqnarray}
along with the normalization condition  as,
\begin{eqnarray}
 \int e^{2kr_c\varphi}\chi^{(m)}(\varphi)\chi^{(n)}(\varphi)d\varphi = \delta_{mn}
 \label{RS1_normalization}
\end{eqnarray}
with $m_n$ being the mass of nth Kaluza-Klein mode. The zeroth order of the KR wave function has the solution \cite{Mukhopadhyaya:2002jn},
\begin{eqnarray}
 \chi^{(0)} = \sqrt{kr_c}e^{-kr_c\pi}
 \label{RS1_solution}
\end{eqnarray}
which is constant throughout the bulk and experienced an exponential suppression due to the warping nature of extra dimension. 
Moreover, the massive KK modes can be solved by introducing a new variable 
$z_n = \frac{m_n}{k}e^{kr_c\varphi}$ and the solutions are given by zeroth order Bessel function. The overlap of the KR extra dimensional wave function with 
the visible brane actually determines the coupling of the KR field with other matter fields in our universe. As a consequence, the couplings of the 
zeroth mode of KR field with the matter fields are obtained as, $\lambda = \frac{1}{M_{Pl}}e^{-kr_c\pi}$. This is a remarkable fact. 
Although gravity and torsion are treated 
at par on the bulk, with the Planck mass characterizing any dimensional parameter controlling their 
interactions, the coupling of the zero-mode Kalb-Ramond field with matter fields experiences an additional exponential 
suppression over the usual gravity-matter coupling $1/M_p$, via the warp factor when the extra dimension is compactified by the RS 
scheme, unlike to the four dimensional model (as mentioned in the previous section) where the KR field has coupling strengths same 
as the gravity-matter coupling.\\
In the original RS scenario, the two branes are considered to be Minkowskian i.e the effective on-brane cosmological constant is assumed to be zero. 
However one can relax this assumption and as a result, a generalized version of RS model 
has been introduced where the brane geometry is different than the Minkowskian, in particular they can be either dS or AdS depending on 
the effective cosmological constant \cite{Das:2007qn}. In the case when the branes are dS, the five dimensional metric comes as,
\begin{eqnarray}
 ds^2 = \omega\sinh{\bigg(\ln{\frac{c_2}{\omega}} - kr_c\varphi\bigg)} g_{\mu\nu}(x) dx^{\mu} dx^{\nu} + r_c^2d\varphi^2
 \label{RS2_metric1}
\end{eqnarray}
where $c_2 = 1+\sqrt{1+\omega^2}$, $\omega^2 = \frac{\Omega}{3k^2}$ and $\Omega > 0$ is the on-brane cosmological constant. Moreover $g_{\mu\nu}(x)$ 
is the dS metric obtained from $\Omega$. On other hand, when the branes are AdS, the metric comes as,
\begin{eqnarray}
 ds^2 = \omega\cosh{\bigg(\ln{\frac{\omega}{c_1}} + kr_c\varphi\bigg)} g_{\mu\nu}(x) dx^{\mu} dx^{\nu} + r_c^2d\varphi^2
 \label{RS2_metric1}
\end{eqnarray}
where $c_1 = 1+\sqrt{1-\omega^2}$, $\omega^2 = -\frac{\Omega}{3k^2}$ and $\Omega < 0$ is the on-brane cosmological constant. It may be observed that 
in order to get a real solution, the brane cosmological constant is constrained in the AdS case, while there is no such constraint imposed 
for the dS $\Omega$. Now if we consider a bulk Kalb-Ramond field in such generalized RS model and adopt the same Kaluza-Klein decomposition as mentioned 
earlier, then the couplings of the zeroth mode KR field with the other matter fields are found to be modified due to the 
presence of the brane cosmological constant \cite{Das:2010xx}. It is shown that due to the constraints on the magnitude of 
the cosmological constant in AdS 3-brane these couplings continue to be small. 
However the scenario becomes different for de-Sitter 3-brane where the couplings may or may not be small, depending upon the value of $\Omega$. 
In particular, 
for the dS case, the couplings of the KR field acquire a large value for a large cosmological constant. Such situation may occur 
in very early stage of the universe where a model with a large cosmological constant is invoked to explain inflationary phase of the universe. 
Thus the antisymmetric tensor fields which are invisible in the present epoch will be an inseparable part in describing 
Physics at the fundamental scale. To understand the possible effects of KR field during early universe from higher dimensional angle, we have to 
discuss the cosmology of this model from early stage of the universe, which is the subject of the next section. In such higher dimensional 
cosmological scenario, we encounter with the dynamical stabilization of the extra dimensional modulus field where the higher curvature term 
in the gravitational action act as a suitable stabilizing agent.\\

\section{Cosmology with Kalb-Ramond in higher curvature warped spacetime}

\begin{figure}[!h]
\begin{center}
 \centering
 \includegraphics[width=3.5in,height=2.5in]{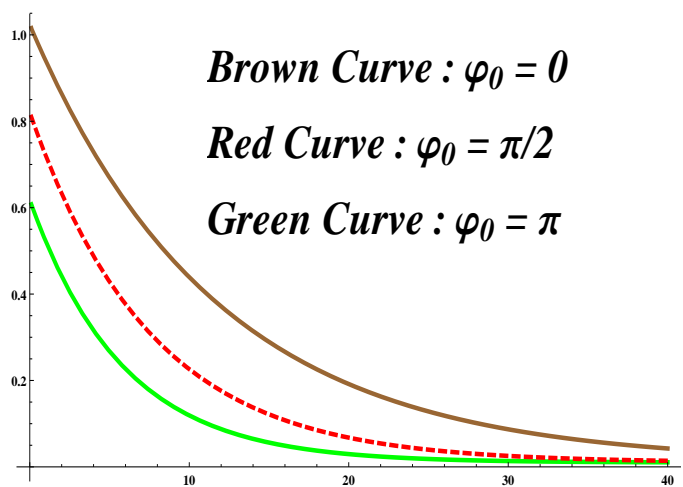}
 \caption{$\chi^{(0)}_{\varphi_0}(t,\varphi)$ vs $t$ for $\gamma = 0.15$, $\kappa v_v = \frac{\sqrt{\Omega_0}}{M^2} \simeq 10^{-7}$, 
 $\frac{m_{\Phi}}{k} = 0.2$ and $\Psi_0 = 36$ (in 4D reduced Planckian unit).}
 \label{plot_observableII}
\end{center}
\end{figure}

We consider a warped braneworld model in presence of Kalb-Ramond field exists in the bulk together with gravity. In such braneworld 
scenario, due to the intervening gravity, the branes must collapse to each other and thus the stabilization issue of the extra dimension 
is an important aspect to address in such model. In order to stabilize the modulus (the distance between the branes), one needs 
to generate an appropriate modulus potential which has indeed a stable configuration. Such kind of modulus 
potential can be generated by considering a massive scalar field in the bulk which has specific boundary values on the branes: this mechanism has been 
introduced by Goldberger and Wise and thus the name goes Goldberger-Wise (GW) stabilization mechanism. here we would like to mention that 
in the original GW mechanism, the stabilization is non-dynamical, but here we will deal with the cosmological evolution and thus we apply the 
dynamical stabilization mechanism where the modulus field dynamically goes to the stable value. In the present context, 
we consider the higher curvature term, in particular the quadratic curvature term, in the bulk as the stabilizing agent. 
The higher curvature terms become dominant in the large curvature regime and thus the RS scenario where the bulk curvature is of the order 
of Planck scale directs us to consider the higher curvature term in addition to the Einstein-Hilbert term in the five dimensional action. 
Thus the action of the model is \cite{Elizalde:2018rmz},
\begin{eqnarray}
 S&=&\int d^4xd\varphi \sqrt{-G} \bigg[\frac{1}{2\kappa_\mathrm{5}^2}\big(R + \alpha R^2\big) - \Lambda + V_h\delta(\varphi) + V_v\delta(\varphi-\pi) 
 - \frac{1}{12}H_{MNL}H^{MNL}\bigg]\nonumber\\
 &=&S_{g} + S_{KR}
 \label{5d actionII}
\end{eqnarray}
where $\alpha$ is a parameter having mass dimension [-2], $\Lambda$ is the bulk cosmological constant which is indeed negative and 
$\kappa_\mathrm{5}^2 = \frac{1}{M^3}$ (with $M$ being the five dimensional Planck scale). 
Moreover $V_h$, $V_v$ are the brane tensions confined to the hidden and visible brane respectively (that is why there are two delta 
functions in the action), $H_{MNL}$ is the KR field strength tensor as usual. Being a closed string mode, the KR field is allowed to propagate 
in the five dimensional bulk, unlike to the electromagnetic and other matter fields which are generally considered to be confined on the brane. 
The overlap of the extra dimensional KR field with the visible brane (i.e the $\varphi = \pi$ brane) actually determines the strength of the KR field 
and hence its interaction with other matter fields in our visible universe. Further, as a 
bulk field, the KR field has Kaluza-Klein modes which obviously get coupled with the extra dimensional modulus. In such situation, it is important to 
discuss the dynamics of the modulus field which in turn controls the evolution of the Kaluza-Klein excitation of the KR field. These issues are 
addressed in the following from the perspective of four dimensional effective theory.\\
The effective action of $S_g$ requires the solution for the five dimensional metric which, in the present scenario, is given by,
\begin{equation}
 ds^2 = e^{-\frac{1}{\sqrt{3}}\kappa_\mathrm{5}\Phi(\varphi)} \bigg[e^{-2kr_c|\varphi|} \eta_{\mu\nu} dx^{\mu} dx^{\nu} + r_c^2d\varphi^2\bigg]
 \label{grav.sol1.F(R)II}
\end{equation}
 where $r_c$ is the interbrane separation, $\varphi$ is the extra dimensional angular coordinate and recall $k = \sqrt{\frac{-\Lambda}{24M^3}}$. Moreover 
 $\Phi(\varphi) = \langle\Phi\rangle + \xi(\varphi)$ with 
 $\langle\Phi\rangle = \frac{2}{\kappa_\mathrm{5}\sqrt{3}}\ln{\big[\sqrt{9 - 40\kappa^2\alpha\Lambda} - 2\big]}$ and $\xi(\varphi)$ has the following 
 expression,
 \begin{equation}
 \xi(\varphi) = e^{2kr_c|\varphi|} \big[Ae^{\nu kr_c|\varphi|} + Be^{-\nu kr_c|\varphi|}\big]
 \label{sol.scalar.fieldII}
\end{equation}
with $\nu = \sqrt{4 + m_{\Phi}^2/k^2}$ and 
$m_{\Phi}^2 = \frac{1}{8\alpha} \big[\sqrt{9 - 40\kappa^2\alpha\Lambda}\big]\big[\sqrt{9 - 40\kappa^2\alpha\Lambda} - 2\big]^{-\frac{2}{3}}$. 
Further $A$, $B$ are integration constants which can be obtained from the boundary conditions: $\xi(0)=v_h$ and $\xi(\pi)=v_h$. The 
five dimensional metric solution is obtained by using the conformal correspondence of an F(R) theory to a scalar-tensor theory. With 
the help of such conformal connection, one can find the solution in the scalar-tensor theory and then transform back the solutions in the 
original F(R) model by the inverse conformal transformation. In the present context, the scalar-tensor model is the RS scenario 
with a massive scalar field in the bulk (which is symbolized by $\Phi$ appeared from the quadratic curvature term). It may be mentioned that 
$\Phi$ is endowed within a stable potential having the minimum at $\langle\Phi\rangle$ and moreover the $m_{\Phi}$ is the mass of the 
scalar field (the explicit expressions of $\langle\Phi\rangle$ and $m_{\Phi}$ have been showed earlier). The stable value of the 
scalar potential contributes an effective cosmological constant in the bulk i.e $\Lambda_{eff} = \Lambda + V(\langle\Phi\rangle)$ 
which is indeed negative i.e the bulk spacetime in the scalar-tensor theory is still AdS in nature. 
Considering $\kappa v_h < 1$ (which has been also considered in \cite{Goldberger:1999uk}), 
it can be shown that the energy-momentum tensor of the scalar field 
is lesser than that of the bulk cosmological constant and in view of this condition, the backreaction of the scalar field can be safely ignored in the 
background five dimensional spacetime. Thus the spacetime metric in the ST model is same as Randall-Sundrum metric and consequently the fluctuation 
of the scalar field around $\langle\Phi\rangle$ is given in Eq.(\ref{sol.scalar.fieldII}). Using these solution of ST model, one can obtain 
the metric of the original F(R) model by using the inverse conformal transformation and shown in Eq.(\ref{grav.sol1.F(R)II}) where 
$e^{-\frac{1}{\sqrt{3}}\kappa_\mathrm{5}\Phi(\varphi)}$ is the inverse conformal factor. In order to introduce the radion field, the interbrane separation 
$r_c$ is replaced by a field symbolized by $T(x)$, known as radion field, and here for simplicity this new field is taken as the function only of the 
brane coordinates. Thus the line element turns out to be,
\begin{eqnarray}
 ds^2 = e^{-\frac{1}{\sqrt{3}}\kappa_\mathrm{5}\Phi(x,\varphi)} \bigg[e^{- 2 kT(x)|\varphi|} 
 g_{\mu\nu}(x) dx^{\mu} dx^{\nu} + T(x)^2d\varphi^2\bigg]
 \label{grav.sol2.F(R)}
\end{eqnarray}
where $g_{\mu\nu}(x)$ is the induced on-brane metric and $\Phi(x,\varphi)$ can be obtained from eqn.(\ref{sol.scalar.fieldII}) by 
replacing $r_c$ to $T(x)$. Plugging back the above solution of $G_{MN}$ into the action $S_{g}[G_{MN}]$ and integrating over extra 
dimensional coordinate ($\varphi$) yields the effective four dimensional on-brane action as follows :
\begin{eqnarray}
 A_{eff}^{1} = \int d^4x \sqrt{-g} \bigg[\frac{1}{2}M_{4}^2R_{(4)} - \frac{1}{2}g^{\mu\nu}\partial_{\mu}\Psi\partial_{\nu}\Psi - U_{rad}(\Psi)\bigg]
 \label{effective action1II}
\end{eqnarray}
where $M_{(4)}^2= \frac{M^3}{k}\bigg[\sqrt{9 - 40\kappa^2\alpha\Lambda} - 2\bigg]^{1/2}$ is the 
effective four dimensional reduced Planck scale in the context of higher curvature warped spacetime, 
$R_{(4)}$ is the on-brane Ricci scalar formed by $g_{\mu\nu}(x)$. Moreover,  
$\Psi(x) = \sqrt{\frac{24M^3}{k}} \big[1 + \frac{20}{\sqrt{3}}\alpha k^2\kappa_\mathrm{5} v_h\big] e^{-k\pi T(x)} = fe^{-k\pi T(x)}$ 
(with $f = \sqrt{\frac{24M^3}{k}} [1 + \frac{20}{\sqrt{3}}\alpha k^2\kappa_\mathrm{5} v_h]$), is the 
canonical radion field and $U_{rad}(\Psi)$ is the radion potential having the following form \cite{Das:2017htt}
\begin{eqnarray}
 U_{rad}(\Psi) = \frac{20}{\sqrt{3}}\frac{\alpha k^5}{M^6} \Psi^4 
 \bigg[(v_h - \frac{\kappa_\mathrm{5} v_h^2}{2\sqrt{3}} + \frac{\kappa_\mathrm{5} v_hv_v}{2\sqrt{3}})(\Psi/f)^\omega - v_v\bigg]^2
 \label{potential.radion.F(R)II}
\end{eqnarray}
where the terms proportional to  $\omega$ ($= \frac{m_{\Phi}^2}{k^2}<1$, which is also consistent with observational bounds) 
are neglected. On projecting the bulk gravity on the brane, the extra dimensional component of the metric appears 
as a scalar field (i.e $\Psi$, the radion field) in the on-brane effective theory. It may be observed that $U_{rad}(\Psi)$ is proportional 
to the higher curvature parameter $\alpha$ and thus vanishes for $\alpha \rightarrow 0$. This is expected, however, because for $\alpha \rightarrow 0$, the 
gravitational action contains only the Einstein-Hilbert term which is not able to generate a radion potential. Thereby, in the present context, 
the radion potential is generated entirely due to the presence of the higher curvature term and hence it can be argued that the sign of the higher curvature 
comes through the radion potential in the effective theory. Moreover, as mentioned earlier, the modulus stabilization depends on the fact that whether 
the radion potential has a stable minimum or not. It can be easily checked that the $U_{rad}(\Psi)$ has a minimum at 
$\langle\Psi\rangle 
= \bigg[\frac{v_v f^{\omega}}{\big(v_h - \frac{\kappa_\mathrm{5} v_h^2}{2\sqrt{3}} + \frac{\kappa_\mathrm{5} v_vv_h}{2\sqrt{3}}\big)}\bigg]^{1/\omega}$, 
which immediately leads to the stable value of the interbrane separation as,
\begin{eqnarray}
 k\pi\langle T(x)\rangle = \frac{4k^2}{m_{\Phi}^2}[\ln{(\frac{v_h}{v_v})} - \frac{\kappa_\mathrm{5} v_v}{2\sqrt{3}}(\frac{v_h}{v_v} - 1)]
 \label{brane separationII}
\end{eqnarray}
The aforementioned expression of $m_{\Phi}$ clearly indicates that $\langle T(x)\rangle$ 
is proportional to $\alpha$, which once again confirms the fact that 
the interbrane separation is stabilized due to the presence of the higher curvature term ($\alpha R^2$) in the bulk. Having determined the effective action 
for $S_g[G_{MN}]$, now we move to calculate the same for the KR field action i.e for 
$S_{KR} = -\frac{1}{12}\int d^4xd\varphi \sqrt{-G}\big[H_{MNL}H^{MNL}\big]$. Plugging the Kaluza-Klein mode decomposition 
$B_{\mu\nu}(x,\varphi) = \sum B_{\mu\nu}^{(n)}(x)\chi^{(n)}(x,\varphi)$ into 
$S_{KR}$ and integrating over the extra dimensional coordinate yields the four dimensional effective theory of the KR field as follows,
\begin{eqnarray}
 A_{eff}^{(2)} = -\frac{1}{12} \int d^4x \sqrt{-g}
 \bigg[g^{\mu\alpha}g^{\nu\beta}g^{\lambda\gamma}H_{\mu\nu\lambda}^{(n)}H_{\alpha\beta\gamma}^{(n)} 
 + 3m_n^2 g^{\mu\alpha}g^{\nu\beta}B_{\mu\nu}^{(n)}B_{\alpha\beta}^{(n)}\bigg]
 \label{effective action2II}
\end{eqnarray}
provided $\chi^{(n)}(x,\phi)$ satisfies the following equation of motion, 
\begin{eqnarray}
 \frac{\partial\chi^{(n)}}{\partial t} \frac{\partial\chi^{(m)}}{\partial t} 
 - \frac{1}{T^2(t)}e^{-2kT(t)\varphi}\chi^{(n)}\frac{\partial^2\chi^{(m)}}{\partial\varphi^2} = m_n^2\chi^{(n)}\chi^{(m)}
 \label{wave function equation}
\end{eqnarray}
along with the normalization condition  as,
\begin{eqnarray}
 \int_0^{\pi} d\varphi e^{2kT(t)\varphi}\chi^{(n)}\chi^{(m)} = \frac{1}{T^2(t)}\delta_{mn}
 \label{wave function normalization}
\end{eqnarray}
where $m_n$ denotes the mass of nth Kaluza-Klein mode. The extra dimensional KR wave function, in particular the overlap of 
$\chi^{(n)}(x,\varphi)$ with the visible brane plays a crucial role in determining the couplings of KR field with various 
Standard Model fields in our visible universe. Furthermore, Eq.(\ref{wave function equation}) clearly indicates that the $\chi^{(n)}(x,\varphi)$ 
gets coupled with the five dimensional modulus field. Coming back to the effective action, Eqs.(\ref{effective action1II}) and (\ref{effective action2II}) 
represent the effective actions for $S_{g}[G_{MN}]$ and for $S_{KR}$ respectively. Thus the full form of the four dimensional effective action 
can be written as,
\begin{eqnarray}
 A_{eff}&=&A_{eff}^{(1)} + A_{eff}^{(2)}\nonumber\\
 &=&\int d^4x\sqrt{-g}\bigg[\frac{1}{2}M_{(4)}^2R^{(4)} - \frac{1}{2}g^{\mu\nu}\partial_{\mu}\Psi\partial_{\nu}\Psi - U_{rad}(\Psi) 
 + g^{\mu\alpha}g^{\nu\beta}g^{\lambda\gamma}H_{\mu\nu\lambda}^{(0)}H_{\alpha\beta\gamma}^{(0)}\bigg]
 \label{full effective action II}
\end{eqnarray}
Here we consider the zeroth Kaluza-Klein mode of the KR field for which $m_{n=0} = 0$. The general form of the effective action 
contains the excited KK modes as well, 
which in turn breaks the Kalb-Ramond gauge invariance of the on-brane effective action. The same argument also holds for the electromagnetic field 
in the situation where the em field is considered to be a bulk field. However, in the present context, we consider only 
the zeroth Kaluza-Klein mode and thus the Kalb-Ramond gauge invariance retains in the four dimensional effective action. The four dimensional 
effective action has three independent fields : $g_{\mu\nu}(x)$, $\Psi(x)$ and $B_{\mu\nu}(x)$ and besides these three fields, there is 
the extra dimensional KR wave function $\chi^{(n)}(x,\phi)$ obeying Eq.(\ref{wave function equation}). The evolution of $B_{\mu\nu}(x)$ 
(in the cosmological context) determines the energy density of the on-brane KR field with the expansion of the universe, while $\chi^{(n)}(x,\phi)$ 
at $\varphi = \pi$ fixes the couplings of the KR field with other matter fields on our visible brane. 
Thus to describe the footprint of the KR field 
on our universe, we need to work out for both the $B_{\mu\nu}(x)$ and $\chi^{(n)}(x,\phi)$. Due to the presence of the potential term, the 
radion field acquires a certain dynamics, which in turn affects the evolution of the KR field wave function (as $\Psi$ and 
$\chi^{(n)}(x,\phi)$ are coupled through Eq.(\ref{wave function equation})). So it is important to investigate whether the cosmological evolution of such 
coupled fields results to a negligible footprint of the KR field in the present energy scale of the universe. Moreover as we are working 
in the Randall-Sundrum model, the resolution of the gauge hierarchy problem is also a crucial aspect to address. So we have to check whether 
the evolution of the modulus field leads to a stabilized value of the interbrane separation 
and consequently solves the gauge hierarchy problem. 
For this purpose, we try to solve the Friedmann equations obtained from $A_{eff}$ and thus the on-brane metric ansatz is taken as FRW one i.e
\begin{eqnarray}
 ds_{(4)}^2 = g_{\mu\nu}(x)dx^{\mu}dx^{\nu} = -dt^2 + b^2(t)\big[dx^2 + dy^2 + dz^2\big]
 \label{4d metricII}
\end{eqnarray}
with $b(t)$ being the on-brane scale factor. In presence of the KR field, the off-diagonal Einstein equations become non-trivial and as a consequence, 
only one component of the KR field, in particular $H_{123}^{(0)} = h_4$ comes as a non-zero component. With this information, the diagonal Freidmann 
equations for $A_{eff}$ becomes,
\begin{eqnarray}
 0&=&-3H_b^2 + \frac{1}{2}\dot{\Psi}^2 + \frac{20}{\sqrt{3}}\frac{\alpha k^5}{M^6}v_v^2 \Psi^4 \bigg[F\Psi^\omega - 1\bigg]^2 + \frac{1}{2}h_4h^4\nonumber\\
 0&=&2\dot{H}_b + 3H_b^2 + \frac{1}{2}\dot{\Psi}^2 - \frac{20}{\sqrt{3}}\frac{\alpha k^5}{M^6}v_v^2 \Psi^4 \bigg[F\Psi^\omega - 1\bigg]^2 
 + \frac{1}{2}h_4h^4
 \label{einstein equation2II}
\end{eqnarray}
where $H_{b}=\frac{\dot{b}}{b}$ is known as on-brane Hubble parameter and 
$F = \frac{1}{v_vf^{\omega}}\big(v_h - \frac{\kappa_\mathrm{5} v_h^2}{2\sqrt{3}} + \frac{\kappa_\mathrm{5} v_vv_h}{2\sqrt{3}}\big)$. 
Moreover, the field equations 
for the scalar field and for the KR field are given by,
\begin{eqnarray}
 \nabla_{\mu}H^{\mu\nu\lambda(0)} = \frac{1}{\sqrt{-g}}\partial_{\mu}\bigg[\sqrt{-g}H^{\mu\nu\lambda(0)}\bigg] = 0
 \label{KR equationII}
\end{eqnarray}
and
\begin{eqnarray}
 \ddot{\Psi} + 3H\dot{\Psi} + \frac{80}{\sqrt{3}}\frac{\alpha k^5}{M^6}v_v^2 \Psi^3 \bigg[F\Psi^\omega - 1\bigg]^2 = 0
 \label{radion equationII}
\end{eqnarray}
respectively. After a little bit juggling, the above field equations leads to the evolution of the KR field energy density 
as $\frac{d}{dt}\rho_{KR} = -6H_b\rho_{KR}$ which can be solved to yield $\rho_{KR} = \frac{\Omega_0}{b^6}$. Thus the KR field energy density 
decreases with the expansion of the universe at a faster rate in comparison to matter and radiation energy density respectively. However 
at the same time, the evolution of $\rho_{KR}$ also indicates that the KR field amplitude is large and may play a significant role during 
the early universe. This directs us to discuss the evolution of the KR field from very early universe when the investigation of inflation 
becomes also important. Keeping this in mind, we solve the field equations for the radion field and for the scale factor 
at early universe and for this purpose we consider the slow roll approximating according to which the potential term of the radion is much 
greater than that of the kinetic term i.e $U_{rad}(\Psi) \gg \frac{1}{2}\dot{\Psi}^2$. Under the slow roll condition, the field equations 
can be solved and the solutions are given by,
\begin{eqnarray}
 \Psi(t) = \frac{\Psi_0}{\bigg[F\Psi_0^{\omega}-\big(F\Psi_0^{\omega}-\frac{\sqrt{\Omega_0}}{b_0^3\xi_0^2}-1\big)\exp
 {\big(-8\omega v_v\sqrt{\frac{5}{3\sqrt{3}}\frac{\alpha k^5}{M^6}}(t-t_0)\big)}\bigg]^{1/\omega}}
 \label{sol of radionII}
\end{eqnarray}
and
\begin{eqnarray}
 b(t) = C \bigg[1 + \sqrt{3\Omega_0}(t-t_0)\bigg]^{1/3} \exp{\bigg[2v_v\sqrt{\frac{5}{3\sqrt{3}}\frac{\alpha k^5}{M^6}} \big(g_1(t)-g_2(t)\big)\bigg]} ,
 \label{sol of scaleII}
\end{eqnarray}
where $\Psi_0$, $C$ are integration constants and recall 
$F = \frac{1}{v_vf^{\omega}}\big(v_h - \frac{\kappa_\mathrm{5} v_h^2}{2\sqrt{3}} + \frac{\kappa_\mathrm{5} v_vv_h}{2\sqrt{3}}\big)$. 
Furthermore $g_1(t)$ has the following expressions: 
\begin{eqnarray}
 g_1(t)&=&-\frac{F\Psi_0^{\omega}}{(F\Psi_0^{\omega}-1)} 
 \frac{\Psi_0^2}{16\omega v_v\sqrt{\frac{5}{3\sqrt{3}}\frac{\alpha k^5}{M^6}}} 
 2F1\bigg(1,1,2+\frac{2}{\omega},\frac{F\Psi_0^{\omega}}{F\Psi_0^{\omega}-1}
 \exp{\big(8\omega v_v\sqrt{\frac{5}{3\sqrt{3}}\frac{\alpha k^5}{M^6}}(t-t_0)\big)}\bigg)\nonumber\\
 &\exp&{\bigg(8\omega v_v\sqrt{\frac{5}{3\sqrt{3}}\frac{\alpha k^5}{M^6}}(t-t_0)\bigg)} 
 \bigg(F\Psi_0^{\omega}-(F\Psi_0^{\omega}-1) 
 \exp{\big(-8\omega v_v\sqrt{\frac{5}{3\sqrt{3}}\frac{\alpha k^5}{M^6}}(t-t_0)\big)}\bigg)^{-2/\omega}
 \label{g1II}
\end{eqnarray}
where $2F1$ symbolizes the hypergeometric function. On the other hand, $g_2(t)$ is given by,
\begin{eqnarray}
 g_2(t)&=&-\frac{\Psi_0^{\omega}}{(F\Psi_0^{\omega}-1)} \frac{\Psi_0^2}{16\omega v_v\sqrt{\frac{5}{3\sqrt{3}}\frac{\alpha k^5}{M^6}}} 
 2F1\bigg(1,1,1+\frac{2}{\omega},\frac{F\Psi_0^{\omega}}{F\Psi_0^{\omega}-1}
 \exp{\big(8\omega v_v\sqrt{\frac{5}{3\sqrt{3}}\frac{\alpha k^5}{M^6}}(t-t_0)\big)}\bigg)\nonumber\\
 &\exp&{\bigg(8\omega v_v\sqrt{\frac{5}{3\sqrt{3}}\frac{\alpha k^5}{M^6}}(t-t_0)\bigg)} \bigg(F\Psi_0^{\omega}-(F\Psi_0^{\omega}-1) 
 \exp{\big(-8\omega v_v\sqrt{\frac{5}{3\sqrt{3}}\frac{\alpha k^5}{M^6}}(t-t_0)\big)}\bigg)^{1-2/\omega}.
 \label{g2II}
\end{eqnarray}
In absence of the higher curvature term i.e for $\alpha = 0$, the above solutions become 
$\Psi(t) = \frac{\Psi_0}{\big[1 + \frac{\sqrt{\Omega_0}}{b_0^3\xi_0^2}\big]^{1/\omega}}=\Psi(t_0)$ and $H \propto \frac{1}{b^3}$ respectively. This is 
due to the fact that without the higher curvature term, the radion potential vanishes leading to the radion field as a non-dynamical one and moreover 
the evolution of the Hubble parameter is controlled entirely by the KR field for which the $H(t)$ goes as $1/b^3$. On other hand, for 
$\Omega  = 0$, the solutions of $\Psi(t)$ and $b(t)$ converge to that of the higher curvature RS model in absence of the KR field 
\cite{Banerjee:2017lxi}. Coming back to the 
solutions of the present model where both the higher curvature and the KR field  are present, Eq.(\ref{sol of radionII}) shows that the radion field 
asymptotically reaches to a constant value which is indeed its vacuum expectation value (vev) i.e
\begin{eqnarray}
 \Psi(t\gg t_0) = f\bigg[\frac{v_v}{v_h}\bigg]^{1/\omega} = <\Psi>
 \nonumber
\end{eqnarray}
Thus, in the present context, the evolution of the interbrane separation ($T(t)$) can be demonstrated as : $T(t)$ increases 
(as $\Psi(t)\propto e^{-k\pi T(t)}$) with the expansion of the visible brane scale factor and gradually goes to its stable value 
($k\pi\langle T\rangle = \frac{4k^2}{m_{\Phi}^2}[\ln{(\frac{v_h}{v_v})} - \frac{\kappa_\mathrm{5} v_v}{2\sqrt{3}}(\frac{v_h}{v_v} - 1)]$) asymptotically. 
Thereby we can argue that the extra dimensional modulus gets dynamically stabilized for $t \gg t_0$. The stabilized modulus is connected 
to the gauge hierarchy solution, in particular $\langle T\rangle$ has to acquire $k\pi\langle T\rangle = 36$ in order to solve 
the gauge hierarchy problem. To check whether this condition is satisfied in the present model, we need to scan the parameters of the model, which we will 
do later.\\
The scale factor in Eq.(\ref{sol of scaleII}) corresponds to an inflationary scenario at the early universe if the condition 
\begin{eqnarray}
2(F\Psi_0^{\omega}-1) \Psi_0^2 v_v\sqrt{\frac{5}{3\sqrt{3}}\frac{\alpha k^5}{M^6}} > \sqrt{\Omega_0}\big(1-\frac{1}{\sqrt{3}}\big)
\label{conditionII}
\end{eqnarray}
holds true. To see this, one needs to determine the acceleration of $b(t)$ at the limit $t \rightarrow t_0$. The above condition actually 
reflects the interplay between the higher curvature and the Kalb-Ramond field strengths in determining whether the early universe undergoes through 
an inflationary stage or not. This is expected because the quadratic curvature term in the gravitational action triggers an accelerating effect while 
the KR field produces a decelerating effect in the expansion of the universe. However in the higher curvature RS model where the KR field is also present in 
the bulk, we stick to the condition (\ref{conditionII}) to allow an inflationary stage in the early universe. The scale factor $b(t)$ also allows an end of 
the inflation at $\Psi = 2\sqrt{2}$ (in four dimensional reduced Planckian unit), in particular the inflationary 
era continues as long as the radion field remains greater than 
$\Psi_f = 2\sqrt{2}$ (in four dimensional reduced Planckian unit). 
This along with the aforementioned radion field solution yields the duration of inflation $t_f - t_0$ as,
\begin{eqnarray}
 t_f-t_0 = \frac{1}{8\omega v_v\sqrt{\frac{5}{3\sqrt{3}}\frac{\alpha k^5}{M^6}}}
 \ln\bigg[\frac{F\Psi_0^{\omega} - 1 - \frac{\sqrt{\Omega_0}}{b_0^3\Psi_0^2}}
 {F\Psi_0^{\omega} - \frac{\Psi_0^{\omega}}{\Psi_f^{\omega}}}\bigg]
 \label{durationII}
\end{eqnarray}
which is indeed finite. Having determined the solutions of the field equations, now we move to confront the model with the latest Planck results, in 
particular we calculate the spectral index ($n_s$), tensor to scalar ratio ($r$) in the present context and match the theoretical expectations 
of $n_s$ and $r$ with the Planck constraints given by $n_s = 0.9649 \pm 0.0042$ and $r < 0.064$. For the four dimensional effective action we are 
working with, the spectral index and tensor to scalar are defined as,
\begin{eqnarray}
 n_s = 1 - 6\epsilon_b\bigg|_{h.c} - 2\frac{\dot{\epsilon_b}}{H_b\epsilon_b}\bigg|_{h.c}~~~~~~~,~~~~~~~~r = 16\epsilon_b\bigg|_{h.c}
 \label{definition}
\end{eqnarray}
where $\epsilon_b = -\frac{\dot{H}_b}{H_b^2}$ and the observable quantities are defined at the time of horizon crossing when 
the perturbation momentum $k$ satisfies $k = aH$. Clearly the horizon crossing time varies with the momentum $k$, however to determine the 
inflationary observable quantities, we use the CMB scale momentum for which the e-folding of the inflationary duration becomes $N \simeq 55-60$. 
Using the slow roll field equations and after some simplifications, one will get the final form of $n_s$ and $r$ as follows:
\begin{eqnarray}
 n_s&=&1 - \frac{U_1}{U_2}\nonumber\\
 r&=&8\bigg[\frac{16p^2v_v^4\xi_0^6(F\Psi_0^{\omega}-1)^4 + \frac{3\Omega_0}{b_0^6}\bigg(pv_v^2\Psi_0^4(F\Psi_0^{\omega}-1)^2
 +\frac{\Omega_0}{2b_0^6}\bigg)}
 {\bigg(pv_v^2\Psi_0^4(F\Psi_0^{\omega}-1)^2 + \frac{\Omega_0}{2b_0^6}\bigg)^2}\bigg]
 \label{spectra index}
\end{eqnarray}
where $U_1$ and $U_2$ have the following expressions:
\begin{eqnarray}
 U_1&=&\bigg[384p^3v_v^6\Psi_0^8(F\Psi_0^{\omega}-1)^6 
 + \frac{18\Omega_0}{b_0^6}\bigg(pv_v^2\Psi_0^4(F\Psi_0^{\omega}-1)^2 + \frac{\Omega_0}{2b_0^6}\bigg)^2 
 - \frac{144\Omega_0}{b_0^6}p^2v_v^4\Psi_0^6(F\Psi_0^{\omega}-1)^4\nonumber\\
 &-&\frac{6\Omega_0}{b_0^6}\bigg(16p^2v_v^4\Psi_0^6(F\Psi_0^{\omega}-1)^4 + \frac{3\Omega_0}{b_0^6}
 \bigg(pv_v^2\Psi_0^4(F\Psi_0^{\omega}-1)^2 + \frac{\Omega_0}{2b_0^6}\bigg)\bigg)\bigg]
 \nonumber
\end{eqnarray}
and
\begin{eqnarray}
 U_2&=&\bigg(pv_v^2\Psi_0^4(F\Psi_0^{\omega}-1)^2 + \frac{\Omega_0}{2b_0^6}\bigg)\bigg(16p^2v_v^4\Psi_0^6(F\Psi_0^{\omega}-1)^4 
 + \frac{3\Omega_0}{b_0^6}\bigg(pv_v^2\Psi_0^4(F\Psi_0^{\omega}-1)^2 + \frac{\Omega_0}{2b_0^6}\bigg)\bigg)
 \nonumber
\end{eqnarray}
respectively. The $n_s$ and $r$ depend on the parameters $v_v$, $\Omega_0$ and $\Psi_0$ and to fix these parameters we use the Planck constraints. 
Here we take $\kappa v_v = \frac{\sqrt{\Omega_0}}{M^2} \simeq 10^{-7}$ which is also consistent with the condition that is necessary for neglecting the 
backreaction of the bulk scalar field in the background five dimensional spacetime. With such estimations of $v_v$ and $\Omega_0$, the spectral index 
and tensor to scalar ratio become simultaneously compatible with the Planck observations for the regime $34 \lesssim \Psi_0 \lesssim 38$ 
(in 4D reduced Planckian unit). Moreover the duration of inflation is of the order 
$t_f - t_0 = \Delta t \sim 10^{-10}\mathrm{GeV}^{-1}$ if the ratio $\frac{m_{\Phi}}{k}$ (the ratio of the bulk scalar field mass 
to the bulk curvature) is taken as $0.2$. Such consideration of $\frac{m_{\Phi}}{k}$ resolves the gauge hierarchy problem concomitantly. 
For $\Psi_0 = 36$ (in 4D reduced Planckian unit) and $\frac{m_{\Phi}}{k} = 0.2$, the e-folding number of the inflationary duration comes as 
$N \simeq 58$, which confirms that the values of $n_s$ and $r$ we have determined are valid for the CMB scale momentum $k \sim 0.1\mathrm{Mpc}^{-1}$.\\
At this stage it deserves mentioning that the higher curvature braneworld model without the Kalb-Ramond field is able to provide 
an on-brane inflationary scenario with a graceful exit, but is not viable in respect to the Planck 2018 observational results. In particular, for 
the braneworld model without the KR field, the spectral index and tensor to scalar ratio follow Eq.(\ref{spectra index}) with $\Omega_0 = 0$; 
such expressions of $n_s$ and $r$ (with $\Omega_0 = 0$) are found to be compatible with the Planck 2015 results ($n_s = 0.968 \pm 0.006$ and 
$r < 0.14$) \cite{Ade:2015lrj}, but, as just mentioned, not with the 
latest Planck 2018 constraints ($n_s = 0.9649 \pm 0.0042$ and $r < 0.064$) \cite{Akrami:2018odb,Banerjee:2017lxi}. 
However the presence of the Kalb-Ramond field in the five dimensional 
bulk makes the higher curvature braneworld model 
compatible even with the Planck 2018 observations. This indicates the advantage of the Kalb-Ramond field 
in the higher curvature higher dimensional model in comparison to that of without the KR field.\\
Coming to the present model, the equation of motion for the zeroth mode of the extra dimensional KR wave function 
follows from Eq.(\ref{wave function equation}) by putting 
$n = 0$. However as mentioned earlier, the overlap of the KR wave function with the visible brane fixes the coupling strengths of the KR field 
with the other matter fields and that is why we are interested to solve $\chi^{(0)}(t,\varphi)$ near the vicinity of $\varphi = \pi$ where 
the equation of $\chi^{(0)}(t,\varphi)$ can be expressed as,
\begin{eqnarray}
 \bigg(\frac{\partial\chi_v^{(0)}}{\partial t}\bigg)^2 
 - \frac{1}{T^2(t)}e^{-2k\pi T(t)}\chi_v^{(0)}\frac{\partial^2\chi_v^{(0)}}{\partial\varphi^2} = 0
 \label{zeroth mode wave function equation1}
\end{eqnarray}
with $\chi_v^{(0)}$ denotes the KR wave function near the visible brane. Eq.(\ref{zeroth mode wave function equation1}) can be solved by the method of 
separation of variables i.e we write $\chi_v^{(0)}(t,\varphi) = f_1(t)f_2({\varphi})$. Plugging back this expression into 
Eq.(\ref{zeroth mode wave function equation1}), one gets
\begin{eqnarray}
 T^2(t)e^{2k\pi T(t)} \frac{1}{f_1^2}\bigg(\frac{df_1}{dt}\bigg)^2 = \frac{1}{f_2}\frac{d^2f_2}{d\varphi^2}
 \label{zeroth mode wave function equation2}
\end{eqnarray}
Clearly the left hand side of the above equation is a function of $t$, while the right hand side depends on $\varphi$ alone. Thus both the sides 
can be separately equated to a constant as,
\begin{eqnarray}
 T^2(t)e^{2k\pi T(t)} \frac{1}{f_1^2}\bigg(\frac{df_1}{dt}\bigg)^2 = \gamma^2
 \label{separation equation1}
\end{eqnarray}
and
\begin{eqnarray}
 \frac{1}{f_2}\frac{d^2f_2}{d\varphi^2} = \gamma^2
 \label{separation equation2}
\end{eqnarray}
 where $\gamma$ is the constant of separation. Eq.(\ref{separation equation2}) can be solved to obtain 
 $f_2(\varphi) = e^{-\gamma\varphi}$ and the other one i.e Eq.(\ref{separation equation1}) is solved numerically. Thus the 
 $\chi_v^{(0)}(t,\varphi)$ is given by $\chi_v^{(0)}(t,\varphi) = e^{-\gamma\varphi}f_1(t)$. Similarly near any other constant 
 hypersurface, for example $\varphi = \varphi_0$, the KR wave function behaves as 
 $\chi_{\varphi_0}^{(0)}(t,\varphi) = e^{-\gamma\varphi}f_{\varphi_0}(t)$ where $f_{\varphi_0}(t)$ obeys the equation: 
 $T^2(t)e^{2k\varphi_0T(t)} \frac{1}{f_1^2}\bigg(\frac{df_1}{dt}\bigg)^2 = \gamma^2$ 
 (obviously $f_{\varphi_0 = \pi}(t) = f_1(t)$). The solution of $\chi_{\varphi_0}^{(0)}(t,\varphi)$ leads to the numerical plot for the time 
 evolution of the KR wave function on the $\varphi = \varphi_0$ hypersurface, which is depicted in 
 Fig.[\ref{plot_observableII}] for several values of $\varphi_0$. It reveals that the zeroth mode of the KR wave 
 function $\chi^{(0)}(t,\varphi)$ decreases with time in the whole five dimensional bulk, i.e., for $0 \leq \varphi \leq \pi$. 
 However, for a fixed $t$, $\chi^{(0)}(t,\varphi)$ has different values on the hidden and visible brane and such 
 hierarchial nature of $\chi^{(0)}(t,\varphi)$ (between the two branes) is controlled by the constant $\gamma$.\\
 
 For $T(t) = \langle T\rangle$, the zeroth mode of KR wave function acquires a constant value throughout the bulk and given by 
 $\chi^{(0)}(t,\varphi)\bigg|_{T=\langle T\rangle} = \sqrt{\frac{k}{\langle T\rangle}}e^{-k\pi\langle T\rangle}$. This result is in agreement 
 with \cite{Mukhopadhyaya:2002jn} where $\chi^{(0)}(t,\varphi)$ is found to be constant at the stabilized RS set-up. Using this expression of 
 $\chi^{(0)}(t,\varphi)\bigg|_{T=\langle T\rangle}$, we get the couplings of the KR field with U(1) gauge 
 field and fermion field on the visible brane as follows:
 \begin{eqnarray}
 \lambda_{KR-U(1)} = \lambda_{KR-fer} = \frac{1}{M_{Pl}}e^{-k\pi\langle T\rangle}
 \label{coupling1}
\end{eqnarray}
 Thereby the coupling strengths of the KR field with other matter fields get exponential suppressed over the usual gravity-matter 
 coupling $1/M_{Pl}$. The solution of the gauge hierarchy problem requires $k\pi\langle T\rangle = 36$ which in turn makes the suppression factor 
 $10^{-16}$ (recall, in the present context the gauge hierarchy problem is indeed solved for $\frac{m_{\Phi}}{k} = 0.2$ and 
 $\Psi_0 = 36$ in 4D reduced Planckian unit, as confirmed earlier). Hence the KR field couplings in our present universe gets suppressed by the factor $10^{-16}$ over $1/M_{Pl}$. This is an 
 important result in the context of Randall-Sundrum braneworld scenario. In the four dimensional model as we discussed in the previous section, 
 the couplings of the KR field is $1/M_{Pl}$ i.e same as the gravity-matter coupling : this theoretical expectation does not go well 
 with the Gravity Probe-B experiment. 
 However in the five dimensional RS model, the warping nature of the extra dimension makes an exponential suppression to the the KR field couplings 
 over the gravity-matter coupling, which indicates that the KR field has practically no footprint even in some present day gravity experiment and thus 
 the finding is in agreement with the Gravity Probe-B experiment. Thus in comparison to the four dimensional model of Sec.[\ref{sec_dynamical4d}], 
 the higher curvature warped braneworld scenario may serve a better explanation 
 about why the present universe carries practically no observable signatures of antisymmetric Kalb-Ramond field.\\
 
 Although at present the KR field amplitude reduces significantly and almost does not affect the universe evolution, but, as we have shown, it 
 has a significant contribution during early universe. Thus the important question that remains is:

\begin{itemize}
 \item Sitting in present day universe, how do we confirm the existence of the Kalb-Ramond field which has considerably low energy density (with respect 
 to the other components) in our present universe but has a significant impact during early universe ? 
\end{itemize}

The answer to this question may be encrypted in some late time phenomena which carries the information of early universe. One of such phenomena can be 
the ``Cosmological Quantum Entanglement''. Keeping this in mind, here we try to address the possible effects of KR field on cosmological 
particle production as well as on quantum entanglement for a massive scalar field propagating in a four dimensional FRW spacetime.\\
 
 \section{Cosmological quantum entanglement with Kalb-Ramond field}\label{sec_entanglement}
\begin{figure}[!h]
\begin{center}
 \centering
 \includegraphics[width=3.0in,height=2.0in]{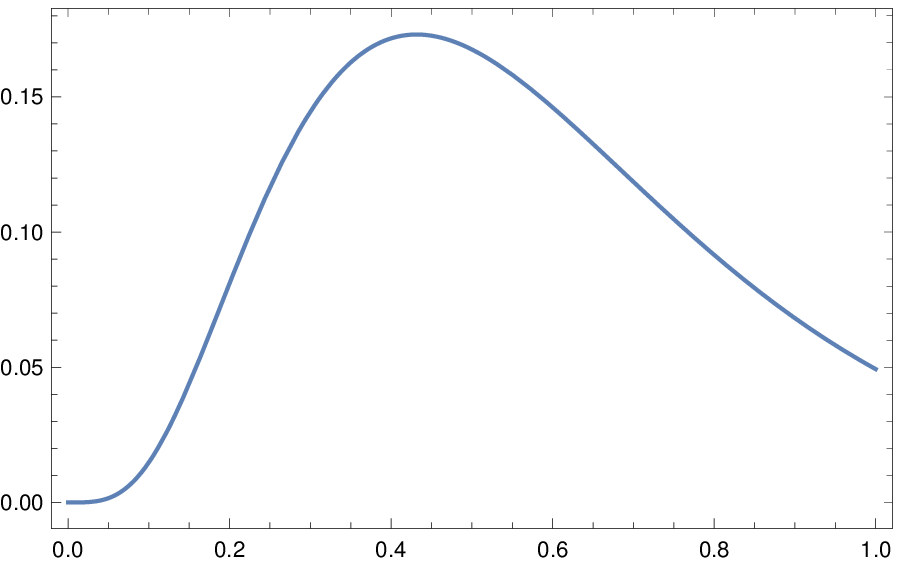}
 \includegraphics[width=3.0in,height=2.0in]{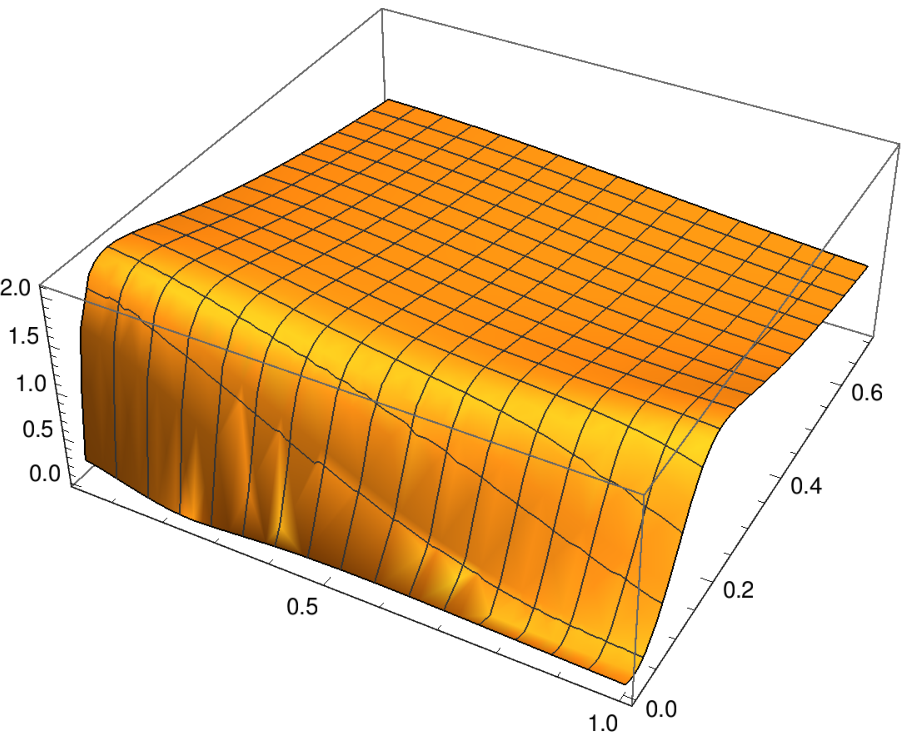}
 \caption{$Left~part$ : $S$ (along y axis) vs. $m$ (along x axis) 
 for $\xi = 1/6$ ; $k = 0.01$ in absence of KR field. $Right~part$ : 
 3D plot of $S$ with respect to mass ($0 \leq m \leq 1$ along x axis) 
and KR field energy density ($0 \leq \alpha h_0 \leq 0.7$ along y axis) for $\xi = 1/6$ ; $k = 0.01$. The quantities $m$, 
$k$ and $\alpha h_0$ (having mass dimensions [+1], [+1] and [+2] respectively) are taken in the reduced Planckian unit.}
 \label{plot entanglement1}
\end{center}
\end{figure}

\begin{figure}[!h]
\begin{center}
 \centering
 \includegraphics[width=3.0in,height=2.0in]{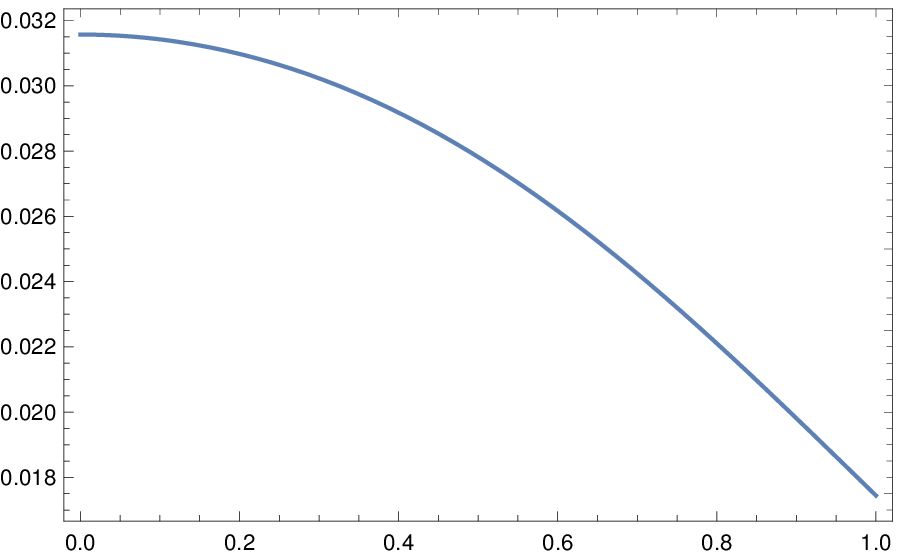}
 \includegraphics[width=3.0in,height=2.0in]{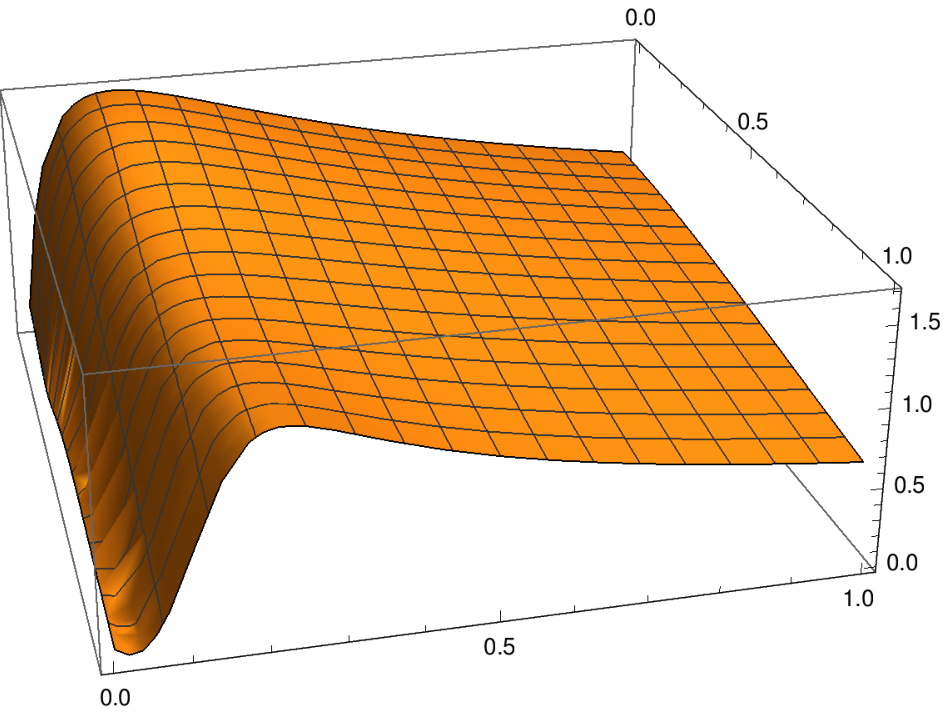}
 \caption{$Left~part$ : $S$ (along y axis) vs. $m$ (along x axis) for $\xi = 0$ ; $k = 0.01$ in absence of KR field. $Right~part$ : 
 3D plot of $S$ with respect to mass ($0 \leq m \leq 1$ along x axis, in Planckian unit) 
and KR field energy density ($0 \leq \alpha h_0 \leq 0.7$ along y axis for $\xi = 0$ ; $k = 0.01$. The quantities $m$, 
$k$ and $\alpha h_0$ (having mass dimensions [+1], [+1] and [+2] respectively) are taken in the reduced Planckian unit.}
 \label{plot entanglement2}
\end{center}
\end{figure}

 The scalar field coupled Kalb-Ramond action is \cite{Paul:2020bdy},
 \begin{eqnarray}
 S = \int d^4x \sqrt{-g} \bigg[\frac{1}{2}g^{\mu\nu}\partial_{\mu}\partial_{\nu}\Phi - \frac{1}{2}m^2\Phi^2 - \frac{1}{2}\xi R\Phi^2 
 - \frac{1}{12}H_{\mu\nu\alpha}H^{\mu\nu\alpha} - \frac{\alpha}{2}f(\Phi)H_{\mu\nu\alpha}H^{\mu\nu\alpha}\bigg]
 \label{action1}
\end{eqnarray}
$\Phi$ is a massive scalar field which has a non-minimal coupling to the scalar curvature with $\xi$ be the coupling strength. The particular 
values $\xi = 0$ and $\xi = 1/6$ denote the minimal and the conformal coupling of the scalar field respectively. Moreover the coupling 
between $\Phi$ and the KR field is symbolized by $f(\Phi)$. In absence of the KR field, the case $m = 0$ and $\xi = 1/6$ yields 
a conformally invariant scalar field action, however the coupling with the KR field in the present context breaks the conformal invariance of 
the scalar field even for $m = 0$ and $\xi = 1/6$, which has interesting consequences on the particle production and the entanglement entropy. 
In such scenario, the scalar field is treated as a quantum field i.e $\Phi(x^{\mu})$ will be 
quantized by imposing a suitable commutation relation, while the $g_{\mu\nu}$ and $B_{\mu\nu}$ are considered to be the background classical fields. 
The FRW metric ansatz will fulfill our purpose i.e we take,
\begin{eqnarray}
 ds^2 = a^2(\eta) \big[d\eta^2 - dx^2 - dy^2 - dz^2\big].
 \label{conformal metric}
\end{eqnarray}
which is written in the conformal coordinates with $\eta$ be the conformal time and $a(\eta)$ is the scale factor of the universe. The 
off-diagonal Einstein equation in the FRW spacetime become non-trivial leading to only one non-zero component of the KR field 
and that is given by $H_{123}$ ($= h_4$, say). With this information, the KR field can be solved to obtain 
$\rho_{KR} = \frac{1}{2}h_4h^4 \propto \frac{1}{a^6}$ i.e the KR field energy density decreases as $1/a^6$ with the expansion of the universe. 
Due to the time dependent 
gravitational field, the couplings of the scalar field (with the gravity and the KR field) causes 
the scalar particle production in the asymptotic future from a distant past vacuum. In order to calculate the number of produced particles and consequently 
the entanglement entropy fo the scalar field, we need to calculate the Bogoliubov coefficients at the asymptotic future. In the metric 
(\ref{conformal metric}), the scalar field equation is,
\begin{eqnarray}
 \bigg[\Box_{g} + m^2 + \xi R\bigg]\Phi + \alpha \rho_{KR} f'(\Phi) = 0,
 \label{scalar field equation1}
\end{eqnarray}
with $\Box_g$ is the d'Alembertian operator formed by $g_{\mu\nu}$. To get an explicit form of the above equation, we take
\begin{eqnarray}
f(\Phi) = \frac{\Phi^2}{2}~~~~~~~~,~~~~~~~~~~a(\eta) = 1 - \frac{\sigma^2}{2\big(\eta^2 + \sigma^2\big)},
 \label{scale factor}
\end{eqnarray}
The above form of the scale factor corresponds to a non-singular symmetric 
bounce at $\eta = 0$. At this stage it deserves mention that here, in the present paper, 
we choose this certain form of the scale factor in order to make the calculations easier. However we will show that our 
final argument regarding the testbed of KR field through entanglement entropy remains valid for the class of scale 
factor, $irrespective~of~a~particular~form$, which exhibits a symmetric bounce. It is convenient to work in the Fourier space and thus we 
expand the scalar field in Fourier modes as 
\begin{eqnarray}
 \Phi(\vec{x},\eta) = \int d^3k \Phi_{\vec{k}}(\vec{x},\eta) 
 = \int d^3k \bigg[\frac{1}{(2\pi)^{3/2}} e^{i\vec{k}.\vec{x}} \frac{\chi_k(\eta)}{a(\eta)}\bigg],
 \label{fourier decomposition}
\end{eqnarray}
where we also introduce the auxiliary field $\chi_k(\eta) = a(\eta)\Phi_k(\eta)$. Such auxiliary field along 
the forms of $f(\Phi)$ and $a(\eta)$ immediately lead to the field equation 
for $\chi_k(\eta)$,
\begin{eqnarray}
 \frac{d^2\chi_k}{d\eta^2} + \bigg[\omega_k^2 - V_k(\eta)\bigg]\chi_k(\eta) = 0,
 \label{scalar field equation4}
\end{eqnarray}
where $\omega_k^2 = k^2 + m^2 + \alpha h_0$ and
\begin{eqnarray}
 V_k(\eta) = m^2\big[1 - a^2(\eta)\big] - \big(\xi - \frac{1}{6}\big)R(\eta)a^2(\eta) + \alpha h_0\bigg[1 - \frac{1}{a^4(\eta)}\bigg].
 \label{potential}
\end{eqnarray}
Clearly the scalar field behaves as a harmonic oscillator with a time dependent mass. The time dependency is reflected through the $V_k(\eta)$ which 
consists of three terms proportional to $m^2$ (the mass of the scalar field $\Phi$), $\xi$ (the coupling between $\Phi$ and the Ricci scalar) and 
$h_0$ (the KR field strength) respectively. Thus $V_k(\eta)$ is non-zero even for $m = 0$ and $\xi = 1/6$ due to the presence of the KR field, which 
is a consequence of the fact that the coupling between the KR and the scalar field acts as one of the conformal symmetry broken agents in the 
present context. In order to quantize the scalar field, $\chi_k(\eta)$ is replaced by the field operator i.e $\hat{\chi}_k(\eta)$ and we impose the 
following equal time commutation relation on $\hat{\Psi}_k(\vec{x},\eta) = e^{i\vec{k}.\vec{x}}\hat{\chi}_k(\eta)$ as,
\begin{eqnarray}
 \bigg[\hat{\Psi}_k(\vec{x},\eta) , \hat{\pi}_k(\vec{x},\eta)\bigg] = i\delta(\vec{x} - \vec{y}),
 \label{commutation}
\end{eqnarray}
with $\hat{\pi}_k(\vec{x},\eta) = \frac{d\hat{\Psi}_k(\vec{x},\eta)}{d\eta}$ be the canonical momentum conjugate to 
$\hat{\Psi}_k(\vec{x},\eta)$. The quantum Hamiltonian of $\hat{\Psi}_k(\vec{x},\eta)$ is same as harmonic oscillator Hamiltonian 
except the fact that the effective mass depends on time and given by 
$m_{eff}^2(\eta) = \big(\xi - \frac{1}{6}\big) R(\eta)a^2(\eta) + m^2a^2(\eta) + \frac{\alpha h_0}{a^4}$. Thus the Hamiltonian of the scalar field 
explicitly depends on time: due to such 
explicit time dependency in the Hamiltonian, the vacuum of the scalar field also changes with time and hence if the scalar field starts from 
a vacuum at the distant past, it will no more in vacuum in the asymptotic future i.e the scalar particle production occurs. Moreover the Heisenberg 
equation for the $\hat{\chi}_k(\eta)$ resembles with Eq.(\ref{scalar field equation4}). The important point that deserves mentioning here is that the 
$V_k(\eta)$ asymptotically goes to zero at $\eta \rightarrow \pm \infty$. Thus the asymptotic nature of ``in-mode'' and ``out-mode'' functions 
are given by,
\begin{eqnarray}
 \zeta_k^{(in)}(\eta \rightarrow -\infty) = \frac{1}{\sqrt{2\omega_k}}e^{-i\omega_k\eta}\nonumber\\
 \zeta_k^{(out)}(\eta \rightarrow +\infty) = \frac{1}{\sqrt{2\omega_k}}e^{-i\omega_k\eta}
 \label{correct mode function}
\end{eqnarray}
$\zeta_k^{(in)}(\eta)$ and $\zeta_k^{(out)}(\eta)$ are known as ``in-mode'' and ``out-mode'' solution respectively. With these boundary conditions, 
the ``in-mode'' solution at time $\eta$ becomes,
\begin{eqnarray}
 \zeta_k^{(in)}(\eta) = \frac{1}{\sqrt{2\omega_k}}e^{-i\omega_k\eta} 
 + \frac{1}{i\omega_k}\int_{-\infty}^{\eta} d\eta_1 V_k(\eta_1)\bigg[e^{i\omega(\eta-\eta_1)} - e^{-i\omega(\eta-\eta_1)}\bigg]\chi_k(\eta_1)
 \label{in mode solution}
\end{eqnarray}
The overlap of $\zeta_k^{(out)}(\eta)$ and $\zeta_k^{*(out)}(\eta)$ on $\zeta_k^{(in)}(\eta)$ provide the Bogoliubov coefficients $\alpha_k$ and 
$\beta_k$ at $\eta$, however here we are interested in calculating the particle number at asymptotic future and thus $\alpha_k$, $\beta_k$ 
come as,
\begin{eqnarray}
 \alpha_k&=&1 + i\int_{-\infty}^{\infty} d\eta \zeta_k^{(out)*}(\eta \rightarrow \infty) V_k(\eta) \zeta_k^{(in)}(\eta).\nonumber\\
 \beta_k&=&-i\int_{-\infty}^{\infty} d\eta \zeta_k^{(out)}(\eta \rightarrow \infty) V_k(\eta) \zeta_k^{(in)}(\eta).
 \label{bogolyubov 4}
\end{eqnarray}
Using the explicit forms of $a(\eta)$ and $V_k(\eta)$, the Bogoliubov coefficients in the present context can be determined and are given by,
\begin{eqnarray}
 \big|\alpha_k\big|^2&=&1 + \frac{\pi^2}{(k^2 + m^2 + \alpha h_0)} \bigg[\frac{7m^2\sigma}{16} 
 + \frac{6(7 - 5\sqrt{2}) \big(\xi - \frac{1}{6})}{\sigma} - \frac{141\alpha h_0\sigma}{32\sqrt{2}}\bigg]^2\nonumber\\
 \big|\beta_k\big|^2&=&\frac{\pi^2}{(k^2 + m^2 + \alpha h_0)} \bigg[\frac{m^2\sigma (2\sigma\sqrt{k^2+m^2+\alpha h_0} - 7)e^{-2\omega\sigma}}{16} 
 + \frac{\big(\xi - \frac{1}{6}\big)}{16\sigma} 
 \bigg((384\sigma\omega + 672)e^{-2\omega\sigma} - 480\sqrt{2}e^{-\sqrt{2}\omega\sigma}\bigg)\nonumber\\ 
 &+&\frac{\alpha h_0\sigma e^{-\sqrt{2}\omega\sigma}}{192} 
 \bigg(423\sqrt{2} + 462\sigma\omega + 60\sqrt{2}\sigma^2\omega^2 + 4\sigma^3\omega^3\bigg)\bigg]^2 .
 \label{bogolyubov 7d}
\end{eqnarray}
 The coefficient $\big|\beta_k\big|^2$ determines the particle number $\langle n_k\rangle_{out}$ having momentum $k$ 
in the asymptotic future. It may be observed that $m = 0$ and $\xi = 1/6$ do not yield $\big|\beta_k\big|^2 = 0$, 
unlike to the case when the KR field is absent. This is a reflection of the fact that the presence of KR field actually spoils the conformal invariance 
of a massless scalar field propagating in FRW spacetime. In particular, the $\langle n_k\rangle_{out}$ exhibits a maximum with respect to the KR parameter 
$h_0$ irrespective of the conformal or weak coupling. Here we would like to mention that this behaviour of $\langle n_k\rangle_{out}$ is not only confined 
to the particular form of the scale factor we considered earlier, but also holds true for the class of 
symmetric bounce scale factor irrespective of any specific form. The demonstration for acquiring a maximum of $\langle n_k\rangle_{out}$ (with respect 
to $h_0$) goes as: for $\frac{\alpha h_0}{(k^2 + m^2)} \ll 1$, the energy of the scalar particle can be approximated as 
$\omega_k \simeq \sqrt{k^2 + m^2}$ i.e independent of $h_0$. Now due to the coupling $f(\Phi)$, the energy supplied from the KR field to the scalar 
field increases with increasing $h_0$. This along with the fact that the energy of each scalar particle is independent of $h_0$ explains 
why the particle production enhances as the KR field energy density increases for $\frac{\alpha h_0}{(k^2 + m^2)} \ll 1$. On the other hand, 
for $\frac{\alpha h_0}{(k^2 + m^2)} \gg 1$, the energy of scalar particle is proportional to $\sqrt{h_0}$, in particular 
$\omega_k \propto \sqrt{h_0}$. Therefore 
it becomes more difficult to excite the scalar particle and consequently the particle production decreases with increasing $h_0$ 
for $\frac{\alpha h_0}{(k^2 + m^2)} \gg 1$. Thus as a whole, $\frac{d\langle n_k\rangle_{out}}{d(\alpha h_0)} > 0$ 
for $\frac{\alpha h_0}{(k^2 + m^2)} \ll 1$ while we get $\frac{d\langle n_k\rangle_{out}}{d(\alpha h_0)} < 0$ in the regime 
$\frac{\alpha h_0}{(k^2 + m^2)} \gg 1$, which entails that $\langle n_k\rangle_{out}$ must has a maximum in between these two limits.\\
The various out modes get quantum entangled to each other. To demonstrate this clearly, we expand the field state in terms 
of the basi of the out modes as,
\begin{eqnarray}
 | state \rangle = | 0_{in}\rangle = \sum c_n |n~\rangle_{\vec{k}}^{out} |n~\rangle_{-\vec{k}}^{out}
 \label{entanglement 1}
\end{eqnarray}
Since we are working in the Heisenberg picture, the field state is taken as independent of time. However Eq.(\ref{entanglement 1}) 
indicates that the field state is not separable with respect to out modes, which tells that the out modes get quantum entangled to each other. Such 
entanglement may be quantified by von-Neumann entropy ($S$) defined through the conception of reduced density operator as follows:
\begin{eqnarray}
S = -Tr\bigg[\hat{\rho}_{\vec{k}}^{red}~\log_2(\hat{\rho}_{\vec{k}}^{red})\bigg] 
\label{entanglement 2}
\end{eqnarray}
in $k_B = 1$ (Boltzmann constant) unit, 
where $\hat{\rho}_{\vec{k}}^{red}$ is the reduced density operator for $\vec{k}$th out modes and has the form 
$\hat{\rho}_{\vec{k}}^{red} = \sum \langle m|~\hat{\rho}~|m\rangle_{\vec{-k}}^{out}$ 
with $\hat{\rho}$ be the full density operator of the scalar field and given by 
$\hat{\rho} = |state\rangle \langle state| = |0_{in}\rangle \langle0_{in}|$. 
Going through some simple steps, one obtains the von-Neumann entropy in terms of Bogoliubov coefficients as follows:
\begin{eqnarray}
 S = \log_2\bigg[\frac{\gamma^{\gamma/(\gamma - 1)}}{(1 - \gamma)}\bigg]
 \label{entanglement 5}
\end{eqnarray}
with $\gamma = \bigg| \frac{\beta_k}{\alpha_k} \bigg|^2$ and the expressions of $|\alpha_k|^2$, $|\beta_k|^2$ are obtained earlier.\\
Using Eq.(\ref{entanglement 5}), we give the following plots : (1) Left part of 
Figure[\ref{plot entanglement1}] is the variation of entropy ($S$) with respect to mass ($m$) of the scalar field for $\xi = 1/6$ 
(i.e for conformal coupling) in absence of the KR field coupling parameter $\alpha$, (2) Right part of 
Figure[\ref{plot entanglement1}] is the 3D plot exploring the variation of $S$ with respect to mass ($0 \leq m \leq 1$ along x axis, 
in reduced Planckian unit) 
and KR field energy density ($0 \leq \alpha h_0 \leq 0.7$ along y axis, in reduced Planckian unit) for $\xi = 1/6$, 
(3) Left and right parts of Figure[\ref{plot entanglement2}] 
give the corresponding plots respectively for $\xi = 0$ i.e for the weak coupling case.

The figures lead to the following arguments 
in regard to the effects of the coupling parameters $\xi$ (the non-minimal coupling parameter of the scalar field to the Ricci scalar) and 
$\alpha$ (the coupling parameter between the KR field and the scalar field) on entanglement entropy.\\
\begin{itemize}
 \item {\underline{The effects of $\xi$}} on the entanglement entropy can be understood from the left panels of Fig.[\ref{plot entanglement1}] and 
 Fig.[\ref{plot entanglement2}], where $\alpha$ has the fixed value zero. The left panels clearly demonstrate that in the conformal coupling 
 case i.e for $\xi = 1/6$, the entanglement entropy vanishes for $m = 0$, while in the case of weak coupling (i.e $\xi = 0$) the entanglement entropy 
 acquires a non-zero value even at $m = 0$. This is a consequence of the fact that for $\xi = 1/6$, the action of a massless scalar field 
 becomes conformally invariant in four dimensional spacetime, unlike to $\xi = 0$ for which the corresponding conformal invariance is broken. 
 Moreover it is evident that the maximum value of the entanglement entropy is larger for $\xi = 1/6$ in comparison to that of the weak coupling case.
 
 \item {\underline{The effects of $\alpha$}} on the entanglement entropy can be understood from the left and right panels of Fig.[\ref{plot entanglement1}] 
 where $\xi$ has a fixed value $1/6$, or 
 from the left and right panels of Fig.[\ref{plot entanglement2}] having $\xi = 0$ 
 (recall the left and right panels of the figures are plotted for $\alpha = 0$ and $\alpha \neq 0$ respectively). 
 Fig.[\ref{plot entanglement1}] demonstrates that 
 in absence of the interaction between the Kalb-Ramond and scalar field (i.e for $\alpha = 0$), the entropy vanishes at $m = 0$, 
 while a non-zero KR field coupling parameter (i.e $\alpha \neq 0$) leads to a non-zero entropy 
 even for $m = 0$. Again this is related to the conformal symmetry of the scalar field, in particular for $\xi = 1/6$, $m = 0$ and $\alpha = 0$, 
 the scalar field becomes conformally invariant in 4 dimensional spacetime, however the condition $\alpha \neq 0$ breaks 
 the conformal invariance of a massless scalar field even for $\xi = 1/6$ and thus the corresponding entanglement entropy becomes non-zero. Moreover 
 Fig.[\ref{plot entanglement1}] also depicts that without the KR field, the entanglement entropy is bounded by $S \lesssim 0.17$ (in 
 the unit of $k_B = 1$, see the left panel), while due to the presence of KR field the upper bound of entropy goes beyond $0.17k_B$ 
and reaches up to $S \lesssim 2k_B$ (see the right panel). Similarly Fig.[\ref{plot entanglement2}] reveals that for $\xi = 0$, the maximum value 
of the von-Neumann entropy is given by $S \lesssim 0.032k_B$ in absence of the KR field, however the Kalb-Ramond field indeed affects the situation, 
in particular the maximum entanglement entropy reaches upto $S \lesssim 1.5k_B$ due to $\alpha \neq 0$. 
Therefore the maximum value of the entanglement entropy becomes larger due to the coupling between 
Kalb-Ramond and scalar field, in comparison to the case when the coupling parameter $\alpha$ is zero.
\end{itemize}

The above arguments indicate that for $\xi = 1/6$, if the entanglement entropy is found to lie 
within $0.17k_B \lesssim S \lesssim 2k_B$, then it may provide a possible testbed for the existence of Kalb-Ramond field in our universe. 
Similarly Figure[\ref{plot entanglement2}] reveals that for $\xi = 0$, 
if the von-Neumann entropy situates in between $0.032k_B$ and $1.5k_B$, then one can infer about the possible presence of KR field. 
Moreover this argument is not only confined to the scale factor we considered earlier, but also valid for the general 
class of symmetric bounce scale factor. 
Thus as a whole, for a non-singular symmetric universe, irrespective of 
 conformal and weak coupling case, we get $max[S(k, m, \alpha \neq 0)] > max[S(k, m, \alpha = 0)]$ and if the entanglement entropy is 
 found to lie within these two upper bounds (i.e within $max[S(k, m, \alpha \neq 0)]$ and $max[S(k, m, \alpha = 0)]$), 
 then it may provide a possible testbed for the existence of Kalb-Ramond field in our universe. 
 However the measurement procedure of the entanglement entropy is still a problem. If the actual method of measurement of entanglement entropy in a 
 cosmological background takes a shape in future, these predictions will certainly be useful to test the existence of a Kalb-Ramond field.\\
 
 \section{Conclusion}
 The paper deals with various aspects of higher rank antisymmetric tensor fields, particularly the second rank antisymmetric 
 Kalb-Ramond field, in modified theories of gravity. We try to address the following questions in this article:\\
 \begin{itemize}
  \item Why the present universe is practically free from any noticeable footmarks of higher rank antisymmetric tensor fields, despite 
  having the signatures of scalar, vector, fermion as well as symmetric rank 2 tensor field in the form of gravity ?
  
  \item What are the possible roles of the Kalb-Ramond field during early universe ?
  
  \item If the Kalb-Ramond field has considerable impact during early universe, then an immediate question will be - 
  Sitting in present day universe, how do we confirm the existence of the Kalb-Ramond field which has considerably low energy density (with respect 
 to the other components) in our present universe but has a significant impact during early universe ?
 \end{itemize}
 We provide some possible explanations of these questions from modified theories of gravity, both in four and five dimensional spacetime.\\
 In the context of 4 dimensional higher curvature gravity theory, 
 it turns out that the KR field energy density decreases with the expansion of our universe 
 at a faster rate in comparison to radiation and matter components. Thus as the Universe evolves and cools down, 
 the contribution of the KR field on the evolutionary process reduces significantly, 
 and at present it almost does not affect the universe evolution. However the KR field has a significant 
 contribution during early universe, in particular, it affects the beginning of inflation as well as 
 increases the amount of primordial gravitational radiation and consequently enlarges the value of tensor to scalar ratio in respect to the case 
 when the KR field is absent. Such effects of the KR field during early universe have been discussed for various forms of F(R) model like - 
 the Starobinsky $F(R)$, the logarithmic corrected $R^2$ model, the cubic $F(R)$ etc. However in regard to the coupling strength, 
 the KR field couplings with other matter fields turn out to be $\frac{1}{M_{Pl}}$ in four dimensional context, which is same as the usual 
 gravity-matter coupling. This leads to the expectation that the KR field may show its signatures in some present day 
 gravity experiment. But no experimental evidences so far support this expectation and an example is the Gravity Probe B experiment. 
 Thereby the four dimensional model may not serve the full explanation of why the present universe practically carries 
 no noticeable footmarks of the Kalb-Ramond field.\\
 Thus we move forward to investigate the phenomena of the suppression of the KR field from some higher dimensional spacetime model. As a higher dimensional 
 model, we consider a 5 dimensional higher curvature warped braneworld model 
 in presence of Kalb-Ramond field exists in the bulk together with gravity. The higher curvature term generates a stable radion potential in the 
 effective four dimensional on-brane theory, which ensures the stabilization of the extra dimensional modulus. The important point that deserves 
 mentioning that the warping nature of the 
 extra dimension makes an exponential suppression to the the on-brane KR field couplings over the gravity-matter coupling $1/M_{Pl}$, in particular, 
 the resolution of the gauge hierarchy problem leads to the suppression factor of the KR field couplings in our present universe as 
 $10^{-16}$ over $1/M_{Pl}$. This may explain why the KR field shows practically no observable signatures 
 even in some present day gravity experiment and thus 
 the finding is in agreement with the Gravity Probe-B experiment. Thereby in comparison to the four dimensional model, 
 the higher curvature warped braneworld scenario may serve a better explanation about why the present 
 universe is free from the observable footmarks of antisymmetric Kalb-Ramond field. Moreover in view of the higher dimensional warped braneworld model, 
 we may also argue that the suppression of the KR field 
 in our present universe and the resolution of the gauge hierarchy problem are intimately connected.\\
 The cosmological evolution of the KR field both from the perspective of four and five dimensional spacetime model leads to a well motivate 
 question - ``How do we confirm the existence of the Kalb-Ramond field which has considerably low energy density in our 
 present universe but has a significant impact during early universe ?'' For this purpose, we address the possible effects of the KR field 
 on cosmological quantum entanglement entropy 
 for a massive scalar field, which may encrypt the information of early universe. The scalar field is considered to be non-minimally 
 coupled with the Ricci scalar and moreover the KR field also couples to the scalar field. In such scenario, we calculate the entanglement entropy 
 of the scalar field and as a result, the maximum value of the entanglement entropy is found to be larger due to the presence of the KR field in 
 comparison to the case when the KR field is absent i.e $max[S(k, m, \alpha \neq 0)] > max[S(k, m, \alpha = 0)]$ 
 (where $\alpha$ is the KR field coupling parameter, $m$ is the scalar field mass and $k$ denotes the momentum of the scalar particle). 
 Furthermore this argument holds irrespective of the conformal or 
 weak coupling between the scalar field and the Ricci scalar. Thus 
 if the entanglement entropy is found to lie within these two upper bounds (i.e within $max[S(k, m, \alpha \neq 0)]$ 
 and $max[S(k, m, \alpha = 0)]$), then it may provide a possible testbed for the existence of Kalb-Ramond field in our universe.

 \section{Brief discussions on future perspectives}
 Here we have studied various aspects of antisymmetric tensor fields, in particular the second rank Kalb-Ramond field, in modified gravity 
 theories. What remains to be done in this context is to study the possible effects of the KR field in the formation 
 of Primordial Black Hole (PBH). Since the KR field energy density is supposed to be large during early universe, the KR field may 
 has significant contributions in PBH formation. In Sec.[\ref{sec_entanglement}], we showed that KR field indeed affects the phenomena which 
 carries the information of early universe, like the cosmological quantum entanglement of a scalar field coupled to the KR field. In this arena, we 
 can progress our research further to get some deeper insights of quantum cosmology from the existence KR field. 
 Moreover the models we discussed in this review are the inflationary models which, in fact, suffers 
 the well known singularity problem. Due to this reason, the bouncing cosmology gets a lot of attention in theoretical cosmology, as it can 
 remove the initial Big-Bang singularity. Thus it is important to study the evolution of Kalb-Ramond field in non-singular 
 bounce models. In particular, the tensor perturbation in F(R) bounce models are, in general, found to be comparable with 
 the scalar perturbation and thus the tensor to scalar ratio becomes order of unity which is not consistent with the Planck constraints. 
 May be the presence of non-minimally coupled KR field (i.e non-minimally coupled with the curvature) in F(R) bounce model 
 suppress the amplitude of the tensor perturbation and make the observable quantities consistent with the Planck observations. These issues 
 may be further studied in future.


\begin{thebibliography}{99}
  
\bibitem{Guth:1980zm}
A.~H.~Guth,
Phys.\ Rev.\ D {\bf 23} (1981) 347. doi:10.1103/PhysRevD.23.347

\bibitem{Linde:1981mu}
A.~D.~Linde,
Phys.\ Lett.\  {\bf 108B} (1982) 389 [Adv.\ Ser.\ Astrophys.\
Cosmol.\  {\bf 3} (1987) 149]. doi:10.1016/0370-2693(82)91219-9



\bibitem{Albrecht:1982wi}
A.~Albrecht and P.~J.~Steinhardt,
Phys.\ Rev.\ Lett.\  {\bf 48} (1982) 1220 [Adv.\ Ser.\ Astrophys.\
Cosmol.\  {\bf 3} (1987) 158]. doi:10.1103/PhysRevLett.48.1220


\bibitem{Kinney:2003xf}
W.~H.~Kinney,
NATO Sci. Ser. II \textbf{123} (2003), 189-243
doi:10.1007/978-94-010-0076-5-5
[arXiv:astro-ph/0301448 [astro-ph]].

\bibitem{Langlois:2004de}
D.~Langlois,
[arXiv:hep-th/0405053 [hep-th]].


\bibitem{Riotto:2002yw}
A.~Riotto,
ICTP Lect. Notes Ser. \textbf{14} (2003), 317-413
[arXiv:hep-ph/0210162 [hep-ph]].



\bibitem{Barrow:1993ah}
J.~D.~Barrow and P.~Saich,
Class. Quant. Grav. \textbf{10} (1993), 279-283
doi:10.1088/0264-9381/10/2/009


\bibitem{Barrow:1994nx}
J.~D.~Barrow and J.~P.~Mimoso,
Phys. Rev. D \textbf{50} (1994), 3746-3754
doi:10.1103/PhysRevD.50.3746


\bibitem{Mimoso:1994wn}
J.~P.~Mimoso and D.~Wands,
Phys. Rev. D \textbf{51} (1995), 477-489
doi:10.1103/PhysRevD.51.477
[arXiv:gr-qc/9405025 [gr-qc]].


\bibitem{Baumann:2009ds}
D.~Baumann,
doi:10.1142/9789814327183-0010
[arXiv:0907.5424 [hep-th]].

\bibitem{Nojiri:2007as}
S.~Nojiri and S.~D.~Odintsov,
Phys. Lett. B \textbf{657} (2007), 238-245
doi:10.1016/j.physletb.2007.10.027
[arXiv:0707.1941 [hep-th]].

\bibitem{Sriramkumar:2009kg}
L.~Sriramkumar,
[arXiv:0904.4584 [astro-ph.CO]].


\bibitem{Langlois:2010xc}
D.~Langlois,
Lect. Notes Phys. \textbf{800} (2010), 1-57
doi:10.1007/978-3-642-10598-2-1
[arXiv:1001.5259 [astro-ph.CO]].



\bibitem{Brandenberger:1997yf}
R.~H.~Brandenberger,
[arXiv:astro-ph/9711106 [astro-ph]].


\bibitem{Wang:2013zva}
Y.~Wang,
Commun. Theor. Phys. \textbf{62} (2014), 109-166
doi:10.1088/0253-6102/62/1/19
[arXiv:1303.1523 [hep-th]].



\bibitem{Brandenberger:2012zb}
R.~H.~Brandenberger,
arXiv:1206.4196 [astro-ph.CO].



\bibitem{Brandenberger:2016vhg}
R.~Brandenberger and P.~Peter,
arXiv:1603.05834 [hep-th].


\bibitem{Battefeld:2014uga}
 D.~Battefeld and P.~Peter,
 Phys.\ Rept.\ {\bf 571} (2015) 1
 doi:10.1016/j.physrep.2014.12.004
 [arXiv:1406.2790 [astro-ph.CO]].


\bibitem{Novello:2008ra}
M.~Novello and S.~E.~P.~Bergliaffa,
 ``Bouncing Cosmologies,''
Phys.\ Rept.\ {\bf 463} (2008) 127
doi:10.1016/j.physrep.2008.04.006
[arXiv:0802.1634 [astro-ph]].


\bibitem{Cai:2014bea}
Y.~F.~Cai,
Sci.\ China Phys.\ Mech.\ Astron.\  {\bf 57} (2014) 1414
doi:10.1007/s11433-014-5512-3
[arXiv:1405.1369 [hep-th]].

\bibitem{Ijjas:2018qbo}
A.~Ijjas and P.~J.~Steinhardt,
Class. Quant. Grav. \textbf{35} (2018) no.13, 135004
doi:10.1088/1361-6382/aac482
[arXiv:1803.01961 [astro-ph.CO]].

\bibitem{Cai:2013vm}
Y.~F.~Cai, R.~Brandenberger and P.~Peter,
Class. Quant. Grav. \textbf{30} (2013), 075019
doi:10.1088/0264-9381/30/7/075019
[arXiv:1301.4703 [gr-qc]].


\bibitem{Battefeld:2004mn}
T.~J.~Battefeld, S.~P.~Patil and R.~Brandenberger,
Phys. Rev. D \textbf{70} (2004), 066006
doi:10.1103/PhysRevD.70.066006
[arXiv:hep-th/0401010 [hep-th]].


\bibitem{Peter:2016kan}
P.~Peter and S.~D.~P.~Vitenti,
Mod. Phys. Lett. A \textbf{31} (2016) no.21, 1640006
doi:10.1142/S021773231640006X
[arXiv:1603.02342 [gr-qc]].


\bibitem{Cai:2012va}
Y.~F.~Cai, D.~A.~Easson and R.~Brandenberger,
JCAP \textbf{08} (2012), 020
doi:10.1088/1475-7516/2012/08/020
[arXiv:1206.2382 [hep-th]].

\bibitem{Odintsov:2015zza}
S.~D.~Odintsov and V.~K.~Oikonomou,
Phys. Rev. D \textbf{92} (2015) no.2, 024016
doi:10.1103/PhysRevD.92.024016
[arXiv:1504.06866 [gr-qc]].


\bibitem{Cai:2016thi}
Y.~Cai, Y.~Wan, H.~G.~Li, T.~Qiu and Y.~S.~Piao,
JHEP \textbf{01} (2017), 090
doi:10.1007/JHEP01(2017)090
[arXiv:1610.03400 [gr-qc]].



\bibitem{Cai:2017tku}
Y.~Cai, H.~G.~Li, T.~Qiu and Y.~S.~Piao,
Eur. Phys. J. C \textbf{77} (2017) no.6, 369
doi:10.1140/epjc/s10052-017-4938-y
[arXiv:1701.04330 [gr-qc]].


\bibitem{Qiu:2010ch}
T.~Qiu and K.~C.~Yang,
JCAP \textbf{11} (2010), 012
doi:10.1088/1475-7516/2010/11/012
[arXiv:1007.2571 [astro-ph.CO]].


\bibitem{Koehn:2015vvy}
M.~Koehn, J.~L.~Lehners and B.~Ovrut,
Phys. Rev. D \textbf{93} (2016) no.10, 103501
doi:10.1103/PhysRevD.93.103501
[arXiv:1512.03807 [hep-th]].


\bibitem{Nojiri:2010wj}
S.~Nojiri and S.~D.~Odintsov,
Phys.\ Rept.\  {\bf 505} (2011) 59
doi:10.1016/j.physrep.2011.04.001
[arXiv:1011.0544 [gr-qc]].

\bibitem{Nojiri:2017ncd}
S.~Nojiri, S.~D.~Odintsov and V.~K.~Oikonomou,
Phys.\ Rept.\  {\bf 692} (2017) 1
doi:10.1016/j.physrep.2017.06.001
[arXiv:1705.11098 [gr-qc]].

\bibitem{Capozziello:2011et}
S.~Capozziello and M.~De Laurentis,
Phys.\ Rept.\  {\bf 509} (2011) 167
doi:10.1016/j.physrep.2011.09.003
[arXiv:1108.6266 [gr-qc]].


\bibitem{Csaki:2004ay}
C.~Csaki,
[arXiv:hep-ph/0404096 [hep-ph]].


\bibitem{Csaki:2005vy}
C.~Csaki, J.~Hubisz and P.~Meade,
[arXiv:hep-ph/0510275 [hep-ph]].


\bibitem{Brax:2003fv}
P.~Brax and C.~van de Bruck,
Class. Quant. Grav. \textbf{20} (2003), R201-R232
doi:10.1088/0264-9381/20/9/202
[arXiv:hep-th/0303095 [hep-th]].


\bibitem{Maartens:2010ar}
R.~Maartens and K.~Koyama,
Living Rev. Rel. \textbf{13} (2010), 5
doi:10.12942/lrr-2010-5
[arXiv:1004.3962 [hep-th]].


\bibitem{Whisker:2008kk}
R.~Whisker,
[arXiv:0810.1534 [gr-qc]].


\bibitem{Brax:2004xh}
P.~Brax, C.~van de Bruck and A.~C.~Davis,
Rept. Prog. Phys. \textbf{67} (2004), 2183-2232
doi:10.1088/0034-4885/67/12/R02
[arXiv:hep-th/0404011 [hep-th]].


\bibitem{Kim:2003pc}
Y.~b.~Kim, C.~O.~Lee, I.~b.~Lee and J.~J.~Lee,
J. Korean Astron. Soc. \textbf{37} (2004), 1-14
doi:10.5303/JKAS.2004.37.1.001
[arXiv:hep-th/0307023 [hep-th]].


\bibitem{Artymowski:2014gea}
M.~Artymowski and Z.~Lalak,
JCAP \textbf{09} (2014), 036
doi:10.1088/1475-7516/2014/09/036
[arXiv:1405.7818 [hep-th]].

\bibitem{Nojiri:2003ft}
S.~Nojiri and S.~D.~Odintsov,
Phys. Rev. D \textbf{68} (2003), 123512
doi:10.1103/PhysRevD.68.123512
[arXiv:hep-th/0307288 [hep-th]].

\bibitem{Odintsov:2019mlf}
S.~D.~Odintsov and V.~K.~Oikonomou,
Phys.\ Rev.\ D {\bf 99} (2019) no.6,  064049
doi:10.1103/PhysRevD.99.064049
[arXiv:1901.05363 [gr-qc]].

\bibitem{Johnson:2019vwi}
J.~P.~Johnson and S.~Shankaranarayanan,
Phys. Rev. D \textbf{100} (2019) no.8, 083526
doi:10.1103/PhysRevD.100.083526
[arXiv:1904.07608 [astro-ph.CO]].

\bibitem{Pinto:2018rfg}
P.~Pinto, L.~Del Vecchio, L.~Fatibene and M.~Ferraris,
JCAP \textbf{11} (2018), 044
doi:10.1088/1475-7516/2018/11/044
[arXiv:1807.00397 [gr-qc]].

\bibitem{Odintsov:2019evb}
S.~D.~Odintsov and V.~K.~Oikonomou,
Phys.\ Rev.\ D {\bf 99} (2019) no.10,  104070
doi:10.1103/PhysRevD.99.104070
[arXiv:1905.03496 [gr-qc]].

\bibitem{Nojiri:2019riz}
S.~Nojiri, S.~D.~Odintsov and V.~K.~Oikonomou,
arXiv:1907.01625 [gr-qc].

\bibitem{Nojiri:2019fft}
S.~Nojiri, S.~D.~Odintsov and V.~K.~Oikonomou,
arXiv:1912.13128 [gr-qc].

\bibitem{Lobo:2008sg}
F.~S.~Lobo,
[arXiv:0807.1640 [gr-qc]].

\bibitem{Gorbunov:2010bn}
D.~Gorbunov and A.~Panin,
Phys. Lett. B \textbf{700} (2011), 157-162
doi:10.1016/j.physletb.2011.04.067
[arXiv:1009.2448 [hep-ph]].

\bibitem{Li:2007xn}
B.~Li and J.~D.~Barrow,
Phys. Rev. D \textbf{75} (2007), 084010
doi:10.1103/PhysRevD.75.084010
[arXiv:gr-qc/0701111 [gr-qc]].

\bibitem{Odintsov:2020nwm}
S.~D.~Odintsov and V.~K.~Oikonomou,
Phys.\ Rev.\ D {\bf 101} (2020) no.4,  044009
doi:10.1103/PhysRevD.101.044009
[arXiv:2001.06830 [gr-qc]].

\bibitem{Odintsov:2020iui}
S.~D.~Odintsov and V.~K.~Oikonomou,
EPL {\bf 129} (2020) no.4,  40001
doi:10.1209/0295-5075/129/40001
[arXiv:2003.06671 [gr-qc]].

\bibitem{Appleby:2007vb}
S.~A.~Appleby and R.~A.~Battye,
Phys. Lett. B \textbf{654} (2007), 7-12
doi:10.1016/j.physletb.2007.08.037
[arXiv:0705.3199 [astro-ph]].

\bibitem{Elizalde:2010ts}
E.~Elizalde, S.~Nojiri, S.~Odintsov, L.~Sebastiani and S.~Zerbini,
Phys. Rev. D \textbf{83} (2011), 086006
doi:10.1103/PhysRevD.83.086006
[arXiv:1012.2280 [hep-th]].

\bibitem{Cognola:2007zu}
G.~Cognola, E.~Elizalde, S.~Nojiri, S.~Odintsov, L.~Sebastiani and S.~Zerbini,
Phys. Rev. D \textbf{77} (2008), 046009
doi:10.1103/PhysRevD.77.046009
[arXiv:0712.4017 [hep-th]].

\bibitem{Li:2007jm}
B.~Li, J.~D.~Barrow and D.~F.~Mota,
Phys. Rev. D \textbf{76} (2007), 044027
doi:10.1103/PhysRevD.76.044027
[arXiv:0705.3795 [gr-qc]].

\bibitem{Odintsov:2018nch}
S.~D.~Odintsov, V.~K.~Oikonomou and S.~Banerjee,
Nucl.\ Phys.\ B {\bf 938} (2019) 935
doi:10.1016/j.nuclphysb.2018.07.013
[arXiv:1807.00335 [gr-qc]].

\bibitem{Carter:2005fu}
B.~M.~Carter and I.~P.~Neupane,
JCAP \textbf{06} (2006), 004
doi:10.1088/1475-7516/2006/06/004
[arXiv:hep-th/0512262 [hep-th]].



\bibitem{Nojiri:2019dwl}
S.~Nojiri, S.~Odintsov, V.~Oikonomou, N.~Chatzarakis and T.~Paul,
Eur. Phys. J. C \textbf{79} (2019) no.7, 565
doi:10.1140/epjc/s10052-019-7080-1
[arXiv:1907.00403 [gr-qc]].


\bibitem{Elizalde:2010jx}
E.~Elizalde, R.~Myrzakulov, V.~Obukhov and D.~Saez-Gomez,
Class. Quant. Grav. \textbf{27} (2010), 095007
doi:10.1088/0264-9381/27/9/095007
[arXiv:1001.3636 [gr-qc]].

\bibitem{Makarenko:2016jsy}
A.~N.~Makarenko,
Int. J. Geom. Meth. Mod. Phys. \textbf{13} (2016) no.05, 1630006
doi:10.1142/S0219887816300063

\bibitem{delaCruzDombriz:2011wn}
A.~de la Cruz-Dombriz and D.~Saez-Gomez,
Class. Quant. Grav. \textbf{29} (2012), 245014
doi:10.1088/0264-9381/29/24/245014
[arXiv:1112.4481 [gr-qc]].

\bibitem{Chakraborty:2018scm}
S.~Chakraborty, T.~Paul and S.~SenGupta,
Phys. Rev. D \textbf{98} (2018) no.8, 083539
doi:10.1103/PhysRevD.98.083539
[arXiv:1804.03004 [gr-qc]].

\bibitem{Kanti:2015pda}
P.~Kanti, R.~Gannouji and N.~Dadhich,
Phys. Rev. D \textbf{92} (2015) no.4, 041302
doi:10.1103/PhysRevD.92.041302
[arXiv:1503.01579 [hep-th]].

\bibitem{Kanti:2015dra}
P.~Kanti, R.~Gannouji and N.~Dadhich,
Phys. Rev. D \textbf{92} (2015) no.8, 083524
doi:10.1103/PhysRevD.92.083524
[arXiv:1506.04667 [hep-th]].

\bibitem{Odintsov:2018zhw}
S.~D.~Odintsov and V.~K.~Oikonomou,
Phys.\ Rev.\ D {\bf 98} (2018) no.4,  044039
doi:10.1103/PhysRevD.98.044039
[arXiv:1808.05045 [gr-qc]].

\bibitem{Saridakis:2017rdo}
E.~N.~Saridakis,
Phys.\ Rev.\ D {\bf 97} (2018) no.6,  064035
doi:10.1103/PhysRevD.97.064035
[arXiv:1707.09331 [gr-qc]].

\bibitem{Cognola:2006eg}
G.~Cognola, E.~Elizalde, S.~Nojiri, S.~D.~Odintsov and S.~Zerbini,
Phys. Rev. D \textbf{73} (2006), 084007
doi:10.1103/PhysRevD.73.084007
[arXiv:hep-th/0601008 [hep-th]].


\bibitem{Bamba:2014mya}
K.~Bamba, A.~N.~Makarenko, A.~N.~Myagky and S.~D.~Odintsov,
Phys. Lett. B \textbf{732} (2014), 349-355
doi:10.1016/j.physletb.2014.04.004
[arXiv:1403.3242 [hep-th]].

\bibitem{Bamba:2014zoa}
K.~Bamba, A.~N.~Makarenko, A.~N.~Myagky and S.~D.~Odintsov,
JCAP \textbf{04} (2015), 001
doi:10.1088/1475-7516/2015/04/001
[arXiv:1411.3852 [hep-th]].



\bibitem{Amoros:2014tha}
J.~Amorós, J.~de Haro and S.~D.~Odintsov,
Phys. Rev. D \textbf{89} (2014) no.10, 104010
doi:10.1103/PhysRevD.89.104010
[arXiv:1402.3071 [gr-qc]].


\bibitem{Nojiri:2016ygo}
S.~Nojiri, S.~D.~Odintsov and V.~K.~Oikonomou,
Phys. Rev. D \textbf{93} (2016) no.8, 084050
doi:10.1103/PhysRevD.93.084050
[arXiv:1601.04112 [gr-qc]].


\bibitem{Odintsov:2015ynk}
S.~D.~Odintsov and V.~K.~Oikonomou,
Int. J. Mod. Phys. D \textbf{26} (2017) no.08, 1750085
doi:10.1142/S0218271817500857
[arXiv:1512.04787 [gr-qc]].

\bibitem{Bamba:2013fha}
K.~Bamba, A.~N.~Makarenko, A.~N.~Myagky, S.~Nojiri and S.~D.~Odintsov,
JCAP \textbf{01} (2014), 008
doi:10.1088/1475-7516/2014/01/008
[arXiv:1309.3748 [hep-th]].


\bibitem{Haro:2015oqa}
J.~Haro, A.~N.~Makarenko, A.~N.~Myagky, S.~D.~Odintsov and V.~K.~Oikonomou,
Phys. Rev. D \textbf{92} (2015) no.12, 124026
doi:10.1103/PhysRevD.92.124026
[arXiv:1506.08273 [gr-qc]].


\bibitem{Helling:2009ia}
R.~C.~Helling,
[arXiv:0912.3011 [gr-qc]].

\bibitem{Elizalde:2020zcb}
E.~Elizalde, S.~D.~Odintsov, V.~K.~Oikonomou and T.~Paul,
Nucl. Phys. B \textbf{954} (2020), 114984
doi:10.1016/j.nuclphysb.2020.114984
[arXiv:2003.04264 [gr-qc]].

\bibitem{ArkaniHamed:1998rs}
N.~Arkani-Hamed, S.~Dimopoulos and G.~R.~Dvali,
Phys. Lett. B \textbf{429} (1998), 263-272
doi:10.1016/S0370-2693(98)00466-3
[arXiv:hep-ph/9803315 [hep-ph]].

\bibitem{Antoniadis:1998ig}
I.~Antoniadis, N.~Arkani-Hamed, S.~Dimopoulos and G.~R.~Dvali,
Phys. Lett. B \textbf{436} (1998), 257-263
doi:10.1016/S0370-2693(98)00860-0
[arXiv:hep-ph/9804398 [hep-ph]].


\bibitem{Randall:1999ee}
L.~Randall and R.~Sundrum,
Phys. Rev. Lett. \textbf{83} (1999), 3370-3373
doi:10.1103/PhysRevLett.83.3370
[arXiv:hep-ph/9905221 [hep-ph]].


\bibitem{Randall:1999vf}
L.~Randall and R.~Sundrum,
Phys. Rev. Lett. \textbf{83} (1999), 4690-4693
doi:10.1103/PhysRevLett.83.4690
[arXiv:hep-th/9906064 [hep-th]].


\bibitem{ArkaniHamed:1998nn}
N.~Arkani-Hamed, S.~Dimopoulos and G.~R.~Dvali,
Phys. Rev. D \textbf{59} (1999), 086004
doi:10.1103/PhysRevD.59.086004
[arXiv:hep-ph/9807344 [hep-ph]].


\bibitem{ArkaniHamed:1999gq}
N.~Arkani-Hamed, S.~Dimopoulos, N.~Kaloper and J.~March-Russell,
Nucl. Phys. B \textbf{567} (2000), 189-228
doi:10.1016/S0550-3213(99)00667-7
[arXiv:hep-ph/9903224 [hep-ph]].


\bibitem{Goldberger:1999uk}
W.~D.~Goldberger and M.~B.~Wise,
Phys. Rev. Lett. \textbf{83} (1999), 4922-4925
doi:10.1103/PhysRevLett.83.4922
[arXiv:hep-ph/9907447 [hep-ph]].


\bibitem{Goldberger:1999un}
W.~D.~Goldberger and M.~B.~Wise,
Phys. Lett. B \textbf{475} (2000), 275-279
doi:10.1016/S0370-2693(00)00099-X
[arXiv:hep-ph/9911457 [hep-ph]].


\bibitem{Chakraborty:2016gpg}
S.~Chakraborty and S.~SenGupta,
Eur. Phys. J. C \textbf{77} (2017) no.8, 573
doi:10.1140/epjc/s10052-017-5138-5
[arXiv:1701.01032 [gr-qc]].


\bibitem{Das:2017htt}
A.~Das, H.~Mukherjee, T.~Paul and S.~SenGupta,
Eur. Phys. J. C \textbf{78} (2018) no.2, 108
doi:10.1140/epjc/s10052-018-5603-9
[arXiv:1701.01571 [hep-th]].


\bibitem{Csaki:2000zn}
C.~Csaki, M.~L.~Graesser and G.~D.~Kribs,
Phys. Rev. D \textbf{63} (2001), 065002
doi:10.1103/PhysRevD.63.065002
[arXiv:hep-th/0008151 [hep-th]].


\bibitem{DeWolfe:1999cp}
O.~DeWolfe, D.~Z.~Freedman, S.~S.~Gubser and A.~Karch,
Phys. Rev. D \textbf{62} (2000), 046008
doi:10.1103/PhysRevD.62.046008
[arXiv:hep-th/9909134 [hep-th]].


\bibitem{Lesgourgues:2000tj}
J.~Lesgourgues, S.~Pastor, M.~Peloso and L.~Sorbo,
Phys. Lett. B \textbf{489} (2000), 411
doi:10.1016/S0370-2693(00)00943-6
[arXiv:hep-ph/0004086 [hep-ph]].


\bibitem{Csaki:1999mp}
C.~Csaki, M.~Graesser, L.~Randall and J.~Terning,
Phys. Rev. D \textbf{62} (2000), 045015
doi:10.1103/PhysRevD.62.045015
[arXiv:hep-ph/9911406 [hep-ph]].


\bibitem{Binetruy:1999ut}
P.~Binetruy, C.~Deffayet and D.~Langlois,
Nucl. Phys. B \textbf{565} (2000), 269-287
doi:10.1016/S0550-3213(99)00696-3
[arXiv:hep-th/9905012 [hep-th]].


\bibitem{Csaki:1999jh}
C.~Csaki, M.~Graesser, C.~F.~Kolda and J.~Terning,
Phys. Lett. B \textbf{462} (1999), 34-40
doi:10.1016/S0370-2693(99)00896-5
[arXiv:hep-ph/9906513 [hep-ph]].


\bibitem{Cline:1999yq}
J.~M.~Cline,
doi:10.1142/9789812792129-0072
[arXiv:hep-ph/0001285 [hep-ph]].


\bibitem{Nojiri:2000gv}
S.~Nojiri and S.~D.~Odintsov,
JHEP \textbf{07} (2000), 049
doi:10.1088/1126-6708/2000/07/049
[arXiv:hep-th/0006232 [hep-th]].


\bibitem{Nojiri:2001ae}
S.~Nojiri, S.~D.~Odintsov and S.~Ogushi,
Phys. Rev. D \textbf{65} (2002), 023521
doi:10.1103/PhysRevD.65.023521
[arXiv:hep-th/0108172 [hep-th]].


\bibitem{Das:2017jrl}
A.~Das, D.~Maity, T.~Paul and S.~SenGupta,
Eur. Phys. J. C \textbf{77} (2017) no.12, 813
doi:10.1140/epjc/s10052-017-5396-2
[arXiv:1706.00950 [hep-th]].


\bibitem{Banerjee:2017lxi}
N.~Banerjee and T.~Paul,
Eur. Phys. J. C \textbf{77} (2017) no.10, 672
doi:10.1140/epjc/s10052-017-5256-0
[arXiv:1706.05964 [hep-th]].


\bibitem{Davoudiasl:1999jd}
H.~Davoudiasl, J.~L.~Hewett and T.~G.~Rizzo,
Phys. Rev. Lett. \textbf{84} (2000), 2080
doi:10.1103/PhysRevLett.84.2080
[arXiv:hep-ph/9909255 [hep-ph]].


\bibitem{Das:2015zxa}
A.~Das and S.~SenGupta,
Eur. Phys. J. C \textbf{76} (2016) no.8, 423
doi:10.1140/epjc/s10052-016-4264-9
[arXiv:1506.05613 [hep-th]].


\bibitem{Tang:2012pv}
Y.~Tang,
JHEP \textbf{08} (2012), 078
doi:10.1007/JHEP08(2012)078
[arXiv:1206.6949 [hep-ph]].


\bibitem{Arun:2014dga}
M.~T.~Arun, D.~Choudhury, A.~Das and S.~SenGupta,
Phys. Lett. B \textbf{746} (2015), 266-275
doi:10.1016/j.physletb.2015.05.008
[arXiv:1410.5591 [hep-ph]].


\bibitem{Das:2013lqa}
A.~Das and S.~SenGupta,
[arXiv:1303.2512 [hep-ph]].


\bibitem{Banerjee:2018kcz}
I.~Banerjee, S.~Chakraborty and S.~SenGupta,
Phys. Rev. D \textbf{99} (2019) no.2, 023515
doi:10.1103/PhysRevD.99.023515
[arXiv:1806.11327 [hep-th]].


\bibitem{Chakraborty:2013ipa}
S.~Chakraborty and S.~Sengupta,
Eur. Phys. J. C \textbf{74} (2014) no.9, 3045
doi:10.1140/epjc/s10052-014-3045-6
[arXiv:1306.0805 [gr-qc]].


\bibitem{Das:2007qn}
S.~Das, D.~Maity and S.~SenGupta,
JHEP \textbf{05} (2008), 042
doi:10.1088/1126-6708/2008/05/042
[arXiv:0711.1744 [hep-th]].


\bibitem{Banerjee:2017jyk}
I.~Banerjee and S.~SenGupta,
Eur. Phys. J. C \textbf{77} (2017) no.5, 277
doi:10.1140/epjc/s10052-017-4857-y
[arXiv:1705.05015 [hep-th]].


\bibitem{Paul:2016itm}
T.~Paul and S.~Sengupta,
Phys. Rev. D \textbf{93} (2016) no.8, 085035
doi:10.1103/PhysRevD.93.085035
[arXiv:1601.05564 [hep-ph]].


\bibitem{Kalb:1974yc}
M.~Kalb and P.~Ramond,
Phys. Rev. D \textbf{9} (1974), 2273-2284
doi:10.1103/PhysRevD.9.2273


\bibitem{Callan:1985ia}
C.~G.~Callan, Jr., E.~J.~Martinec, M.~J.~Perry and D.~Friedan,
Nucl. Phys. B \textbf{262} (1985), 593-609
doi:10.1016/0550-3213(85)90506-1


\bibitem{Buchbinder:2008jf}
I.~L.~Buchbinder, E.~N.~Kirillova and N.~G.~Pletnev,
Phys. Rev. D \textbf{78} (2008), 084024
doi:10.1103/PhysRevD.78.084024
[arXiv:0806.3505 [hep-th]].


\bibitem{Majumdar:1999jd}
P.~Majumdar and S.~SenGupta,
Class. Quant. Grav. \textbf{16} (1999), L89-L94
doi:10.1088/0264-9381/16/12/102
[arXiv:gr-qc/9906027 [gr-qc]].


\bibitem{Mukhopadhyaya:2002jn}
B.~Mukhopadhyaya, S.~Sen and S.~SenGupta,
Phys. Rev. Lett. \textbf{89} (2002), 121101
doi:10.1103/PhysRevLett.89.121101
[arXiv:hep-th/0204242 [hep-th]].

\bibitem{Mukhopadhyaya:2007jn}
B.~Mukhopadhyaya, S.~Sen and S.~SenGupta,
Phys. Rev. D \textbf{76} (2007), 121501
doi:10.1103/PhysRevD.76.121501
[arXiv:0709.3428 [hep-th]].


\bibitem{Das:2014asa}
A.~Das, B.~Mukhopadhyaya and S.~SenGupta,
Phys. Rev. D \textbf{90} (2014) no.10, 107901
doi:10.1103/PhysRevD.90.107901
[arXiv:1410.0814 [hep-th]].


\bibitem{Das:2010xx}
A.~Das and S.~SenGupta,
Phys. Lett. B \textbf{698} (2011), 311-318
doi:10.1016/j.physletb.2011.03.018
[arXiv:1010.2076 [hep-th]].


\bibitem{DiGrezia:2003fe}
E.~Di Grezia and S.~Esposito,
Int. J. Theor. Phys. \textbf{43} (2004), 445-456
doi:10.1023/B:IJTP.0000028877.38700.c5
[arXiv:hep-th/0304058 [hep-th]].


\bibitem{Chakraborty:2014fva}
S.~Chakraborty and S.~SenGupta,
Annals Phys. \textbf{367} (2016), 258-279
doi:10.1016/j.aop.2016.01.023
[arXiv:1412.7783 [gr-qc]].


\bibitem{Elizalde:2018rmz}
E.~Elizalde, S.~D.~Odintsov, T.~Paul and D.~Sáez-Chillón Gómez,
Phys. Rev. D \textbf{99} (2019) no.6, 063506
doi:10.1103/PhysRevD.99.063506
[arXiv:1811.02960 [gr-qc]].


\bibitem{Elizalde:2018now}
E.~Elizalde, S.~D.~Odintsov, V.~K.~Oikonomou and T.~Paul,
JCAP \textbf{02} (2019), 017
doi:10.1088/1475-7516/2019/02/017
[arXiv:1810.07711 [gr-qc]].



\bibitem{Das:2018jey}
A.~Das, T.~Paul and S.~Sengupta,
Phys. Rev. D \textbf{98} (2018) no.10, 104002
doi:10.1103/PhysRevD.98.104002
[arXiv:1804.06602 [hep-th]].


\bibitem{Paul:2018ycm}
T.~Paul and S.~SenGupta,
[arXiv:1811.05778 [gr-qc]].


\bibitem{Aashish:2020mlw}
S.~Aashish, A.~Padhy and S.~Panda,
[arXiv:2005.14673 [gr-qc]].


\bibitem{Aashish:2019ykb}
S.~Aashish and S.~Panda,
Phys. Rev. D \textbf{100} (2019) no.6, 065010
doi:10.1103/PhysRevD.100.065010
[arXiv:1903.11364 [gr-qc]].


\bibitem{Aashish:2019zsy}
S.~Aashish, A.~Padhy and S.~Panda,
Eur. Phys. J. C \textbf{79} (2019) no.9, 784
doi:10.1140/epjc/s10052-019-7308-0
[arXiv:1901.10959 [gr-qc]].


\bibitem{Aashish:2018lhv}
S.~Aashish, A.~Padhy, S.~Panda and A.~Rana,
Eur. Phys. J. C \textbf{78} (2018) no.11, 887
doi:10.1140/epjc/s10052-018-6366-z
[arXiv:1808.04315 [gr-qc]].


\bibitem{Aashish:2018aqn}
S.~Aashish and S.~Panda,
Mod. Phys. Lett. A \textbf{33} (2020) no.1, 2050087
doi:10.1142/S021773232050087X
[arXiv:1806.08194 [gr-qc]].


\bibitem{Aashish:2018jzo}
S.~Aashish and S.~Panda,
Phys. Rev. D \textbf{97} (2018) no.12, 125005
doi:10.1103/PhysRevD.97.125005
[arXiv:1803.10157 [gr-qc]].

\bibitem{Do:2020ojg}
T.~Q.~Do and W.~F.~Kao,
Phys. Rev. D \textbf{101} (2020) no.4, 044014
doi:10.1103/PhysRevD.101.044014


\bibitem{Do:2018zac}
T.~Q.~Do and W.~F.~Kao,
Eur. Phys. J. C \textbf{78} (2018) no.6, 531
doi:10.1140/epjc/s10052-018-6008-5


\bibitem{DeRisi:2007dn}
G.~De Risi,
Phys. Rev. D \textbf{77} (2008), 044030
doi:10.1103/PhysRevD.77.044030
[arXiv:0711.3781 [hep-th]].


\bibitem{DeRisi:2008qw}
G.~De Risi,
[arXiv:0805.1685 [hep-th]].


\bibitem{Hehl:1976kj}
F.~W.~Hehl, P.~Von Der Heyde, G.~D.~Kerlick and J.~M.~Nester,
Rev. Mod. Phys. \textbf{48} (1976), 393-416
doi:10.1103/RevModPhys.48.393



\bibitem{deSabbata:1994wi}
V.~de Sabbata and C.~Sivaram,
``Spin and torsion in gravitation,''
Published in: Singapore, Singapore: World Scientific (1994) 313 p,
doi:10.1142/2358.


\bibitem{Kofinas:2014owa}
G.~Kofinas and E.~N.~Saridakis,
Phys. Rev. D \textbf{90} (2014), 084044
doi:10.1103/PhysRevD.90.084044
[arXiv:1404.2249 [gr-qc]].


\bibitem{Kofinas:2014daa}
G.~Kofinas and E.~N.~Saridakis,
Phys. Rev. D \textbf{90} (2014), 084045
doi:10.1103/PhysRevD.90.084045
[arXiv:1408.0107 [gr-qc]].


\bibitem{Howe:1997pn}
P.~S.~Howe, A.~Opfermann and G.~Papadopoulos,
Commun. Math. Phys. \textbf{197} (1998), 713-727
doi:10.1007/s002200050469
[arXiv:hep-th/9710072 [hep-th]].
 
\bibitem{Howe:1996kj}
P.~S.~Howe and G.~Papadopoulos,
Phys. Lett. B \textbf{379} (1996), 80-86
doi:10.1016/0370-2693(96)00393-0
[arXiv:hep-th/9602108 [hep-th]].

 
\bibitem{Kubyshin:1993wm}
Y.~A.~Kubyshin, V.~O.~Malyshenko and D.~Marin Ricoy,
J. Math. Phys. \textbf{35} (1994), 310-320
doi:10.1063/1.530877
[arXiv:gr-qc/9304047 [gr-qc]].
 
\bibitem{German:1993bq}
G.~German, A.~Macias and O.~Obregon,
Class. Quant. Grav. \textbf{10} (1993), 1045-1053
doi:10.1088/0264-9381/10/5/021
 
\bibitem{Kar:2000ct}
S.~Kar, P.~Majumdar, S.~SenGupta and A.~Sinha,
Eur. Phys. J. C \textbf{23} (2002), 357-361
doi:10.1007/s100520100872
[arXiv:gr-qc/0006097 [gr-qc]].
 
\bibitem{Kar:2001eb}
S.~Kar, P.~Majumdar, S.~SenGupta and S.~Sur,
Class. Quant. Grav. \textbf{19} (2002), 677-688
doi:10.1088/0264-9381/19/4/304
[arXiv:hep-th/0109135 [hep-th]].
 


 
\bibitem{Paul:2020bdy}
T.~Paul and N.~Banerjee,
Class. Quant. Grav. \textbf{37} (2020) no.13, 135013
doi:10.1088/1361-6382/ab8bb9
[arXiv:2004.10111 [gr-qc]].

\bibitem{Odintsov:2017qif}
S.~D.~Odintsov, D.~Sáez-Chillón Gómez and G.~S.~Sharov,
Eur. Phys. J. C \textbf{77} (2017) no.12, 862
doi:10.1140/epjc/s10052-017-5419-z
[arXiv:1709.06800 [gr-qc]].




\bibitem{Odintsov:2017hbk}
S.~D.~Odintsov, V.~K.~Oikonomou and L.~Sebastiani,
Nucl. Phys. B \textbf{923} (2017), 608-632
doi:10.1016/j.nuclphysb.2017.08.018
[arXiv:1708.08346 [gr-qc]].



\bibitem{Hwang:2005hb}
J.~c.~Hwang and H.~Noh,
Phys.\ Rev.\ D {\bf 71} (2005) 063536
doi:10.1103/PhysRevD.71.063536
[gr-qc/0412126].



\bibitem{Noh:2001ia}
H.~Noh and J.~c.~Hwang,
Phys.\ Lett.\ B {\bf 515} (2001) 231
doi:10.1016/S0370-2693(01)00875-9
[astro-ph/0107069].


\bibitem{Hwang:2002fp}
J.~c.~Hwang and H.~Noh,
Phys.\ Rev.\ D {\bf 66} (2002) 084009
doi:10.1103/PhysRevD.66.084009
[hep-th/0206100].

\bibitem{Akrami:2018odb}
Y.~Akrami {\it et al.} [Planck Collaboration],
arXiv:1807.06211 [astro-ph.CO].


\bibitem{Odintsov:2020sqy}
S.~D.~Odintsov, V.~K.~Oikonomou and F.~P.~Fronimos,
[arXiv:2003.13724 [gr-qc]].


\bibitem{Paul:2019hys}
T.~Paul,
EPL \textbf{127} (2019) no.2, 20004
doi:10.1209/0295-5075/127/20004
[arXiv:1905.13033 [gr-qc]].


\bibitem{Nojiri:2019kkp}
S.~Nojiri, S.~D.~Odintsov and E.~N.~Saridakis,
Phys. Lett. B \textbf{797} (2019), 134829
doi:10.1016/j.physletb.2019.134829
[arXiv:1904.01345 [gr-qc]].


\bibitem{Nojiri:2020wmh}
S.~Nojiri, S.~D.~Odintsov, V.~K.~Oikonomou and T.~Paul,
[arXiv:2007.06829 [gr-qc]].


\bibitem{Ade:2015lrj}
P.~A.~R.~Ade \textit{et al.} [Planck],
Astron. Astrophys. \textbf{594} (2016), A20
doi:10.1051/0004-6361/201525898
[arXiv:1502.02114 [astro-ph.CO]].















  
  
  
  
  
  
  
  
  
  
  
  
  
  
  
  
  
  
  
  
  
  
  
  
  
  
  
  
  
  
  
  
  
  
  
  
  
  
  
  
  
  
  
  
  
  
  
  
  
  
  
  
  
  
  
  
  
  
  
  
  
  
  
  
  
  
  
  
  
  
  
  
  
  
  
  
  
  
  
  
  
  
  
  
  
  
  
  
  
  
  
  
  
  
 \end{thebibliography}
\end{document}